\documentclass[12pt]{article}

\usepackage{placeins}
\usepackage{pifont}
\usepackage{graphicx}
\usepackage{booktabs}
\usepackage{multirow}
\usepackage{amsmath}
\usepackage{amssymb}
\usepackage{lmodern}
\usepackage{rotating}
\usepackage{float}
\usepackage[T1]{fontenc}
\usepackage[a4paper,left=2.5cm,right=2.5cm,top=3cm,bottom=3cm]{geometry}
\usepackage[flushleft]{threeparttable}

\usepackage[style=authoryear-comp, backend=bibtex, maxcitenames=2, maxbibnames=99 ,dashed=false]{biblatex}
\bibliography{paper_contplot}

\usepackage{hyperref}

\newcommand{\bigCI}{\mathrel{\text{\scalebox{1.07}{$\perp\mkern-10mu\perp$}}}}

\begin{document}

\title{Visualizing the (Causal) Effect of a Continuous Variable on a Time-To-Event Outcome}
\date{}
\author{Robin Denz and Nina Timmesfeld \\ \\ Ruhr-University Bochum \\ Department of Medical Informatics, Biometry and Epidemiology}

\maketitle

\begin{abstract}
	Visualization is a key aspect of communicating the results of any study aiming to estimate causal effects. In studies with time-to-event outcomes, the most popular visualization approach is depicting survival curves stratified by the variable of interest. This approach cannot be used when the variable of interest is continuous. Simple workarounds, such as categorizing the continuous covariate and plotting survival curves for each category, can result in misleading depictions of the main effects. Instead, we propose a new graphic, the \emph{survival area plot}, to directly depict the survival probability over time and as a function of a continuous covariate simultaneously. This plot utilizes g-computation based on a suitable time-to-event model to obtain the relevant estimates. Through the use of g-computation, those estimates can be adjusted for confounding without additional effort, allowing a causal interpretation under the standard causal identifiability assumptions. If those assumptions are not met, the proposed plot may still be used to depict non-causal associations. We illustrate and compare the proposed graphics to simpler alternatives using data from a large German observational study investigating the effect of the Ankle Brachial Index on survival. To facilitate the usage of these plots, we additionally developed the contsurvplot R-package which includes all methods discussed in this paper.
\end{abstract}

	
	\section{Introduction} \label{chap::introduction}
	
	Visualization is a crucial aspect when reporting the main results of any study. When the aim of the study is to estimate a causal effect of one variable on a specific endpoint, a clear graphical display of the effect can make the results more accessible for the reader \parencite{Zipkin2014}. With a time-to-event endpoint, such as time until recurrence of a disease or death, the most popular approach is to draw Kaplan-Meier survival curves, or confounder-adjusted alternatives, for each group of interest \parencite{Denz2023}. This, however, only works if there is a discrete number of groups. In reality, the variable of interest may be continuous, making the use of Kaplan-Meier survival curves impossible. A few examples for continuous covariates that are frequently of interest are income, duration of exercise or tissue size \parencite{Sachs2020, Wasfy2016}.
	\par\medskip
	As a workaround, researchers often resort to either (1) performing artificial categorization in both the main analysis and the graphical displays, or (2) performing categorization only for the graphics while using the full continuous variable in the main analysis. Dividing the variable into artificial groups at arbitrary or ``optimal'' values \parencite{Mazumdar2000, Giannoni2014} has the advantage that survival curves can be created. It is, however, problematic due to a loss of information, the seemingly arbitrary choice of cut-off points and the improbable assumption of homogeneity of risk inside the categories \parencite{Naggara2011, Bennette2012, Altman2014}. In case (1) the statistical power of the analysis might be reduced and or bias might be introduced. In case (2), the survival curves might show a misleading picture of the treatment effect, which may also not be consistent with the results of the main analysis because different scales were used \parencite{Mazumdar2000}.
	\par\medskip
	Graphics designed for displaying the effect of a continuous covariate on a time-to-event outcome exist, but are rarely used in practice \parencite{Karvanen2009, MeiraMachado2013, Eng2015, Shen2017, Smith2019, Jackson2021}. Most of these approaches fail to represent the treatment effect over time and as a function of the continuous variable simultaneously. Some rely on summary statistics, such as landmark survival probabilities or restricted mean survival times \parencite{Shen2017, Eng2015}, while others rely on rather complex estimation strategies that may not use easily interpretable absolute measures of effect \parencite{Karvanen2009, MeiraMachado2013, Smith2019, Jackson2021}. Furthermore, most of the proposed graphics cannot be adjusted for confounders and can therefore only be used to display causal effects in the absence of confounding. This is rarely the case when the treatment is continuous, because continuous variables are usually not randomly assigned.
	\par\medskip
	In this article, we propose a new type of graphic, the \emph{survival area plot}, that does not suffer from these problems. The survival area plot directly extends the standard Kaplan-Meier plot to the continuous case, by depicting the survival probability as a function of time and a continuous covariate through the use of a colour scaled area. We use g-computation \parencite{Robins1986} to obtain the required probability estimates, allowing the plot to be adjusted for confounders. Under the standard causal identifiability assumptions \parencite{Naimi2015, Hernan2020}, the resulting survival probabilities can be interpreted as counterfactual quantities, resulting in a display of the causal effect of the continuous variable on the time-to-event outcome. To facilitate the use of this method in practice, we also developed the R-package \texttt{contsurvplot} \parencite{Denz2022a}, containing an implementation of the proposed and alternative methods in the R programming language.
	\par\medskip
	First, we give a formal description of the target estimand and the estimation methodology. Next, we introduce the survival area plot and describe alternative ways to visualize the g-computation based estimates. Afterwards, examples of the graphics are presented using data from the German Epidemiological Trial on Ankle Brachial Index \parencite{StudyGroup2002}. Using these examples, we compare the different approaches and justify our practical recommendations.
	
	\section{Proposed Method}
	
	\subsection{Target Estimand}
	
	Before going further, it is crucial to define what ``the causal effect of a continuous variable on a time-to-event outcome'' is. We will first discuss the target estimand in the simpler case of a dichotomous variable. Let $Z$ denote the variable of interest with $Z \in \{0, 1\}$. Let $T$ be the time until the occurrence of an event. In this setup, each individual has two potential failure times: $T^{Z=0}$ being the time until the event occurs under treatment-regime $Z = 0$ and $T^{Z=1}$ being the failure time under $Z = 1$. Our goal is to estimate the population-level counterfactual survival function under both $Z = 0$ and $Z = 1$, defined as:
	
	\begin{equation} \label{eqn::target_estimand}
		S_z(t) = E\left(I\left(T^{Z=z} > t\right)\right).
	\end{equation}
	
	This is the survival probability that would have been observed at time $t$ if the value of $Z$ of all individuals in the population of interest had been set to $z$ (\emph{counterfactual survival probability}) \parencite{Denz2023}. If $Z$ was randomized and the study was performed adequately, this is what a Kaplan-Meier estimator \parencite{Kaplan1958} estimates. The difference between treatment-specific counterfactual quantities ($S_1(t) - S_0(t)$) or their ratio ($S_1(t) / S_0(t)$) may be used to define average causal effects \parencite{Wang2019}.
	\par\medskip
	It is straightforward to generalize this notation to the continuous case \parencite{Galagate2016}. Let $Z$ be a continuous variable that can take any real values $Z \subseteq \mathbb{R}$. Instead of two potential failure times there is now an uncountably infinite set of potential failure times $T^{Z=z}$ for each individual, but equation~\ref{eqn::target_estimand} and its interpretation remain unchanged. Average causal effects may still be defined using differences or ratios, as explained in the e-appendix. Note that this counterfactual quantity is only meaningful if $Z$ corresponds to a well-defined treatment, which may not always be the case with continuous $Z$ \parencite{Hernan2016, Dong2022, VanderWeele2018}.
	
	\subsection{Estimation using G-Computation}
	
	Suppose that setting $Z$ to $z$ corresponds to a well-defined intervention. If $Z$ was randomly assigned, estimating $S_z(t)$ is ``only'' an issue of interpolation. As mentioned above if $Z$ was dichotomous and appropriately randomized, we could obtain an unbiased estimate of $S_z(t)$ using a standard stratified Kaplan-Meier estimator \parencite{Kaplan1958}. This, however, does not work with a continuous $Z$, because of its infinite amount of possible values. A suitable model is necessary to estimate this function. A popular alternative is the Cox proportional hazards regression model \parencite{Cox1972}, given by:
	
	\begin{equation}
		h(t) = h_0(t)\exp(\beta Z),
	\end{equation}
	
	where $\beta$ is the coefficient of the continuous variable and $h_0(t)$ is the baseline-hazard function, which can be estimated from the data \parencite{Breslow1972}. This model can be used to predict the survival probability of an individual with a specific value of $Z$ at a given point in time $t$ using:
	
	\begin{equation}
		S(t|Z=z) = \exp\left(-\int_{0}^{t}h_0(t)e^{\beta Z} dt\right).
	\end{equation}
	
	If $Z$ was randomized and the required assumptions for the Cox model hold, using a Cox model that includes only $Z$ as an independent variable and an estimate of $h_0(t)$ is enough to obtain an unbiased estimate for $S_z(t)$. Without randomization, we also need to take confounding into account. This can be done by including a sufficient set of confounders in the Cox model as additional independent variables (with an appropriate functional form) and using this model to perform \emph{regression standardization}, also known as \emph{g-computation} \parencite{Robins1986, Keil2014}. Here, the value for $Z$ is set to $z$ for each person in the sample. The survival probability at time $t$ is then estimated for each individual person by using the model and the observed values of the included confounders. The mean of these person-specific estimates in a sample of size $n$ is an unbiased estimator of $S_z(t)$. Formally, this can be defined as:
	
	\begin{equation}
		\hat{S}_z(t) = \frac{1}{n} \sum_{i = 1}^{n} S(t|Z=z, X=x_i),
	\end{equation}
	
	where $X$ is a vector of confounders and $x_i$ is the realization of this vector for the $i$-th person. $S(t|Z=z, X=x_i)$ corresponds to equation (3), but including a set of confounders $X$. This method has been used extensively when analyzing time-to-event data \parencite{Makuch1982, Chang1982}. Note that it is not necessary to use a Cox model. Any model capable of making predictions of $S(t|Z=z, X=x_i)$ may be used instead. Possible alternatives include additive hazards models \parencite{Aalen1980}, parametric survival time models \parencite{Jackson2016} or random survival forests \parencite{Ishwaran2008}.
	\par\medskip
	Four fundamental assumptions have to be met to allow a causal interpretation of the obtained values. Firstly, the \emph{no interference assumption} states that the potential survival time of an individual is independent of the value of $Z$ of other individuals \parencite{Naimi2015}. Secondly, \emph{counterfactual consistency} posits that the potential survival time would remain the same under $Z$ regardless of whether $Z$ was set experimentally or not ($P(T = t | Z = z) = P(T^{Z=z} = t | Z = z)$) \parencite{Hernan2016, Pearl2018}. Thirdly, the \emph{conditional exchangeability assumption} asserts that the treatment groups are exchangeable, given a sufficient set of relevant confounders, i.e., $T^{Z=z} \bigCI Z | X$ \parencite{Sarvet2020}. Lastly, the \emph{positivity assumption} states that every individual has a non-zero probability of receiving any value of $Z$ ($P(Z = z | X) > 0$ for all $Z \subseteq \mathbb{R}$). Although it is impossible to observe individuals for every value of $Z$ due to the uncountably infinite number of possible values, this \emph{practical violation} is generally not an issue because of the interpolation performed by the time-to-event model \parencite{Westreich2010, Keil2014, Hernan2020}.
	\par\medskip
	In addition any model-specific assumptions, such as the proportional hazards assumption when using a Cox model \parencite{Cox1972}, also have to be met. This includes the correct specification of the functional form of all included variables and the proper treatment of censored observations. Throughout this article, we assume random right-censoring. A discussion about the implications of dependent censoring is given in the e-appendix. If any of these assumptions are violated, the resulting estimates cannot be interpreted as counterfactual survival probabilities. However, the graphics proposed in this article may still be used to visualize the non-causal marginal association of the continuous covariate and the time-to-event endpoint. This may be useful to communicate results of prediction models, especially when using machine learning methods that are otherwise hard to interpret \parencite{Ishwaran2008, Petch2022}.
	
	\subsection{Computational Details}
	
	Our implementation \parencite{Denz2022a} of g-computation works by first setting $Z$ to $z$ for all individuals included in the sample. Afterwards, the time-to-event model is used to obtain predictions of $S(t|Z=z, X=x_i)$ for each individual at $t$. Internally, the fast \texttt{predictRisk} function from the \texttt{riskRegression} R-package \parencite{Ozenne2017} is used at this step. Finally, those predictions are averaged for each considered $t$. If multiple values of $Z$ are supplied, this procedure is simply repeated. This algorithm can be slow when considering many values of $t$ and $Z$ simultaneously. Nevertheless, it still executes in reasonable time ($< 1$min) for medium sized datasets ($n < 5000$) on a regular computer. The runtime of this procedure may be decreased by executing the computations for different values of $Z$ on multiple cores simultaneously.
	
	\subsection{Visualization of the Results}
	
	Performing the procedure described above for multiple $t$ and several values for $Z$ results in a three-dimensional survival surface, which can be used to create multiple plots based on summary statistics. Most of these plot have already been described in the literature \parencite{Yang2016, Shen2017, Eng2015, Jackson2021}, but the authors introducing them did not use g-computation to obtain the underlying estimates. By using g-computation instead, the values represented by the graphs can be interpreted as counterfactual quantities, if the previously mentioned assumptions hold.
	\par\medskip
	The simplest of these is a \emph{landmark survival probability plot} \parencite{Yang2016, Shen2017}. To create this plot, one or multiple $t$ are chosen. The survival probability at $t$ is then visualized as a function of the continuous covariate. A \emph{survival time quantile plot} follows the same structure, but replaces the landmark survival probability with a chosen survival time quantile. The same type of plot can also be created using the \emph{restricted mean survival time} \parencite{Eng2015}. Another alternative would be to plot standard survival curves, for specific values of $Z$ (\emph{value specific survival curves}) \parencite{Eng2015}.
	\par\medskip
	All of these techniques reduce the surface to two dimensions. Although this can be an effective strategy in some situations, we argue that the most appropriate visualization technique is to represent the whole survival surface instead. It could be argued that the most natural way to present a three-dimensional surface would be a three-dimensional plot \parencite{Smith2019}, but this is not what we advocate for. Many researchers have pointed out the shortcomings of non-interactive three-dimensional plots before \parencite{Sanftmann2012, Wilke2019}. Those include the difficulty in reading off specific values of the surface and the seemingly inconsistent look of the graphic when viewed from different angles \parencite{Wilke2019}.
	\par\medskip
	Instead, the surface can be represented in a two-dimensional plot using color scales. We advocate for two options: \emph{survival contour plots} first introduced by \textcite{Jackson2021} and our own proposal, \emph{survival area plots}. Survival contour plots display the time on the $x$-axis and the continuous covariate on the $y$-axis. The survival probability is categorized and the corresponding area is colored accordingly. The original version proposed by \textcite{Jackson2021} used a kernel-density based estimation strategy, which, contrary to our g-computation based approach, does not allow confounder-adjustment.
	\par\medskip
	Survival area plots keep the structure of Kaplan-Meier plots with $t$ on the $x$-axis and the survival probability on the $y$-axis. Instead of single survival curves for some categories, an area is drawn with the filled color corresponding to values of $Z$. This can be a continuously colored area or divided into bins similar to the survival contour plot. Essentially, survival area plots depict all possible \emph{value specific survival curves} in a given range in the same graphic. Since those change continuously over changing values of $Z$, they create an area instead of discernible lines. Therefore, every infinitely thin part of this area is equal to one specific estimate of $S_z(t)$ over the depicted time range. Because those individual parts can be interpreted as the survival probability that would have been observed if $Z$ was set to $z$ in the target population, the area depicts how this probability changes with changing values of $Z$.
	
	\section{Illustrative Examples}
	
	To illustrate our proposed methodology, we use data from the German Epidemiological Trial on Ankle Brachial Index (getABI) \parencite{StudyGroup2002}. The getABI study is a prospective observational study including 6880 primary care patients aged 65 or older who were followed for up to 8 years. An initial examination was performed for every patient in the beginning of the study in 2001. In this examination, both the Ankle-Brachial-Index (ABI) and other clinically relevant factors, such as past cardiovascular events and socio-economic factors, were collected. The ABI was of special interest in this study, because it is a frequently used marker to diagnose peripheral artery disease (PAD). It is calculated as ``the ratio of the higher of the two systolic pressures (tibial posterior and anterior artery) above the ankle to the average of the right and left brachial artery pressures'' \parencite{Diehm2004}.
	\par\medskip
	We want to estimate the fraction of patients that would have survived between 0 and 8 years if the ABI of every member in the target population had been set to values between 0.35 and 1.5. Because the ABI cannot be ``set'' to a specific value in reality, this is a hypothetical intervention. It is also somewhat imprecise, because it does not refer to any duration of time in which the ABI stays at that level. Although some imprecision in the definition is inevitable \parencite{Hernan2016, VanderWeele2018} it is still important consider these limitations when interpreting the results. Our target population consists of all German primary care patients aged 65 or older with a prior cardiovascular event at baseline. In accord to previous studies, we also only consider patients with an ABI of $<= 1.5$ at baseline, because an ABI $> 1.5$ indicates incompressible arteries \parencite{Meijer1998, McDermott2002}. Applying these restrictions to the getABI data results in a sample size of 508 patients with 143 events.
	\par\medskip
	In practice, the ABI is often categorized, indicating different levels of PAD. An ABI $< 0.9$ is usually considered as ``low'' and an indication of asymptomatic PAD \parencite{Diehm2004, Miguel2017, Krolczyk2022}. Some authors consider an ABI between $0.9$ and $1.3$ as ``normal'' \parencite{Miguel2017}, while others extend this category up to an ABI $1.4$ or $1.5$ \parencite{Diehm2004, Krolczyk2022}. These categorizations are frequently used when analyzing the effect of the ABI on the survival time, which, as stated previously, can lead to a loss of statistical power and might even result in misleading conclusions \parencite{Mazumdar2000}.
	\par\medskip
	To illustrate the difference between the classical and our proposed methods, we first fitted five Cox models. All of these models use death of all causes as response variable, but utilize different specifications of the ABI. The first model includes the ABI as a categorical variable and the relevant confounders age, sex and smoking status. The second one contains only the ABI as independent continuous covariate. The third one adds the relevant confounders to model two. Model (4) and (5) are identical to models (2) and (3), with the only difference being that ABI was modeled using B-Splines \parencite{Perperoglou2019}, allowing a non-linear fit.
	\par\medskip
	
	\begin{table}[!htbp] \centering
		\caption{Hazard Ratios and associated 95\% confidence intervals of different Cox proportional hazards regression models describing the relationship of the ABI on the survival time.}
		\label{tab::models}
		\resizebox{\textwidth}{!}{\begin{tabular}{@{\extracolsep{5pt}}lccccc} 
				\\[-1.8ex]\hline 
				\hline \\[-1.8ex] 
				& \multicolumn{5}{c}{\textit{Dependent variable:}} \\ 
				\cline{2-6} 
				\\[-1.8ex] & \multicolumn{5}{c}{Death of all causes} \\ 
				\\[-1.8ex] & (1) & (2) & (3) & (4) & (5)\\ 
				\hline \\[-1.8ex]
				Age & 1.091 &  & 1.091 &  & 1.089 \\ 
				& (1.056, 1.127) &  & (1.056, 1.127) &  & (1.054, 1.125) \\ 
				& & & & & \\ 
				Sex & 0.988 &  & 1.009 &  & 0.983 \\ 
				& (0.661, 1.477) &  & (0.676, 1.507) &  & (0.657, 1.469) \\ 
				& & & & & \\ 
				Ever Smoker & 2.202 &  & 2.208 &  & 2.225 \\ 
				& (1.410, 3.440) &  & (1.411, 3.454) &  & (1.424, 3.476) \\ 
				& & & & & \\
				ABI (0.5,0.7] & 0.705 &  &  &  &  \\ 
				& (0.298, 1.667) &  &  &  &  \\ 
				& & & & & \\ 
				ABI (0.7,0.9] & 0.524 &  &  &  &  \\ 
				& (0.243, 1.131) &  &  &  &  \\ 
				& & & & & \\ 
				ABI (0.9,1.1] & 0.366 &  &  &  &  \\ 
				& (0.173, 0.773) &  &  &  &  \\ 
				& & & & & \\ 
				ABI (1.1,1.5] & 0.364 &  &  &  &  \\ 
				& (0.167, 0.793) &  &  &  &  \\ 
				& & & & & \\  
				ABI (per 0.1 points) &  & 0.867 & 0.894 &  &  \\ 
				&  & (0.802, 0.937) & (0.826, 0.967) &  &  \\ 
				& & & & & \\ 
				bs(ABI, df = 3)1 &  &  &  & 0.402 & 0.286 \\ 
				&  &  &  & (0.037, 4.343) & (0.026, 3.148) \\ 
				& & & & & \\ 
				bs(ABI, df = 3)2 &  &  &  & 0.098 & 0.152 \\ 
				&  &  &  & (0.022, 0.430) & (0.036, 0.653) \\ 
				& & & & & \\ 
				bs(ABI, df = 3)3 &  &  &  & 0.380 & 0.378 \\ 
				&  &  &  & (0.066, 2.176) & (0.069, 2.082) \\ 
				& & & & & \\ 
				\hline \\[-1.8ex] 
				Observations & 508 & 508 & 508 & 508 & 508 \\ 
				R$^{2}$ & 0.101 & 0.024 & 0.095 & 0.032 & 0.102 \\ 
				Max. Possible R$^{2}$ & 0.966 & 0.966 & 0.966 & 0.966 & 0.966 \\ 
				Log Likelihood & $-$831.596 & $-$852.444 & $-$833.311 & $-$850.397 & $-$831.343 \\ 
				\hline 
				\hline \\[-1.8ex] 
		\end{tabular}}
	\end{table}
	
	The hazard ratios of all models are displayed in table~\ref{tab::models}. Note that these hazard ratios can generally not be endowed with a causal interpretation \parencite{Hernan2010, Aalen2015}. However, it is still apparent that the ABI is associated with the survival time, regardless of the model specification used. One can also see that once age, sex, and smoking status are controlled for, the association becomes smaller. While we can infer these trends from the hazard ratios, it is still difficult to accurately interpret the results, especially when B-Splines were used to model the ABI. Graphical displays of the survival probabilities are necessary to enhance understanding here.
	\par
\begin{figure}[!htb]
		\centering
		\includegraphics[width=1\linewidth]{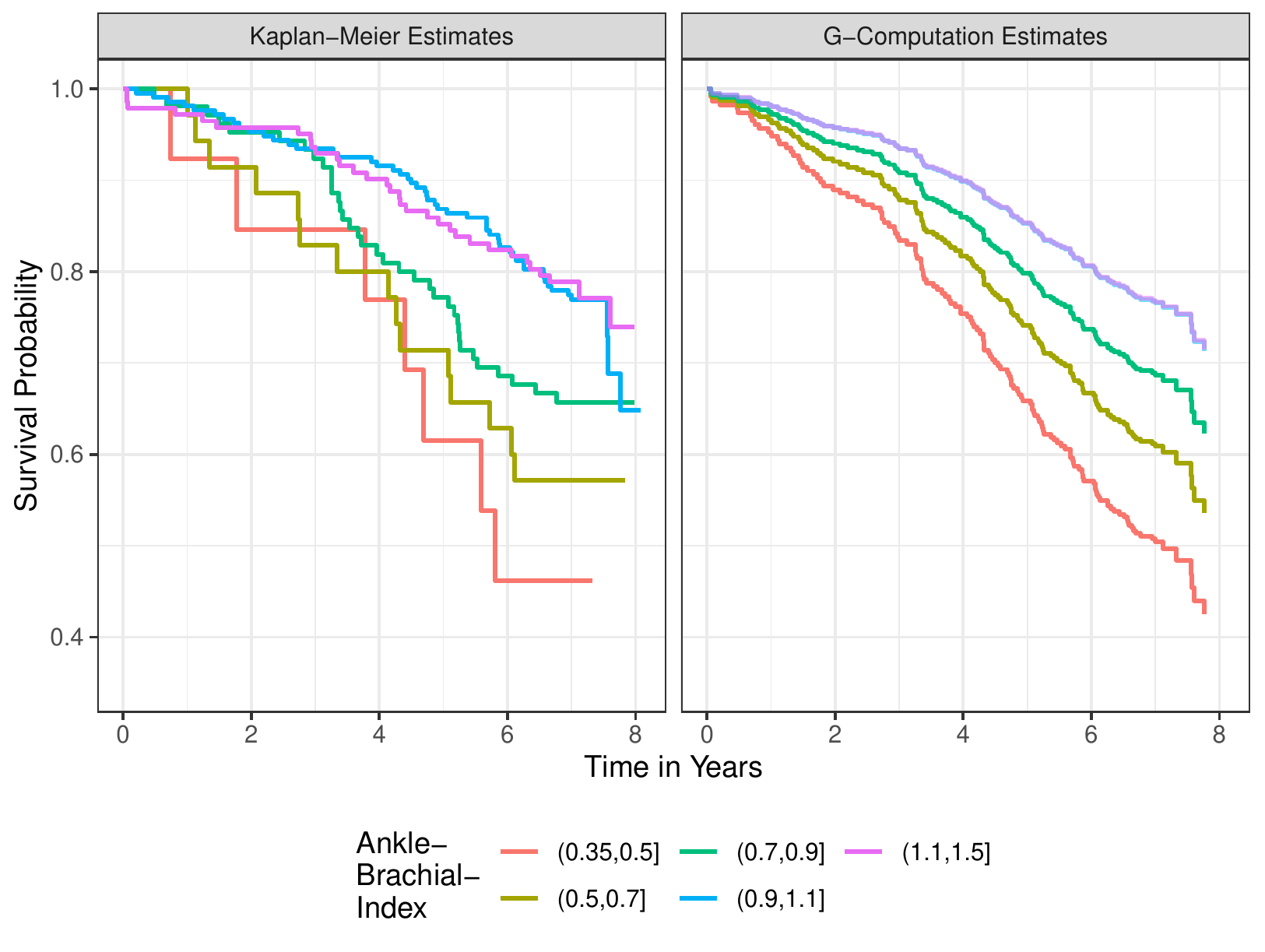}
		\caption{Survival curves stratified by categories of the ABI. Results in the left panel were obtained using a stratified Kaplan-Meier estimator while the results in the right panel were obtained using g-computation and model~(1) displayed in table~1. The survival curves for the highest two categories are displayed as transparent in the right panel, to make it more obvious that those are two curves.}
		\label{fig::surv_categories}
	\end{figure}

\medskip
	Figure~\ref{fig::surv_categories} displays survival curves stratified by the ABI categories defined by \textcite{Diehm2004}. The left panel shows simple Kaplan-Meier estimates, stratified by the ABI categories, while the right panel shows survival curves adjusted for sex, age and smoking status. To adjust these curves, we used model~(1) to perform g-computation. A clear dose-response effect can be seen, regardless of whether the curves were adjusted for confounders or not. The category with the lowest ABI values has the lowest survival probability, while the categories with higher ABI values have a higher survival probability. In the adjusted graph, there is no visible difference between the survival probability of the two upper categories. Looking only at Figure~\ref{fig::surv_categories} or model~1, it remains unclear whether the survival probability remains constant with increasing ABI values after a certain point or if there is a non-linear relationship. It also implies that the risk associated with an ABI of 0.7 is equal to the risk associated with an ABI of 0.9, due to the assumed homogeneity inside each category \parencite{Bennette2012}.
	\par\medskip
	
\begin{figure}[!htb]
		\centering
		\includegraphics[width=1\linewidth]{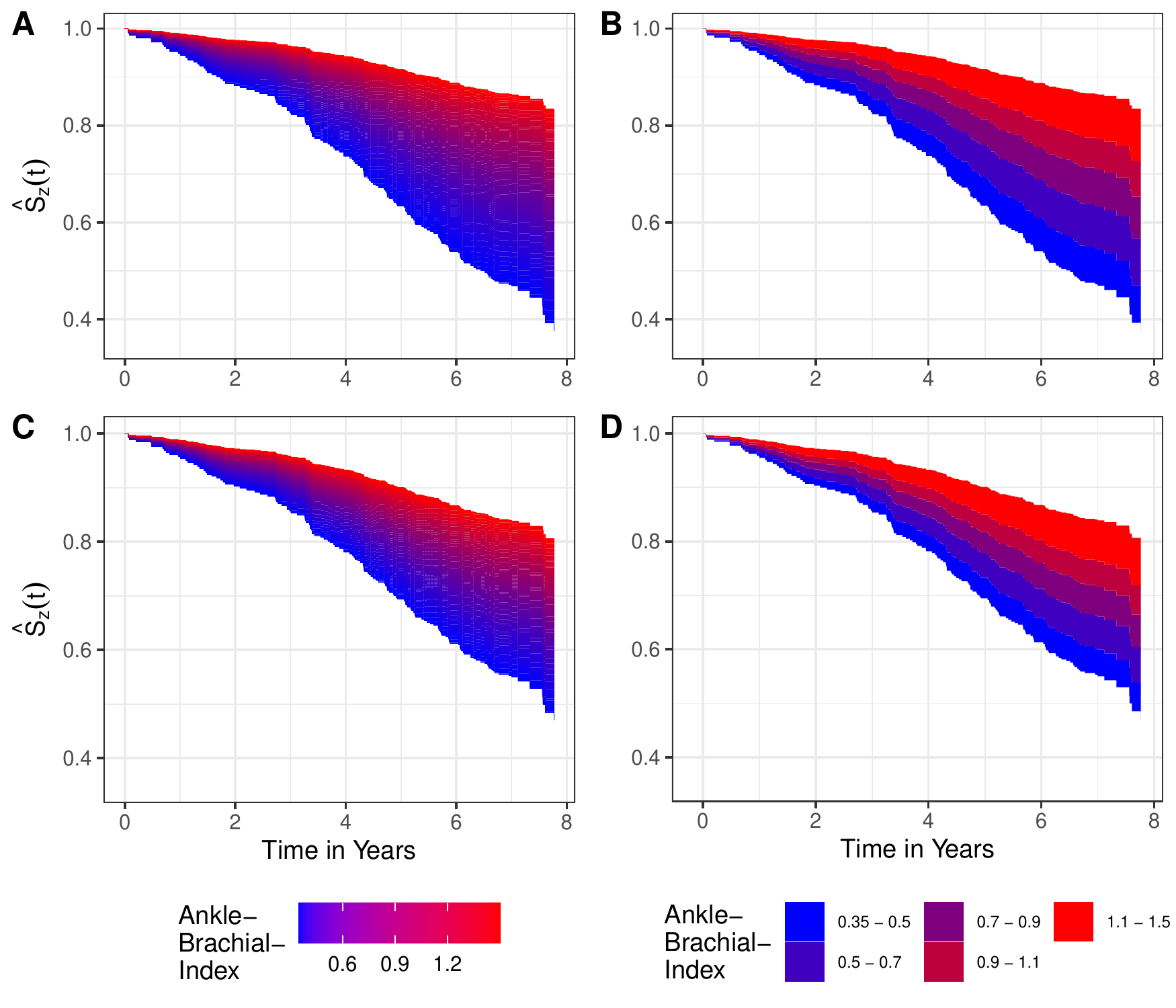}
		\caption{Continuously and discretely colored survival area plots displaying different estimates of the causal effect of the ABI on survival, created using different Cox proportional hazards regression models and g-computation. \\ \textbf{A}: Using model 1 (unadjusted), continuous color shading; \textbf{B}: Using model 1 (unadjusted), discrete color shading; \textbf{C}: Using model 2 (adjusted), continuous color shading; \textbf{D}: Using model 2 (adjusted), discrete color shading.}
		\label{fig::surv_area}
	\end{figure}

	Instead of using the categorical version of the ABI variable, we can also use model (2) and (3) to create survival area plots. Figure~\ref{fig::surv_area} displays two kinds of these plots. On the left side a continuous color scale is used to represent the ABI values, while the right side shows the same plots using a discrete color scale. The version on the left highlights the continuous nature of the causal effect. It is, however, difficult to discern survival probabilities for specific values of the ABI. The right version explicitly shows the range in which the survival probability for the respective categories is in. We argue that both versions give a more accurate depiction of the underlying treatment effect than the simple categorized survival curves, because they correctly depict the change of the survival probability with changing ABI values instead of showing homogeneous probabilities in some categories.
	\par\medskip
	
	In models (2) and (3), we assumed that the ABI has a linear effect on the log scale. The categorized plot shows some evidence that this assumption may be wrong when the ABI values are high ($> 1.1$), which is why we fitted models (4) and (5) where we used B-Splines to model the ABI. Unfortunately, we cannot use these models to create standard survival area plots, because they cannot depict non-monotonic effects. If there is a curved relationship, the corresponding areas would lie on top of each other, rendering the graphic unusable. The only way to still use survival area plots in this case is to divide the plot into multiple facets, where each facet contains only a part of the entire surface. We discuss this in detail in the e-appendix.
	\par
\begin{figure}[!htb]
		\centering
		\includegraphics[width=1\linewidth]{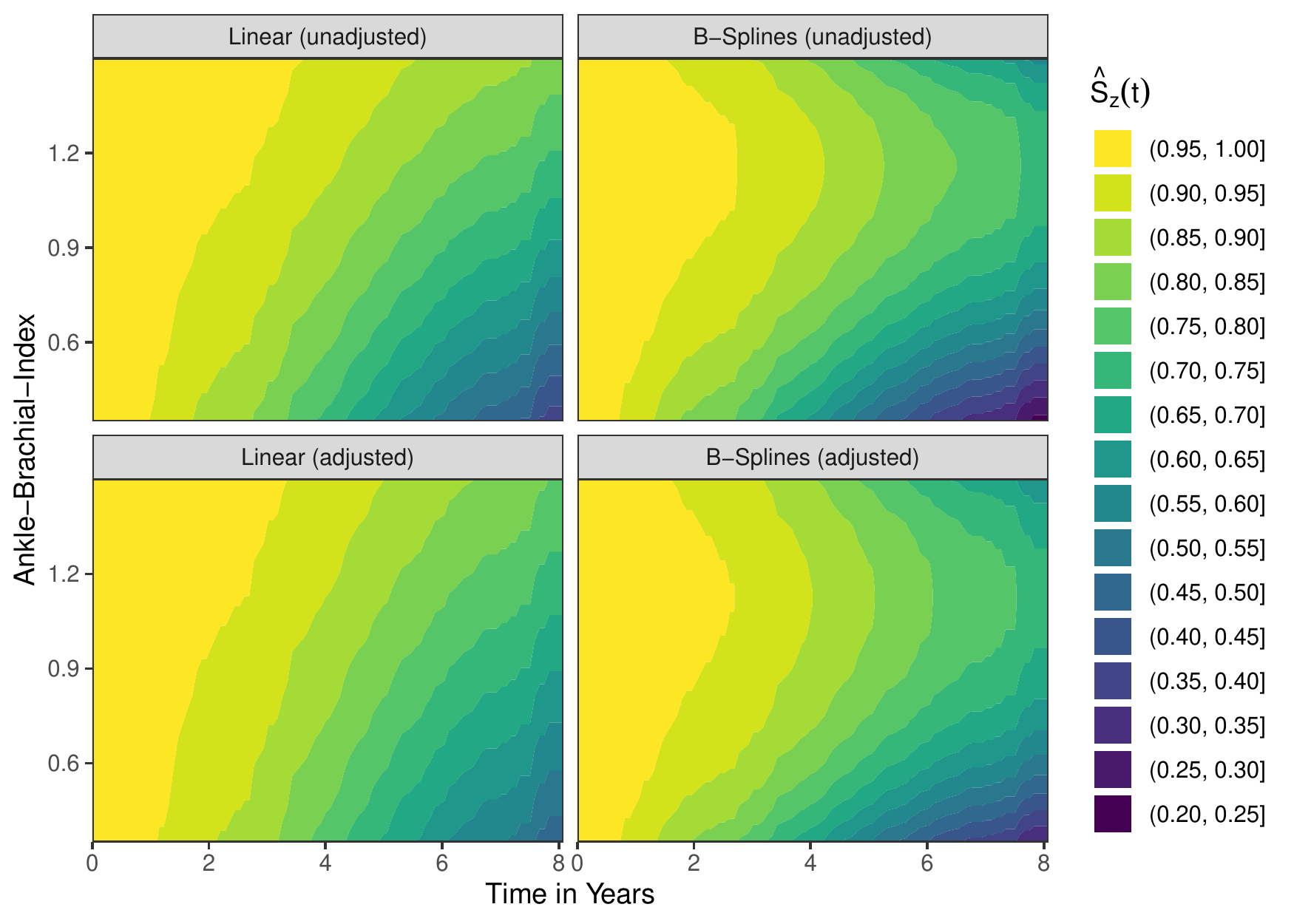}
		\caption{Survival contour plots displaying the causal effect of the ABI on survival, created using four different Cox proportional hazards regression models and g-computation.}
		\label{fig::surv_contour}
	\end{figure}

\medskip
	Instead, survival contour plots can be used. Figure~\ref{fig::surv_contour} shows the corresponding plots for models (2) to (5). As before, there are only small differences between the adjusted and the unadjusted analysis. However, a clear difference can be seen when comparing the survival contour plots created using the linearity assumption with the ones that were created using B-spline models. There seems to be a non-linear relationship, where the survival probability decreases again with ABI values higher than approximately 1. It is unlikely that this is an artifact due to the population size, because there are 290 patients corresponding to 68 events, with an ABI $> 1$.
	
	\FloatBarrier
	
	\section{Discussion}
	
	In this article, we proposed the \emph{survival area plot}, which is a novel method to visualize the effect of a continuous variable on a time-to-event outcome. Additionally, we extended existing methods, such as \emph{survival contour plots} \parencite{Jackson2021} to allow confounder-adjustment through the use of g-computation. We illustrated these methods using data from the German epidemiological trial on Ankle Brachical Index \parencite{StudyGroup2002} and compared them with the standard method of categorized survival curves. Both survival area plots and survival contour plots correctly display the estimated counterfactual survival probability as a function of the continuous variable and time simultaneously. They avoid the problems associated with categorization of continuous variables, such as loss of statistical power, bias and the assumption of homogeneity in intervals \parencite{Altman2014, Bennette2012}. The survival area plot has the advantage that it keeps the well known Kaplan-Meier plot structure and might therefore be easier to understand. It is, however, difficult to draw this kind of plot if the effect of the continuous variable is not monotonic. Survival contour plots do not share this disadvantage, allowing any kind of non-linear effect.
	\par\medskip
	Using g-computation to obtain the required estimates is both a strength and a weakness for multiple reasons. First, it can be considered a strength because it is widely used, relatively simple to understand and implement. For most clinical researcher it is also straightforward to include into their data analysis workflow, as they are already using regression models such as the Cox model. Additionally, it has been shown to be one of the most efficient methods when used properly \parencite{Chatton2020, Denz2023}. However, the method requires that a valid model can be fit to the data, which might be difficult or even impossible in some situations.
	\par\medskip
	Like other methods used for causal inference, untestable assumptions have to be made in order to allow a counterfactual interpretation of the estimates. Among those is the assumption that the variable of interest constitutes a ``well-defined treatment'', which may be violated for some continuous variables such as the Body-Mass-Index. \parencite{VanderWeele2018, Dong2022, Gebremariam2021} This is, however, not always the case. Many variables, such as monthly income or duration of exercise, are both directly manipulable and clearly well-defined treatments. A critical case-by-case assessment of the relevant assumptions is necessary to allow a causal interpretation of the resulting estimates and subsequently of our proposed graphics. If any of the assumptions seem unrealistic, the obtained estimates can only be considered to be marginal associations.
	\par\medskip
	Another limitation of area-based plots is that there is no straightforward way to display uncertainty in them. In standard Kaplan-Meier plots, confidence intervals can be added to give an estimate of the statistical uncertainty of the estimates. This is impossible in graphics based an colored areas. Plots based on summary statistics might be preferable when the uncertainty of the estimates is high. Nevertheless, we still recommend basing these plots on g-computation when possible. This can be done using our \texttt{contsurvplot} R-package \parencite{Denz2022a}, which contains implementations for all plots mentioned in this article.
	\par\medskip
	We only considered a time-to-event scenario with a single type of event in this article. Left-truncation or interval-censoring can be important considerations in practice \parencite{Kleinbaum2012}. It is straightforward to use our proposed methods even in these or other scenarios. All that has to be done is to fit a suitable model which can be used to perform the required g-computation step. That model may include any of the features mentioned. Extending the proposed methodology to the context of time-varying treatments or settings with competing events of interest is more complex, because it would require a different target estimand \parencite{Young2020}. Further work is necessary to extend the proposed methods to those scenarios.
	\par\medskip
	Furthermore, g-computation is only one of many methods to estimate counterfactual survival probabilities \parencite{Denz2023}. It is a single robust method, which means that it relies on the outcome model to obtain unbiased results. To create the plots described in this paper, it is not necessary to rely on this particular estimation method. Any method may be used, as long as it allows estimation of the defined target estimand. Different authors recommend using doubly-robust methods, such as Augmented-Inverse-Probability of Treatment Weighting \parencite{Ozenne2020} or Targeted Maximum Likelihood Estimation \parencite{Cai2020} for similarly defined estimands. Those methods rely on both an outcome model and a treatment-assignment model and produce unbiased results whenever at least one of these models is correctly specified. Using these methods would be a promising alternative to g-computation. However, this is currently not possible, because they do not support continuous exposures. Further research is necessary to extent those estimators to this special case.

\section*{Acknowledgments}

We would like to thank Prof. Dr. Andreas Stang, Marianne Tokic and Prof. Dr. Hans J. Trampisch for their helpful comments and suggestions. Additionally, we want to thank the anonymous reviewers for their extensive comments and suggestions which greatly improved the manuscript and the associated R-package.

\FloatBarrier
\newpage

\printbibliography

\appendix

\newpage

\section{Alternative Graphics}

As discussed in the main text, the three-dimensional survival surface estimated using g-computation can be visualized in many different ways. The manuscript focuses on the \emph{survival area plot} and the \emph{survival contour plot}. Below we give a short description of the other possibilities and illustrate them, again using data from the getABI study. Table~\ref{tab::plots_overview} gives a concise listing of the most important properties of the different visualization techniques. Because some of the plots are based on summary statistics, we first define those quantities.
\par\medskip
Using the same notation as in the main text, we can define the \emph{counterfactual survival time quantile} as:

\begin{equation}
	Q_z(p) = min\left(t | \hat{S}_z(t) \leq p\right),
\end{equation}

where $p$ is some value between 0 and 1, specifying the quantile of interest. This quantity can be interpreted as the time until $p \cdot 100$ percent of all individuals in the target-population would have died if they had received treatment $Z = z$.
\par\medskip
The \emph{counterfactual restricted mean survival time} on the other hand can be defined as:

\begin{equation}
	RMST_{z}(\lambda) = \int_{0}^{\lambda} \hat{S}_z(t)dt,
\end{equation}

where $\lambda$ is some value greater than 0. It can be interpreted as the average time free from an event until time point $\lambda$ that would be observed if all individuals in the target population had received treatment $Z = z$.

\begin{table}[!htb]
	\centering
	\caption{Properties of different visualization techniques based on estimates of $\hat{S}_z(t)$.}
	\medskip
	\label{tab::plots_overview}
	\begin{tabular}{lccclll}
		\toprule
		\multirow{3}{*}{Plot} & \multirow{3}{*}{$x$-axis} & \multirow{3}{*}{$y$-axis} & \multirow{3}{*}{$z$-axis} & \multirow{3}{*}{Color} & Allows & Allows \\
		& & & & & CI & non-linear \\
		& & & & & & effects \\
		\midrule
		Survival Area & $t$ & $\hat{S}_z(t)$ & -- & $Z$ & No & No \\
		\addlinespace[0.12cm]
		Survival Contour & $t$ & $Z$ & -- & $\hat{S}_z(t)$ & No & Yes \\
		\addlinespace[0.12cm]
		Survival Heatmap & $t$ & $Z$ & -- & $\hat{S}_z(t)$ & No & Yes \\
		\addlinespace[0.12cm]
		3D Survival Surface & $t$ & $\hat{S}_z(t)$ & $Z$ & -- & No & Yes \\
		\addlinespace[0.12cm]
		Value Specific & \multirow{2}{*}{$t$} & \multirow{2}{*}{$\hat{S}_z(t)$} & \multirow{2}{*}{--} & \multirow{2}{*}{Possibly $Z$} & \multirow{2}{*}{Yes} & \multirow{2}{*}{Yes} \\
		Survival Curve & & & & & & \\
		\addlinespace[0.12cm]
		Landmark Survival & \multirow{2}{*}{$Z$} & \multirow{2}{*}{$\hat{S}_z(t)$} & \multirow{2}{*}{--} & \multirow{2}{*}{Possibly $t$} & \multirow{2}{*}{Yes} & \multirow{2}{*}{Yes} \\
		Probability & & & & & & \\
		\addlinespace[0.12cm]
		Survival Time Quantile & $Z$ & $\hat{Q}_z(p)$ & -- & Possible $p$ & Yes & Yes \\
		\addlinespace[0.12cm]
		Restricted Mean & \multirow{2}{*}{$Z$} & \multirow{2}{*}{$\widehat{RMST}_z(\lambda)$} & \multirow{2}{*}{--} & \multirow{2}{*}{Possibly $\lambda$} & \multirow{2}{*}{Yes} & \multirow{2}{*}{Yes} \\
		Survival Time & & & & & & \\
		\bottomrule
	\end{tabular}
\end{table}

\begin{figure}[!htb]
	\centering
	\includegraphics[width=0.84\linewidth]{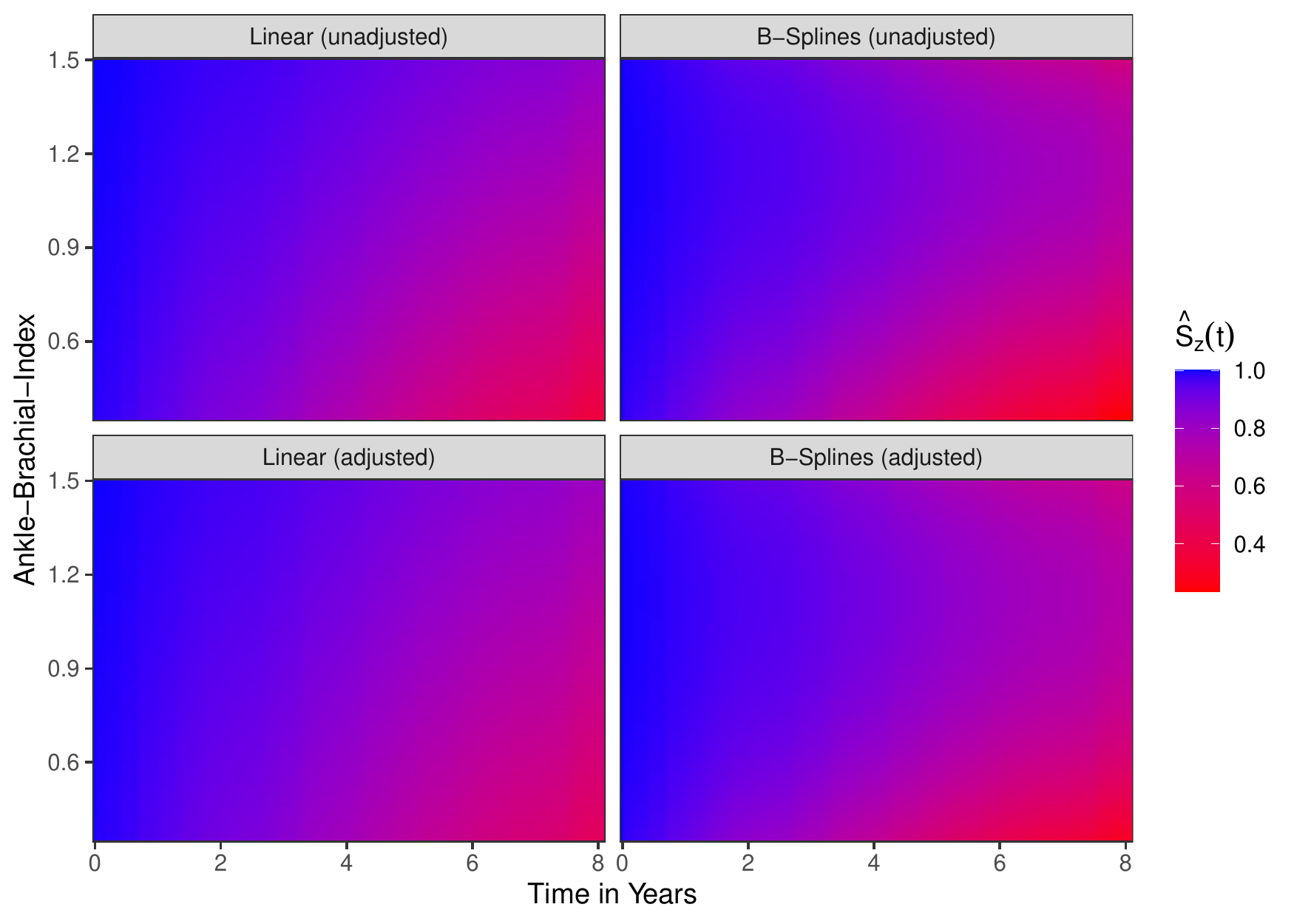}
	\caption{Survival heatmaps for the causal effect of the ABI on survival, created using four different Cox proportional hazards regression models and g-computation.}
\end{figure}

\begin{figure}[!htb]
	\centering
	\includegraphics[width=0.84\linewidth]{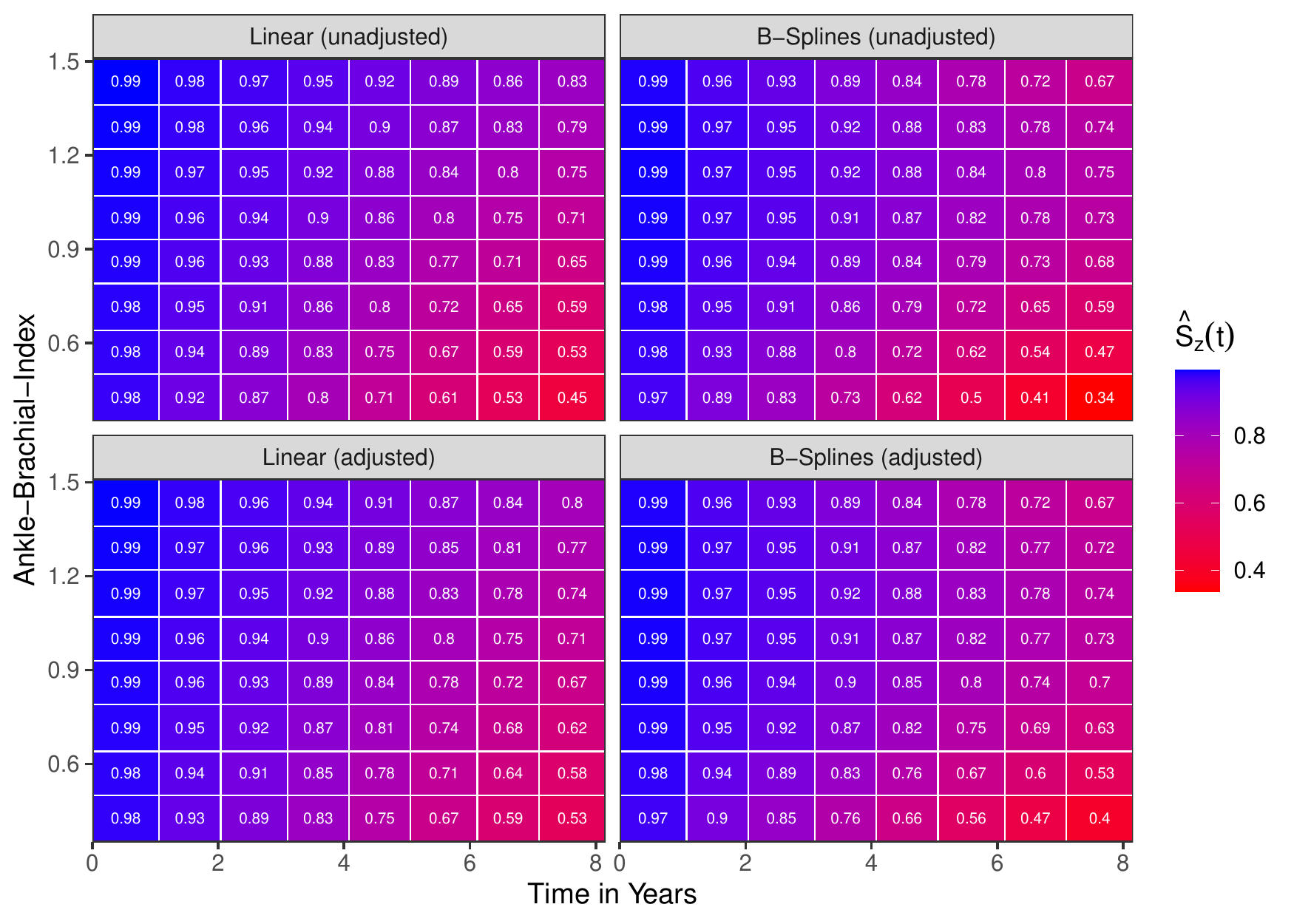}
	\caption{Categorized Survival heatmaps for the causal effect of the ABI on survival, created using four different Cox proportional hazards regression models and g-computation.}
\end{figure}

\begin{figure}[!htb]
	\centering
	\includegraphics[width=0.84\linewidth]{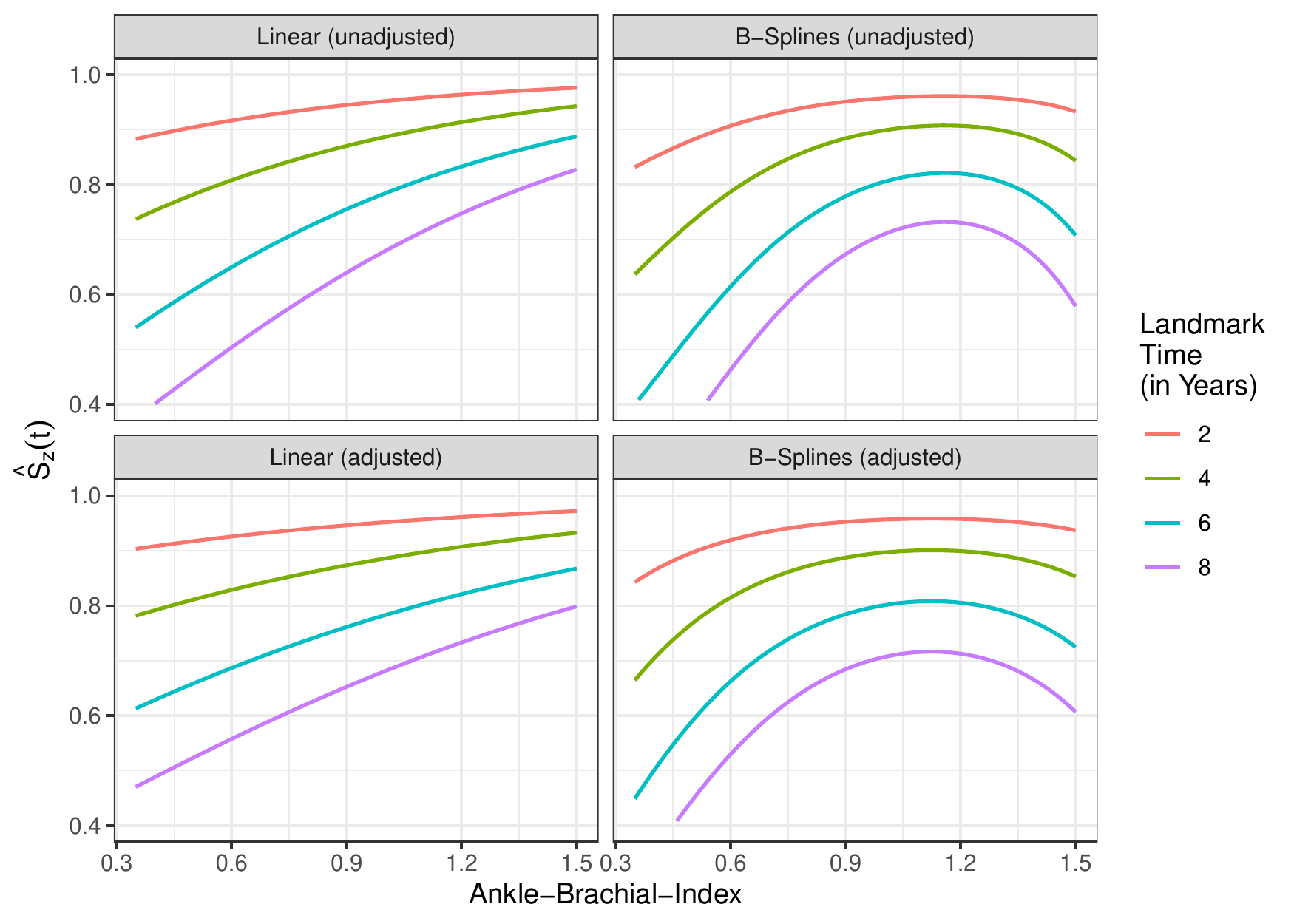}
	\caption{The survival probability at different points in time $t$ as a function of the ABI, created using four different Cox proportional hazards regression models and g-computation.}
	\label{fig::surv_landmark}
\end{figure}

\begin{figure}[!htb]
	\centering
	\includegraphics[width=0.84\linewidth]{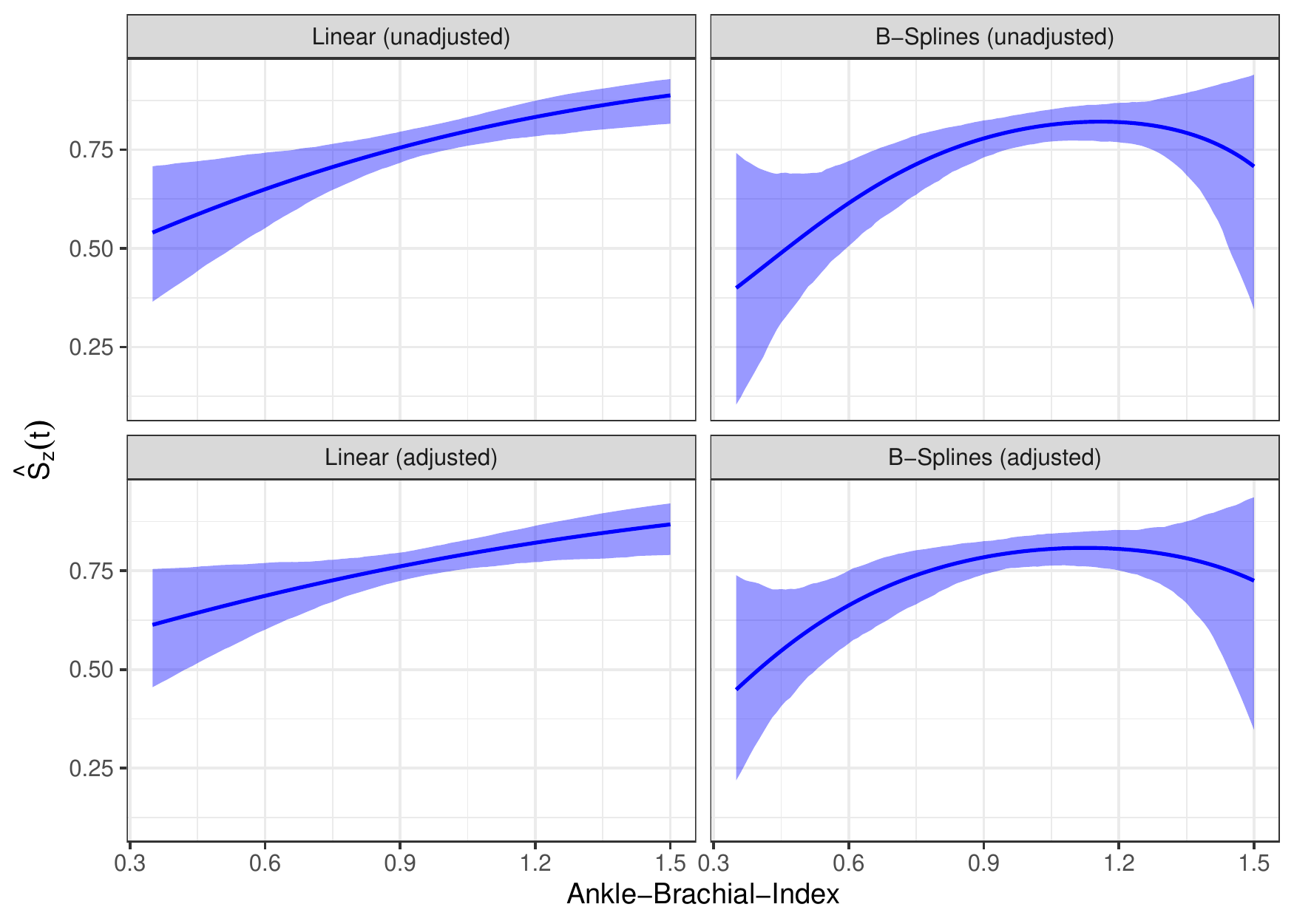}
	\caption{The survival probability at $t = 6$ with bootstrap confidence intervals as a function of the ABI, created using four different Cox proportional hazards regression models and g-computation. The confidence interval was calculated using the bootstrap percentile method with $1000$ bootstrap samples.}
\end{figure}

\begin{figure}[!htb]
	\centering
	\includegraphics[width=0.9\linewidth]{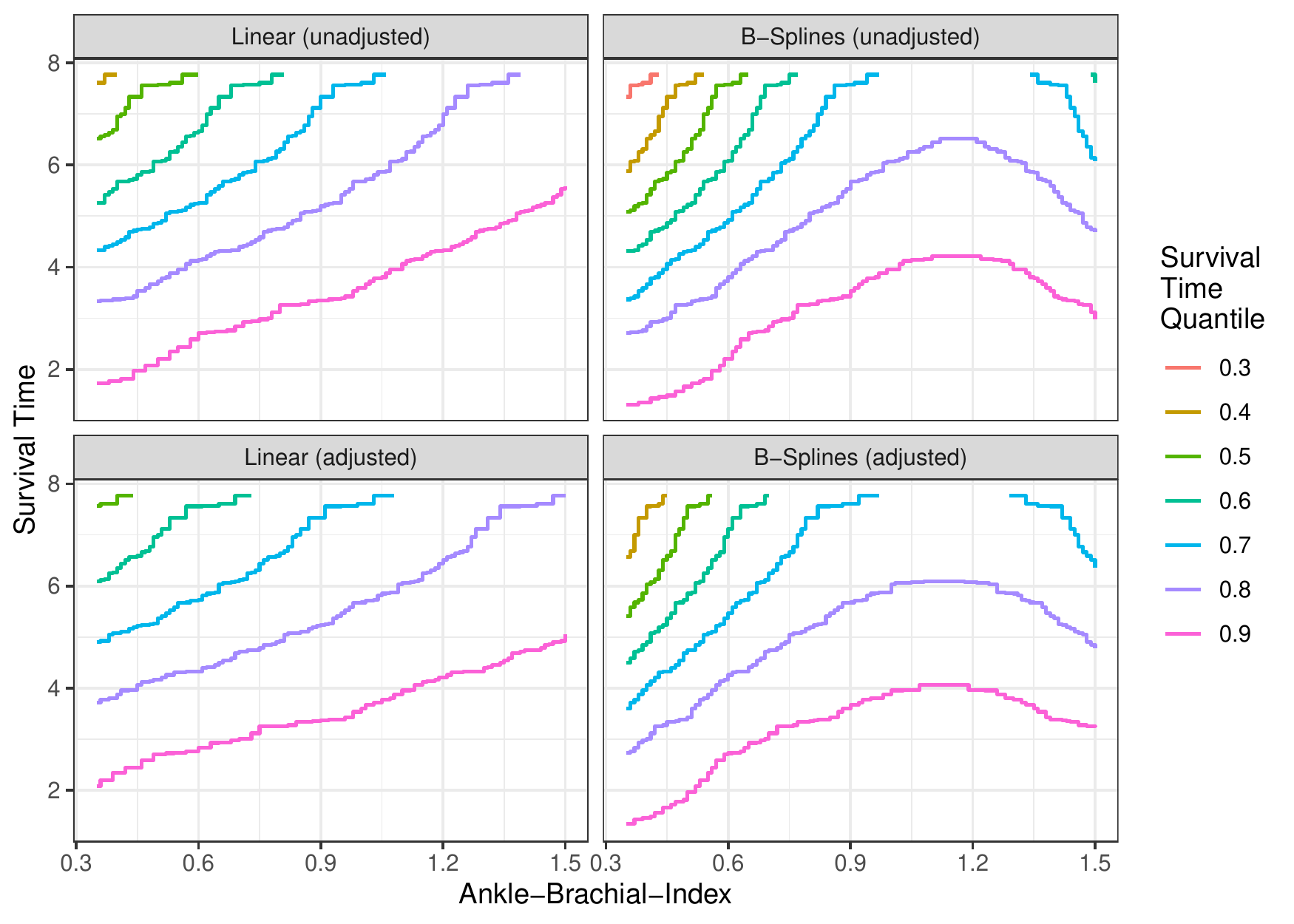}
	\caption{Survival time quantiles as a function of the ABI, created using four different Cox proportional hazards regression models and g-computation.}
\end{figure}

\begin{figure}[!htb]
	\centering
	\includegraphics[width=0.9\linewidth]{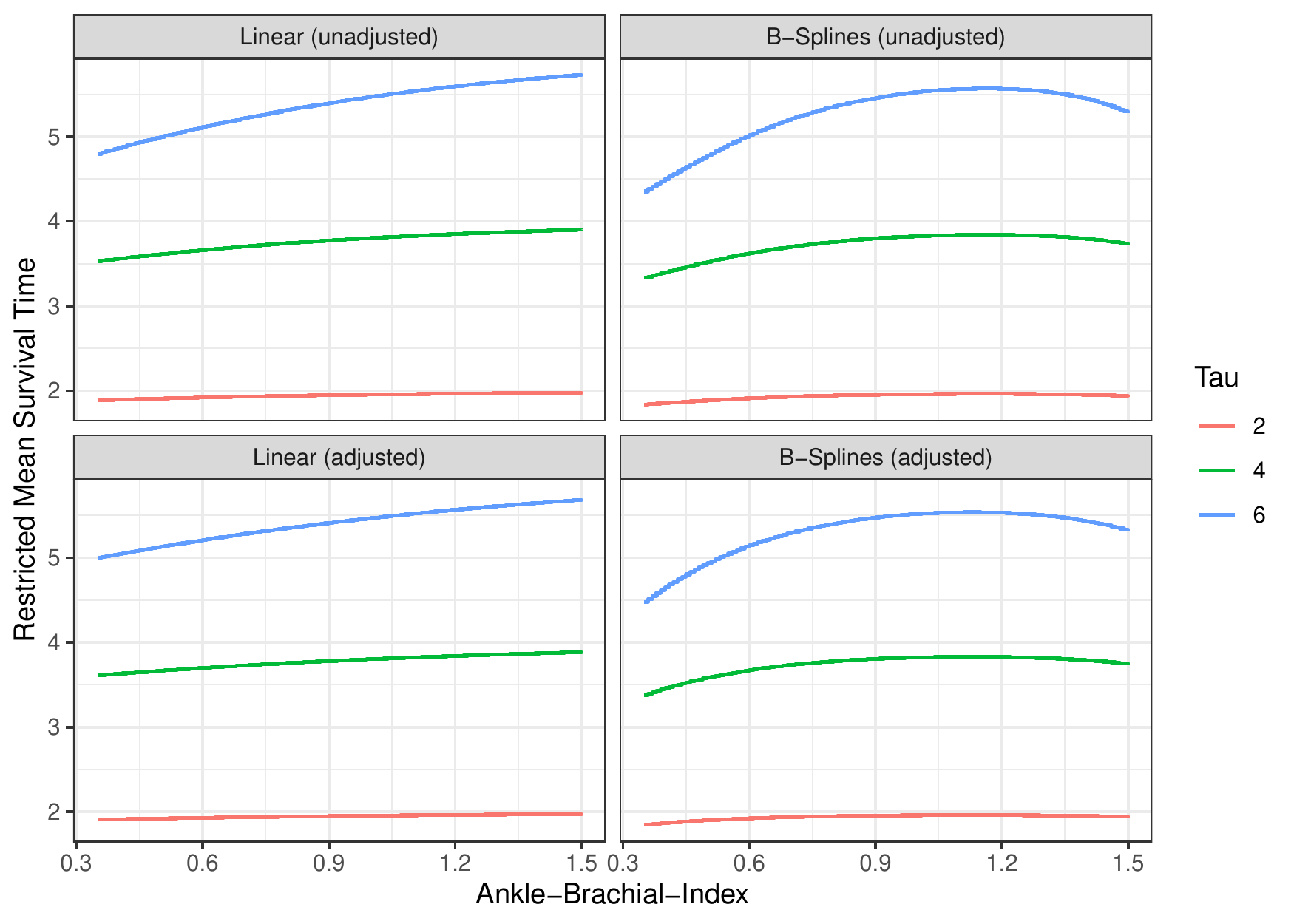}
	\caption{Restricted mean survival times as a function of the ABI, created using four different Cox proportional hazards regression models and g-computation.}
\end{figure}

\begin{figure}[!htb]
	\centering
	\includegraphics[width=0.84\linewidth]{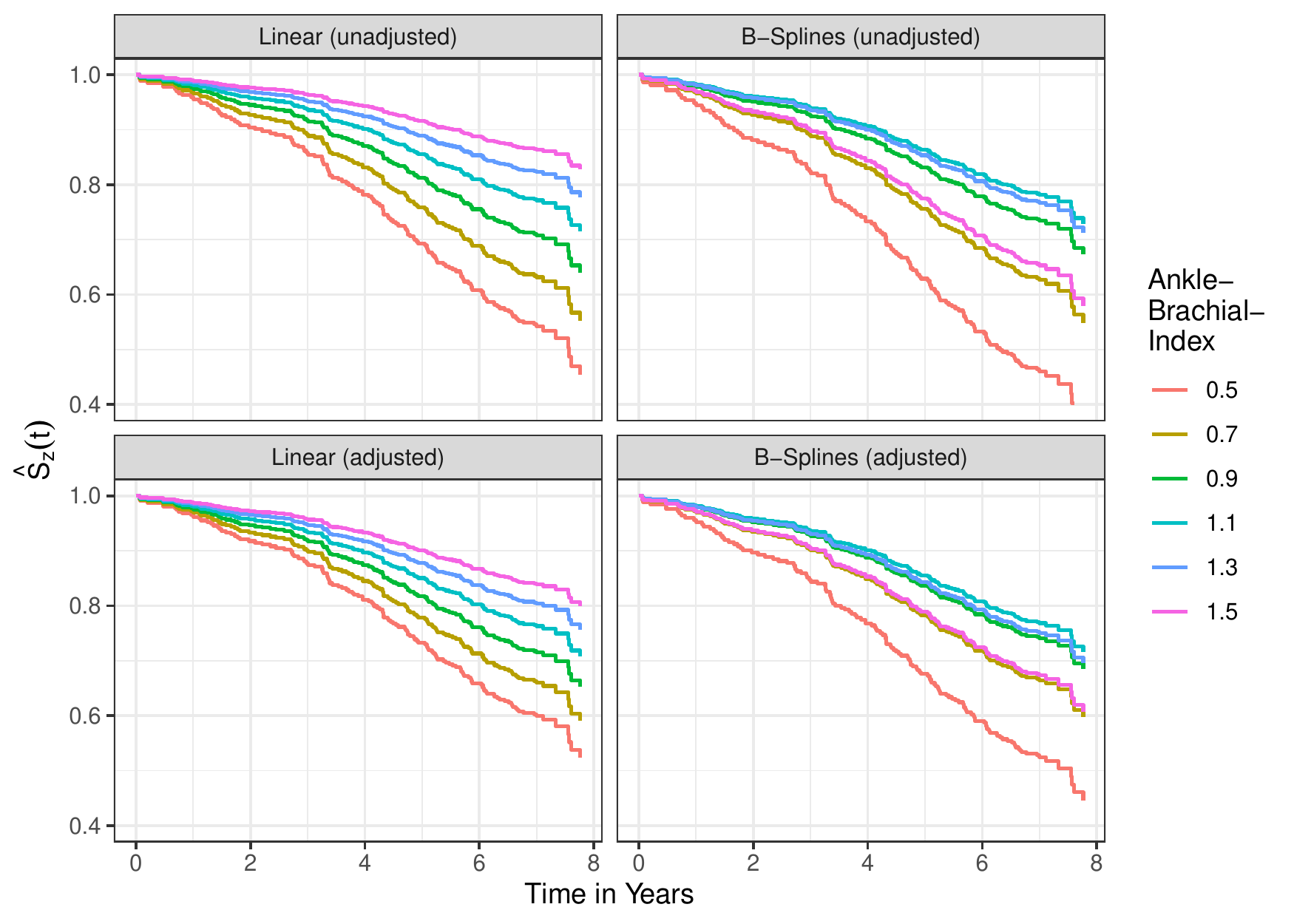}
	\caption{Survival curves for different values of the ABI, created using four different Cox proportional hazards regression models and g-computation.}
\end{figure}

\begin{figure}[!htb]
	\centering
	\includegraphics[width=0.62\linewidth]{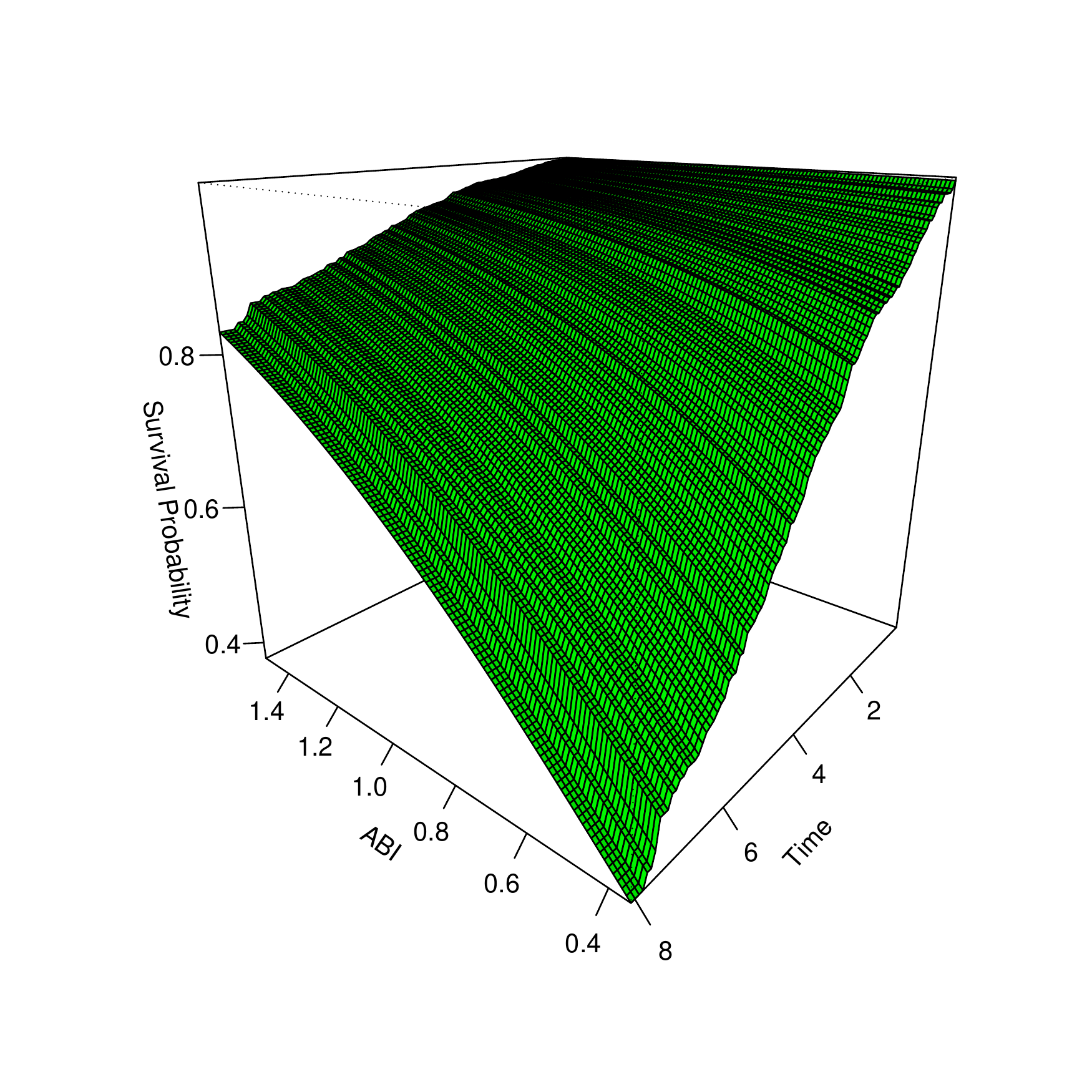}
	\caption{Survival 3D surface plot displaying the effect of the ABI on the survival probability, created using model (2) from table 1 and g-computation.}
\end{figure}

\begin{figure}[!htb]
	\centering
	\includegraphics[width=0.62\linewidth]{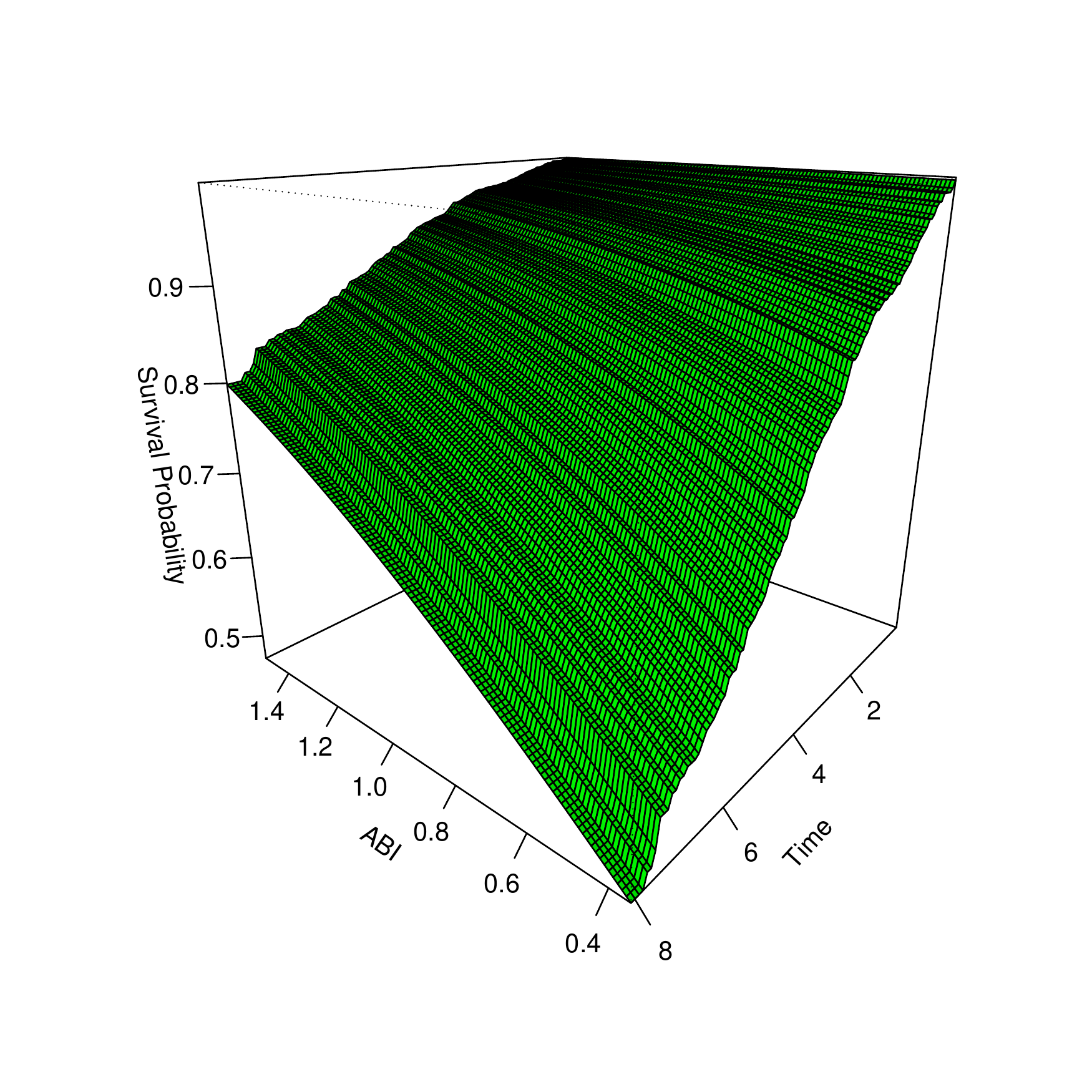}
	\caption{Survival 3D surface plot displaying the effect of the ABI on the survival probability, created using model (3) from table 1 and g-computation.}
\end{figure}

\begin{figure}[!htb]
	\centering
	\includegraphics[width=0.62\linewidth]{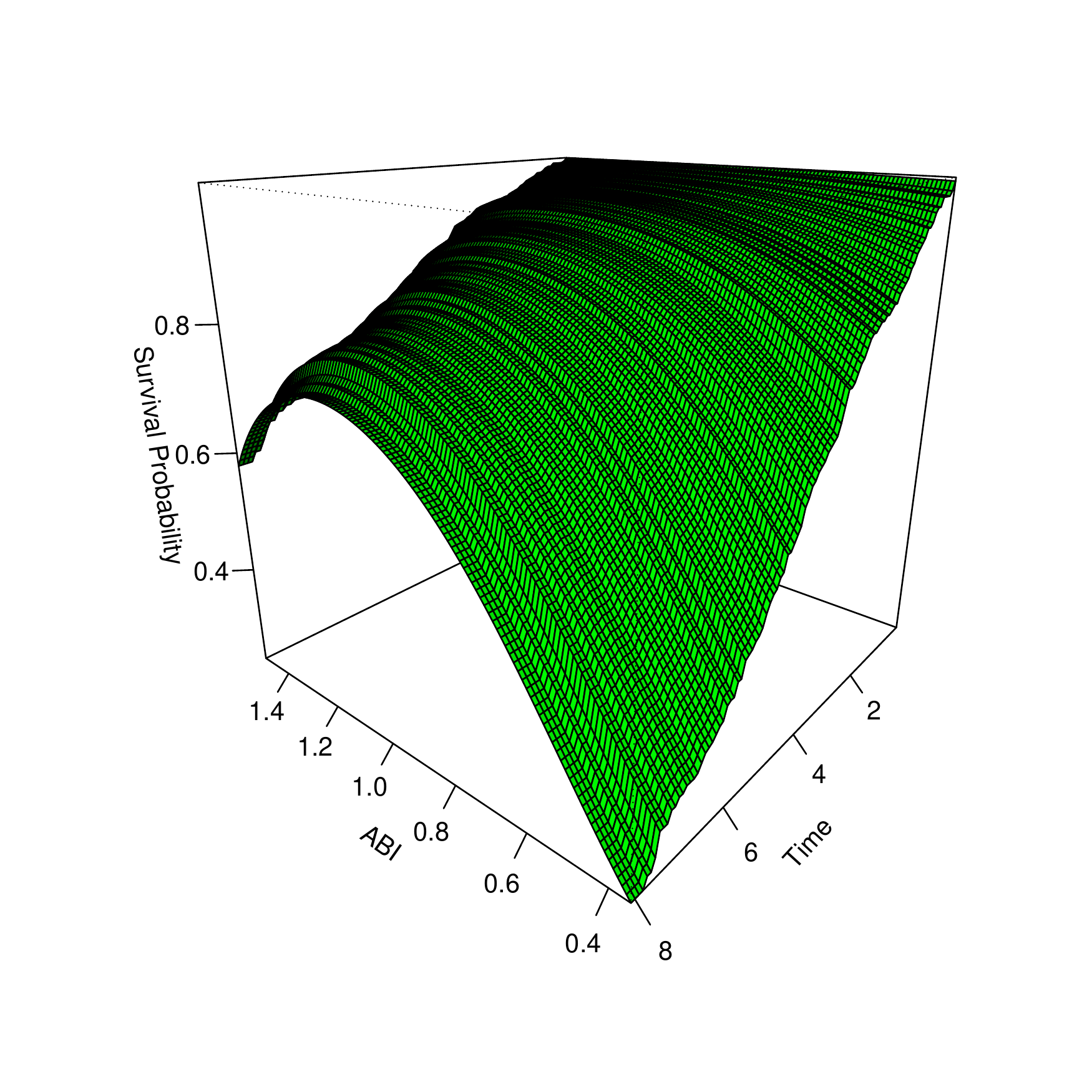}
	\caption{Survival 3D surface plot displaying the effect of the ABI on the survival probability, created using model (4) from table 1 and g-computation.}
\end{figure}

\begin{figure}[!htb]
	\centering
	\includegraphics[width=0.62\linewidth]{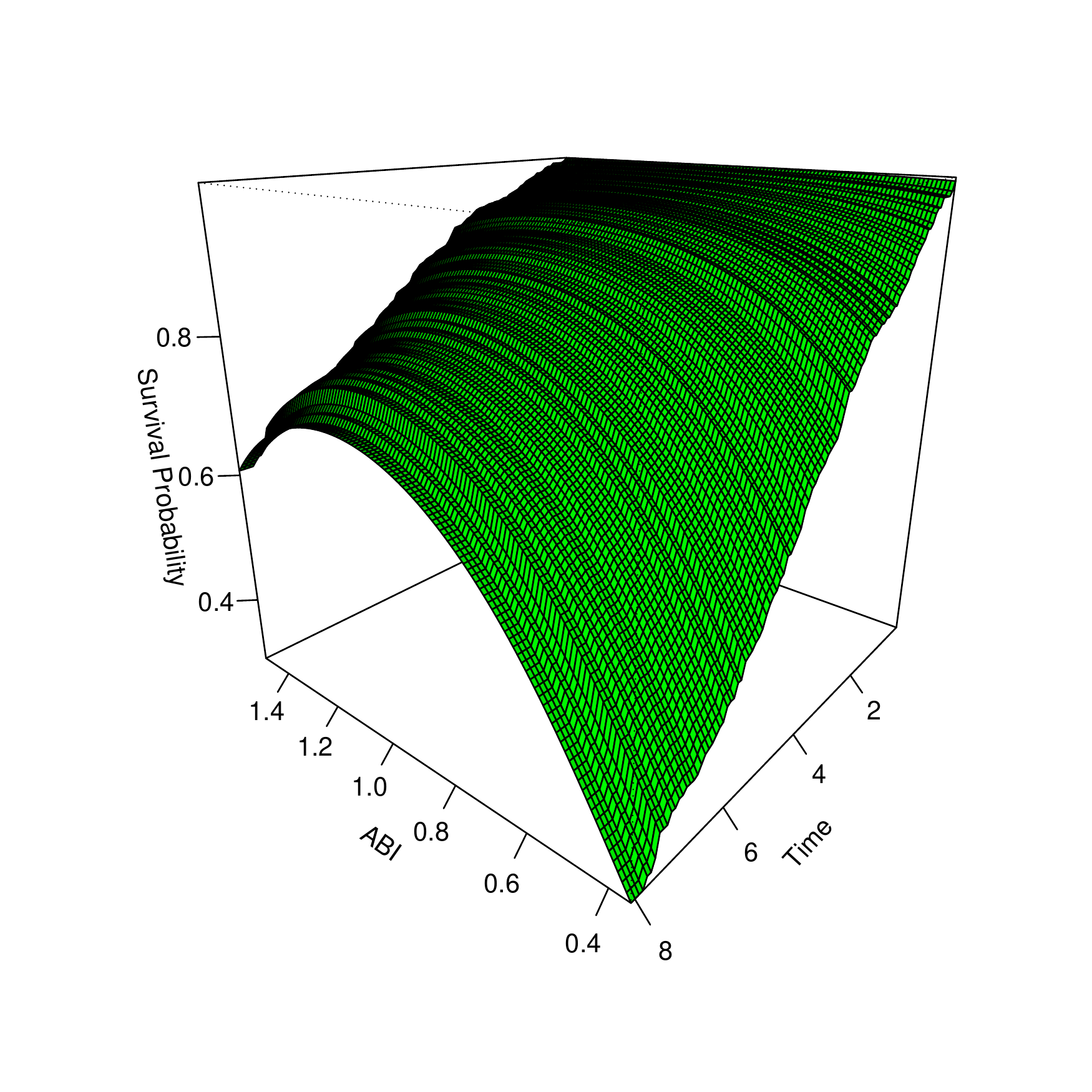}
	\caption{Survival 3D surface plot displaying the effect of the ABI on the survival probability, created using model (5) from table 1 and g-computation.}
\end{figure}

\FloatBarrier
\newpage

\section{Estimating \& Depicting Causal Contrasts}

Instead of relying on the estimand defined in equation~1 in the main text directly, it may be useful to depict causal contrasts based on it instead. One alternative is to use the difference or ratio between the counterfactual survival probability at $t$ if all individuals had received treatment $Z$, and the counterfactual survival probability at $t$ if all individuals had received a specific reference value of $Z$ as treatment, denoted as $\tau$. The difference is then defined as:
\begin{equation}
	\Delta_{\tau}(t, z) = S_{\tau}(t) - S_z(t),
\end{equation}
and the ratio is defined as:
\begin{equation}
	\phi_{\tau}(t, z) = \frac{S_{\tau}(t)}{S_z(t)}.
\end{equation}
In some cases there may be good reasons to choose a specific $\tau$. For example, if the main goal was to estimate the causal effect of a drug dosage on the survival probability, using a dosage of $\tau = 0$ may be a reasonable strategy. If that is not the case, one may instead use the survival probability at $t$ under ``treatment-as-usual'' conditions. We previously defined the actually observed survival time as $T$. This probability is then defined as:
\begin{equation}
	S(t) = P(T > t).
\end{equation}
Therefore the causal difference is defined as:
\begin{equation}
	\Delta_{KM}(t, z) = S(t) - S_z(t),
\end{equation}
and the ratio is defined as:
\begin{equation}
	\phi_{KM}(t, z) = \frac{S(t)}{S_z(t)}.
\end{equation}
In practice, $S(t)$ can be estimated from the crude sample using standard techniques, such as the Kaplan-Meier estimator if $T$ is subject to right-censoring.
\par\medskip
Note that these contrast-based target estimands correspond to different causal questions than the original target estimand defined by equation~1 in the main text. The causal question associated with the original target estimand can be translated to be ``If every individual in the target population had received treatment $Z = z$, what proportion of individuals would have survived up to and including time $t$?''. Table~\ref{tab::causal_questions} shows the causal question associated with each of the target estimands.

\begin{table}[!htb]
	\centering
	\caption{The specific causal questions associated with each of the target estimands.}
	\medskip
	\label{tab::causal_questions}
	\begin{tabular}{p{0.2\linewidth} p{0.8\linewidth}}
		\toprule
		Target Estimand & Causal Question \\
		\midrule
		$S_z(t)$ & If every individual in the target population had received treatment $Z = z$, what proportion of individuals would have survived up to and including time $t$? \\
		& \\
		$\Delta_{\tau}(t, z)$ & What is the difference between the proportion of individuals that would have survived up to and including time $t$ if every person in the target population had received treatment $Z = \tau$ and the proportion of individuals that would have survived up to and including time $t$ if every person in the target population had received treatment $Z = z$? \\
		& \\
		$\phi_{\tau}(t, z)$ & What is the ratio between the proportion of individuals that would have survived up to and including time $t$ if every person in the target population had received treatment $Z = \tau$ and the proportion of individuals that would have survived up to and including time $t$ if every person in the target population had received treatment $Z = z$? \\
		& \\
		$\Delta_{KM}(t, z)$ & What is the difference between the proportion of individuals that have survived up to and including time $t$ under the 'treatment-as-usual' regime and the proportion of individuals that would have survived up to and including time $t$ if every person in the target population had received treatment $Z = z$? \\
		& \\
		$\phi_{KM}(t, z)$ & What is the ratio between the proportion of individuals that have survived up to and including time $t$ under the 'treatment-as-usual' regime and the proportion of individuals that would have survived up to and including time $t$ if every person in the target population had received treatment $Z = z$? \\
		\bottomrule
	\end{tabular}
\end{table}

Tables~\ref{tab::contrast_delta_km}--\ref{tab::contrast_phi_tau} displays estimates of these quantities for $Z = 0.3$ and $Z = 1$ at $t = 4$ and $t = 6$, estimated using all of the models described in the main text. The associated standard errors and 95\% confidence intervals were calculated using the bootstrap percentile method with $1000$ bootstrap samples.
\par\medskip

\begin{table}[!htb]
	\centering
	\caption{Estimates of $\Delta_{KM}$ in the getABI data for some combinations of $t$ and $z$ and all models discussed in the main text. Confidence intervals and standard errors were calculated using $1000$ bootstrap samples.}
	\medskip
	\label{tab::contrast_delta_km}
	\begin{tabular}{lllcccc}
		\toprule
		Model & $t$ & $z$ & $\hat{\Delta}_{KM}$ & $SE\left(\hat{\Delta}_{KM}\right)$ & $CI_{95}$ (Lower) & $CI_{95}$ (Upper) \\
		\midrule
		Linear (unadjusted) & 4 & 0.4 & 0.127 & 0.051 & 0.038 & 0.234 \\ 
		Linear (unadjusted) & 4 & 1.0 & -0.007 & 0.004 & -0.016 & -0.001 \\ 
		Linear (unadjusted) & 6 & 0.4 & 0.210 & 0.076 & 0.069 & 0.355 \\ 
		Linear (unadjusted) & 6 & 1.0 & -0.011 & 0.006 & -0.024 & -0.002 \\ 
		Linear (adjusted) & 4 & 0.4 & 0.088 & 0.041 & 0.017 & 0.173 \\ 
		Linear (adjusted) & 4 & 1.0 & -0.006 & 0.004 & -0.014 & -0.000 \\ 
		Linear (adjusted) & 6 & 0.4 & 0.145 & 0.063 & 0.029 & 0.271 \\ 
		Linear (adjusted) & 6 & 1.0 & -0.009 & 0.005 & -0.021 & -0.001 \\ 
		B-Splines (unadjusted) & 4 & 0.4 & 0.209 & 0.095 & 0.019 & 0.414 \\ 
		B-Splines (unadjusted) & 4 & 1.0 & -0.019 & 0.008 & -0.035 & -0.003 \\ 
		B-Splines (unadjusted) & 6 & 0.4 & 0.329 & 0.127 & 0.036 & 0.562 \\ 
		B-Splines (unadjusted) & 6 & 1.0 & -0.032 & 0.013 & -0.058 & -0.003 \\ 
		B-Splines (adjusted) & 4 & 0.4 & 0.176 & 0.071 & 0.033 & 0.319 \\ 
		B-Splines (adjusted) & 4 & 1.0 & -0.017 & 0.008 & -0.033 & -0.002 \\ 
		B-Splines (adjusted) & 6 & 0.4 & 0.274 & 0.097 & 0.061 & 0.448 \\ 
		B-Splines (adjusted) & 6 & 1.0 & -0.027 & 0.013 & -0.052 & -0.003 \\ 
		\bottomrule
	\end{tabular}
\end{table}

\begin{table}[!htb]
	\centering
	\caption{Estimates of $\phi_{KM}$ in the getABI data for some combinations of $t$ and $z$ and all models discussed in the main text. Confidence intervals and standard errors were calculated using $1000$ bootstrap samples.}
	\medskip
	\label{tab::contrast_phi_km}
	\begin{tabular}{lllcccc}
		\toprule
		Model & $t$ & $z$ & $\hat{\phi}_{KM}$ & $SE\left(\hat{\phi}_{KM}\right)$ & $CI_{95}$ (Lower) & $CI_{95}$ (Upper) \\
		\midrule
		Linear (unadjusted) & 4 & 0.4 & 1.168 & 0.084 & 1.044 & 1.368 \\ 
		Linear (unadjusted) & 4 & 1.0 & 0.992 & 0.004 & 0.982 & 0.999 \\ 
		Linear (unadjusted) & 6 & 0.4 & 1.373 & 0.207 & 1.095 & 1.870 \\ 
		Linear (unadjusted) & 6 & 1.0 & 0.987 & 0.007 & 0.970 & 0.997 \\ 
		Linear (adjusted) & 4 & 0.4 & 1.111 & 0.060 & 1.020 & 1.250 \\ 
		Linear (adjusted) & 4 & 1.0 & 0.993 & 0.004 & 0.984 & 1.000 \\ 
		Linear (adjusted) & 6 & 0.4 & 1.231 & 0.132 & 1.039 & 1.544 \\ 
		Linear (adjusted) & 6 & 1.0 & 0.988 & 0.007 & 0.973 & 0.999 \\ 
		B-Splines (unadjusted) & 4 & 0.4 & 1.312 & 0.215 & 1.022 & 1.895 \\ 
		B-Splines (unadjusted) & 4 & 1.0 & 0.979 & 0.009 & 0.961 & 0.997 \\ 
		B-Splines (unadjusted) & 6 & 0.4 & 1.742 & 0.666 & 1.047 & 3.655 \\ 
		B-Splines (unadjusted) & 6 & 1.0 & 0.961 & 0.016 & 0.930 & 0.996 \\ 
		B-Splines (adjusted) & 4 & 0.4 & 1.250 & 0.133 & 1.039 & 1.564 \\ 
		B-Splines (adjusted) & 4 & 1.0 & 0.981 & 0.009 & 0.964 & 0.997 \\ 
		B-Splines (adjusted) & 6 & 0.4 & 1.547 & 0.329 & 1.084 & 2.377 \\ 
		B-Splines (adjusted) & 6 & 1.0 & 0.966 & 0.015 & 0.937 & 0.996 \\
		\bottomrule
	\end{tabular}
\end{table}

\begin{table}[!htb]
	\centering
	\caption{Estimates of $\Delta_{0.6}$ in the getABI data for some combinations of $t$ and $z$ and all models discussed in the main text. Confidence intervals and standard errors were calculated using $1000$ bootstrap samples.}
	\medskip
	\label{tab::contrast_delta_tau}
	\begin{tabular}{lllcccc}
		\toprule
		Model & $t$ & $z$ & $\hat{\Delta}_{0.6}$ & $SE\left(\hat{\Delta}_{0.6}\right)$ & $CI_{95}$ (Lower) & $CI_{95}$ (Upper) \\
		\midrule
		Linear (unadjusted) & 4 & 0.4 & 0.055 & 0.025 & 0.014 & 0.109 \\ 
		Linear (unadjusted) & 4 & 1.0 & -0.079 & 0.029 & -0.137 & -0.025 \\ 
		Linear (unadjusted) & 6 & 0.4 & 0.086 & 0.034 & 0.026 & 0.154 \\ 
		Linear (unadjusted) & 6 & 1.0 & -0.134 & 0.047 & -0.223 & -0.046 \\ 
		Linear (adjusted) & 4 & 0.4 & 0.037 & 0.019 & 0.006 & 0.080 \\ 
		Linear (adjusted) & 4 & 1.0 & -0.057 & 0.025 & -0.108 & -0.011 \\ 
		Linear (adjusted) & 6 & 0.4 & 0.058 & 0.027 & 0.011 & 0.115 \\ 
		Linear (adjusted) & 6 & 1.0 & -0.096 & 0.041 & -0.175 & -0.021 \\ 
		B-Splines (unadjusted) & 4 & 0.4 & 0.116 & 0.090 & -0.053 & 0.311 \\ 
		B-Splines (unadjusted) & 4 & 1.0 & -0.112 & 0.037 & -0.190 & -0.041 \\ 
		B-Splines (unadjusted) & 6 & 0.4 & 0.171 & 0.120 & -0.090 & 0.408 \\ 
		B-Splines (unadjusted) & 6 & 1.0 & -0.190 & 0.059 & -0.315 & -0.077 \\ 
		B-Splines (adjusted) & 4 & 0.4 & 0.111 & 0.068 & -0.030 & 0.247 \\ 
		B-Splines (adjusted) & 4 & 1.0 & -0.082 & 0.032 & -0.147 & -0.024 \\ 
		B-Splines (adjusted) & 6 & 0.4 & 0.162 & 0.094 & -0.046 & 0.334 \\ 
		B-Splines (adjusted) & 6 & 1.0 & -0.138 & 0.051 & -0.239 & -0.040 \\
		\bottomrule
	\end{tabular}
\end{table}

\begin{table}[!htb]
	\centering
	\caption{Estimates of $\phi_{0.6}$ in the getABI data for some combinations of $t$ and $z$ and all models discussed in the main text. Confidence intervals and standard errors were calculated using $1000$ bootstrap samples.}
	\medskip
	\label{tab::contrast_phi_tau}
	\begin{tabular}{lllcccc}
		\toprule
		Model & $t$ & $z$ & $\hat{\phi}_{0.6}$ & $SE\left(\hat{\phi}_{0.6}\right)$ & $CI_{95}$ (Lower) & $CI_{95}$ (Upper) \\
		\midrule
		Linear (unadjusted) & 4 & 0.4 & 1.073 & 0.041 & 1.017 & 1.170 \\ 
		Linear (unadjusted) & 4 & 1.0 & 0.911 & 0.033 & 0.845 & 0.972 \\ 
		Linear (unadjusted) & 6 & 0.4 & 1.153 & 0.091 & 1.036 & 1.369 \\ 
		Linear (unadjusted) & 6 & 1.0 & 0.829 & 0.060 & 0.717 & 0.943 \\ 
		Linear (adjusted) & 4 & 0.4 & 1.047 & 0.028 & 1.007 & 1.116 \\ 
		Linear (adjusted) & 4 & 1.0 & 0.936 & 0.028 & 0.879 & 0.987 \\ 
		Linear (adjusted) & 6 & 0.4 & 1.092 & 0.056 & 1.015 & 1.228 \\ 
		Linear (adjusted) & 6 & 1.0 & 0.877 & 0.052 & 0.777 & 0.974 \\ 
		B-Splines (unadjusted) & 4 & 0.4 & 1.174 & 0.181 & 0.937 & 1.662 \\ 
		B-Splines (unadjusted) & 4 & 1.0 & 0.876 & 0.041 & 0.789 & 0.955 \\ 
		B-Splines (unadjusted) & 6 & 0.4 & 1.384 & 0.493 & 0.881 & 2.836 \\ 
		B-Splines (unadjusted) & 6 & 1.0 & 0.764 & 0.072 & 0.616 & 0.903 \\ 
		B-Splines (adjusted) & 4 & 0.4 & 1.158 & 0.118 & 0.964 & 1.443 \\ 
		B-Splines (adjusted) & 4 & 1.0 & 0.909 & 0.035 & 0.835 & 0.973 \\ 
		B-Splines (adjusted) & 6 & 0.4 & 1.325 & 0.266 & 0.939 & 2.007 \\ 
		B-Splines (adjusted) & 6 & 1.0 & 0.827 & 0.062 & 0.706 & 0.950 \\  
		\bottomrule
	\end{tabular}
\end{table}

It is difficult to use these point estimates without any visualizations, because they are dependent on both $z$ and $t$. The proposed graphics can be adapted in a straightforward fashion to display estimates of $\Delta_{\tau}, \Delta_{KM}, \phi_{\tau}$ and $\phi_{KM}$ instead of estimates of $S_z(t)$ by simply substituting $\hat{S}_z(t)$ with one of those quantities. Below we give examples for all of these quantities estimated from the getABI study using some of the previously shown types of graphics.
\par\medskip

\begin{figure}[!htb]
	\centering
	\includegraphics[width=1\linewidth]{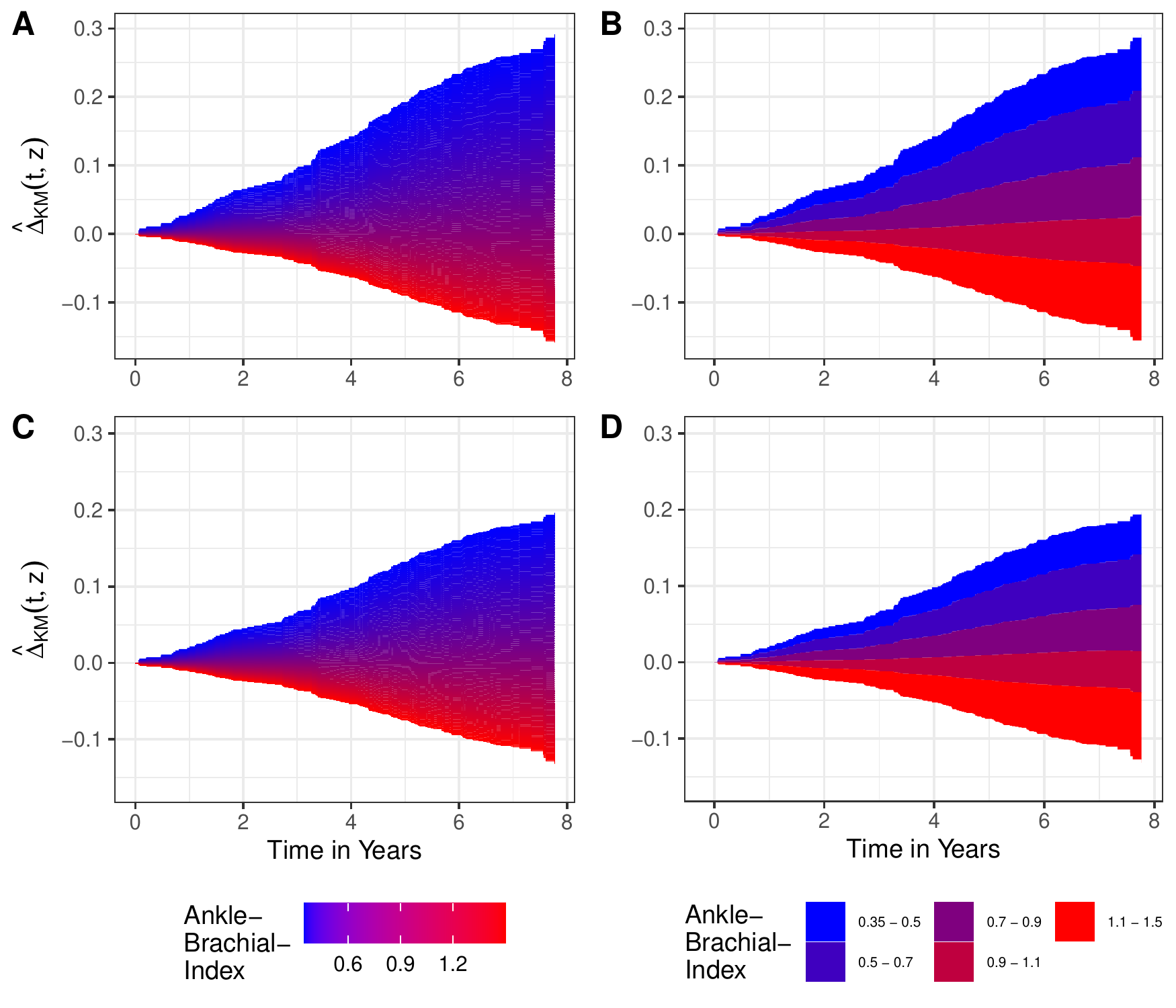}
	\caption{Continuously and discretely colored survival area plots displaying the causal effect of the ABI on survival using $\hat{\Delta}_{KM}$, created using different Cox proportional hazards regression models and g-computation. \\ \textbf{A}: Using model 1 (unadjusted), continuous color shading; \textbf{B}: Using model 1 (unadjusted), discrete color shading; \textbf{C}: Using model 2 (adjusted), continuous color shading; \textbf{D}: Using model 2 (adjusted), discrete color shading.}
\end{figure}

\begin{figure}[!htb]
	\centering
	\includegraphics[width=1\linewidth]{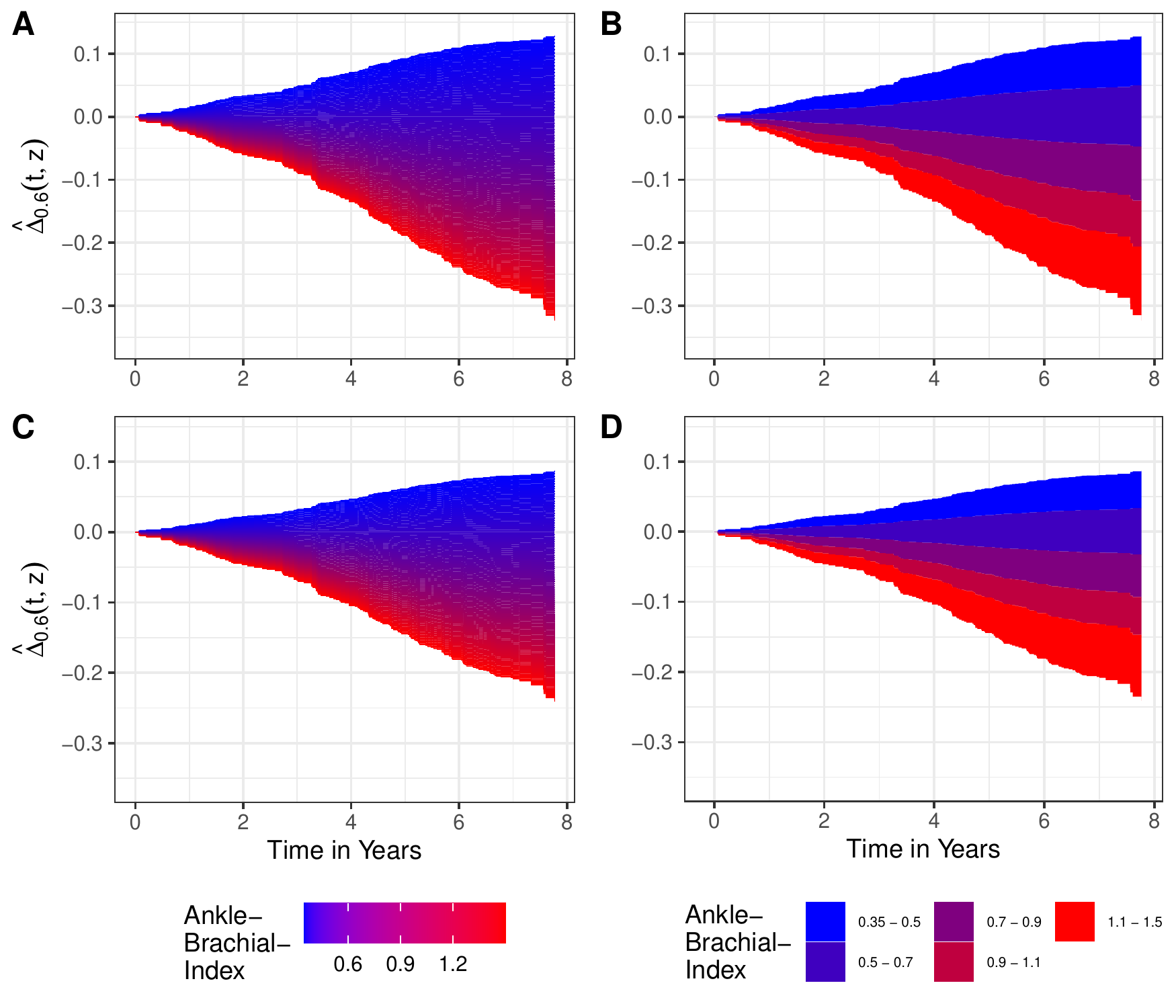}
	\caption{Continuously and discretely colored survival area plots displaying the causal effect of the ABI on survival using $\hat{\Delta}_{0.6}$, created using different Cox proportional hazards regression models and g-computation. \\ \textbf{A}: Using model 1 (unadjusted), continuous color shading; \textbf{B}: Using model 1 (unadjusted), discrete color shading; \textbf{C}: Using model 2 (adjusted), continuous color shading; \textbf{D}: Using model 2 (adjusted), discrete color shading.}
\end{figure}

\begin{figure}[!htb]
	\centering
	\includegraphics[width=1\linewidth]{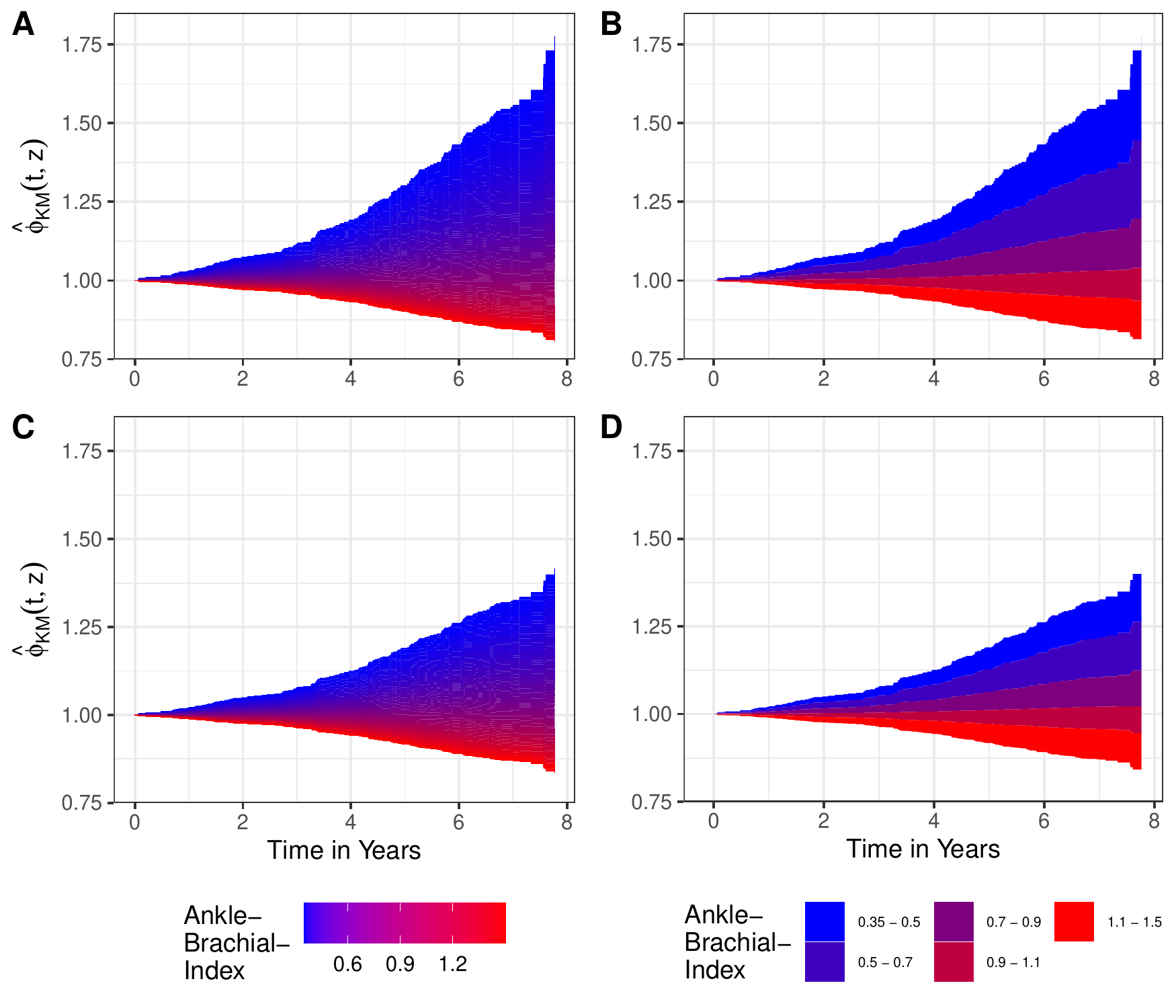}
	\caption{Continuously and discretely colored survival area plots displaying the causal effect of the ABI on survival using $\hat{\phi}_{KM}$, created using different Cox proportional hazards regression models and g-computation. \\ \textbf{A}: Using model 1 (unadjusted), continuous color shading; \textbf{B}: Using model 1 (unadjusted), discrete color shading; \textbf{C}: Using model 2 (adjusted), continuous color shading; \textbf{D}: Using model 2 (adjusted), discrete color shading.}
\end{figure}

\begin{figure}[!htb]
	\centering
	\includegraphics[width=1\linewidth]{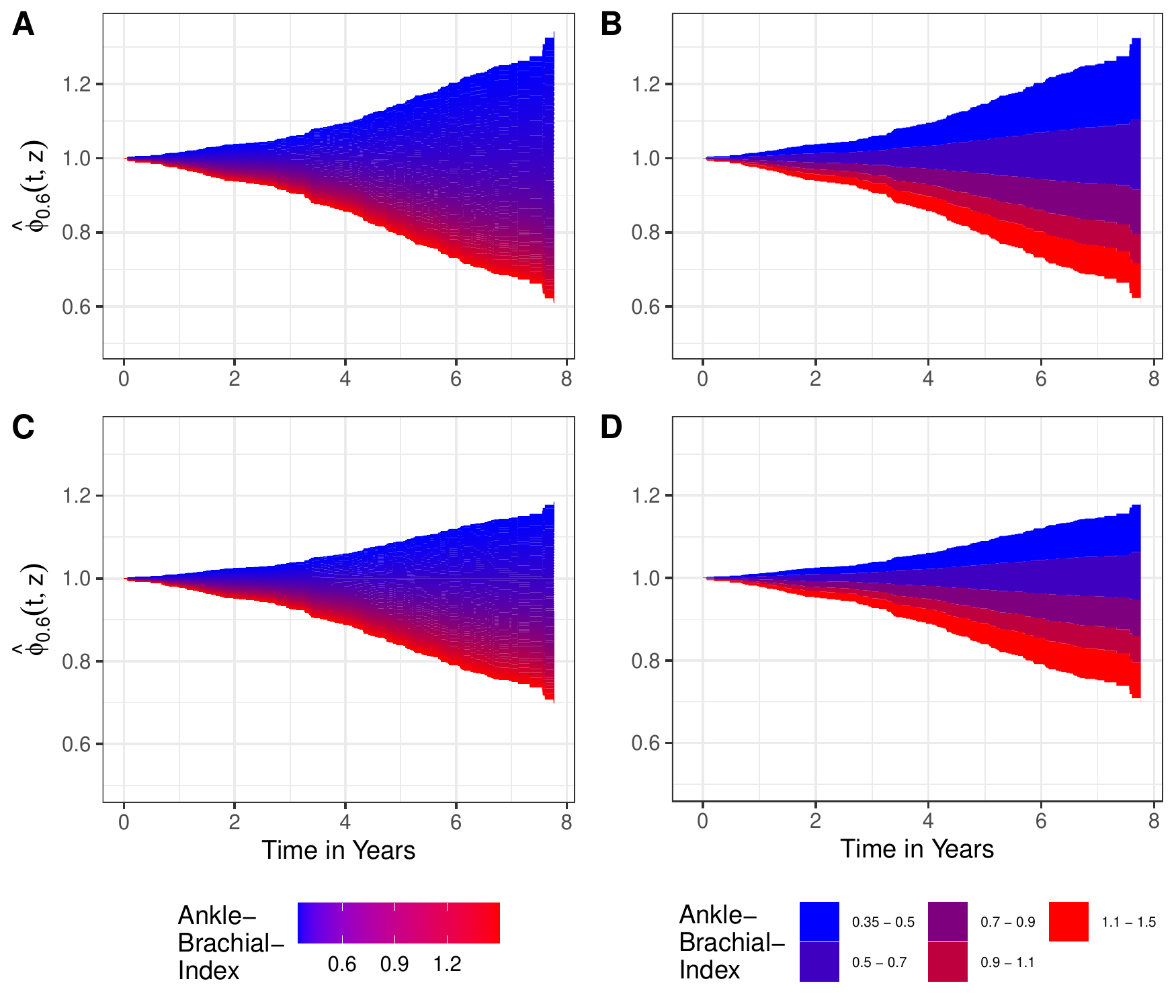}
	\caption{Continuously and discretely colored survival area plots displaying the causal effect of the ABI on survival using $\hat{\phi}_{0.6}$, created using different Cox proportional hazards regression models and g-computation. \\ \textbf{A}: Using model 1 (unadjusted), continuous color shading; \textbf{B}: Using model 1 (unadjusted), discrete color shading; \textbf{C}: Using model 2 (adjusted), continuous color shading; \textbf{D}: Using model 2 (adjusted), discrete color shading.}
\end{figure}

\begin{figure}[!htb]
	\centering
	\includegraphics[width=0.84\linewidth]{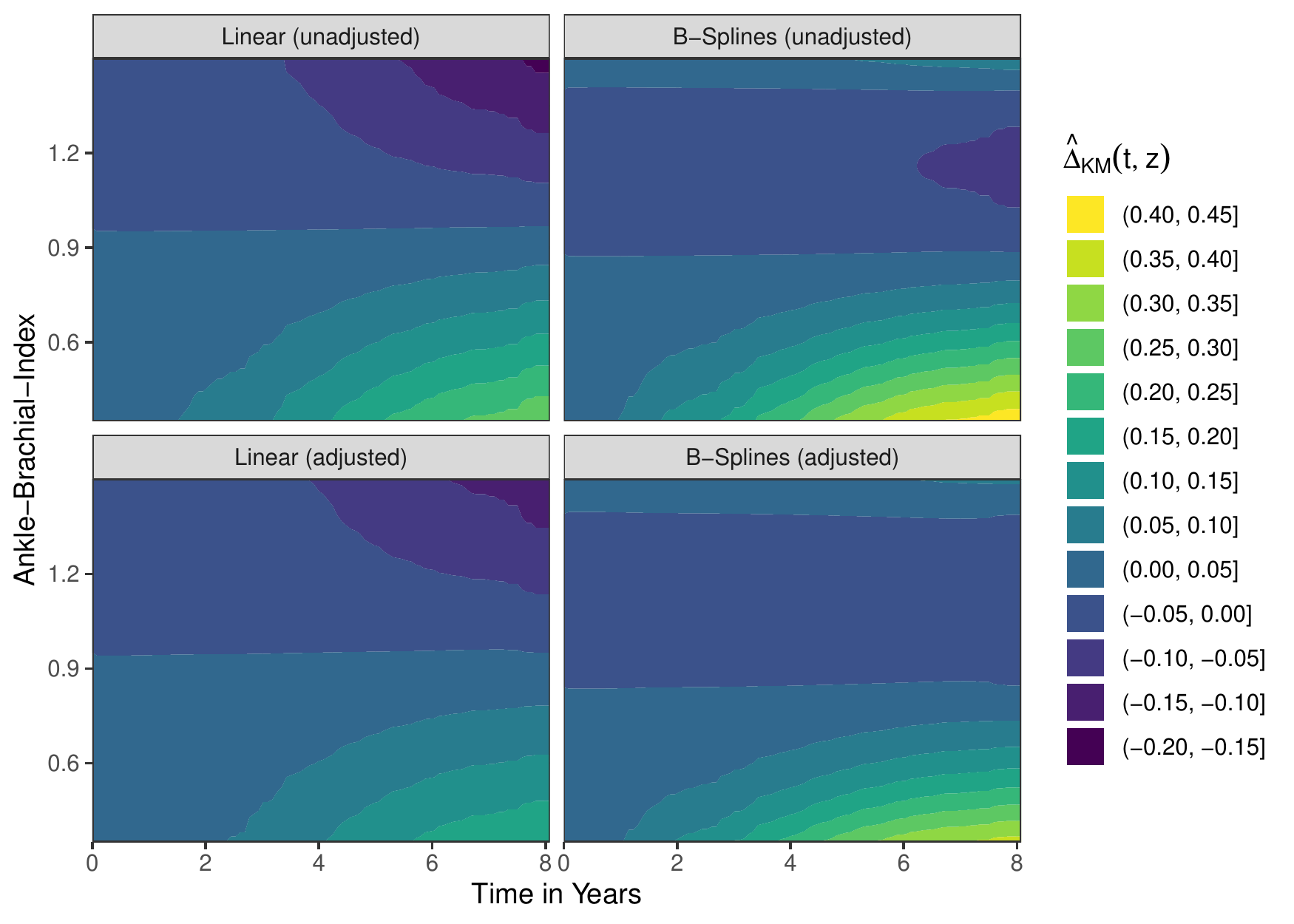}
	\caption{Survival contour plots displaying the causal effect of the ABI on survival using $\hat{\Delta}_{KM}$, created using four different Cox proportional hazards regression models and g-computation.}
\end{figure}

\begin{figure}[!htb]
	\centering
	\includegraphics[width=0.84\linewidth]{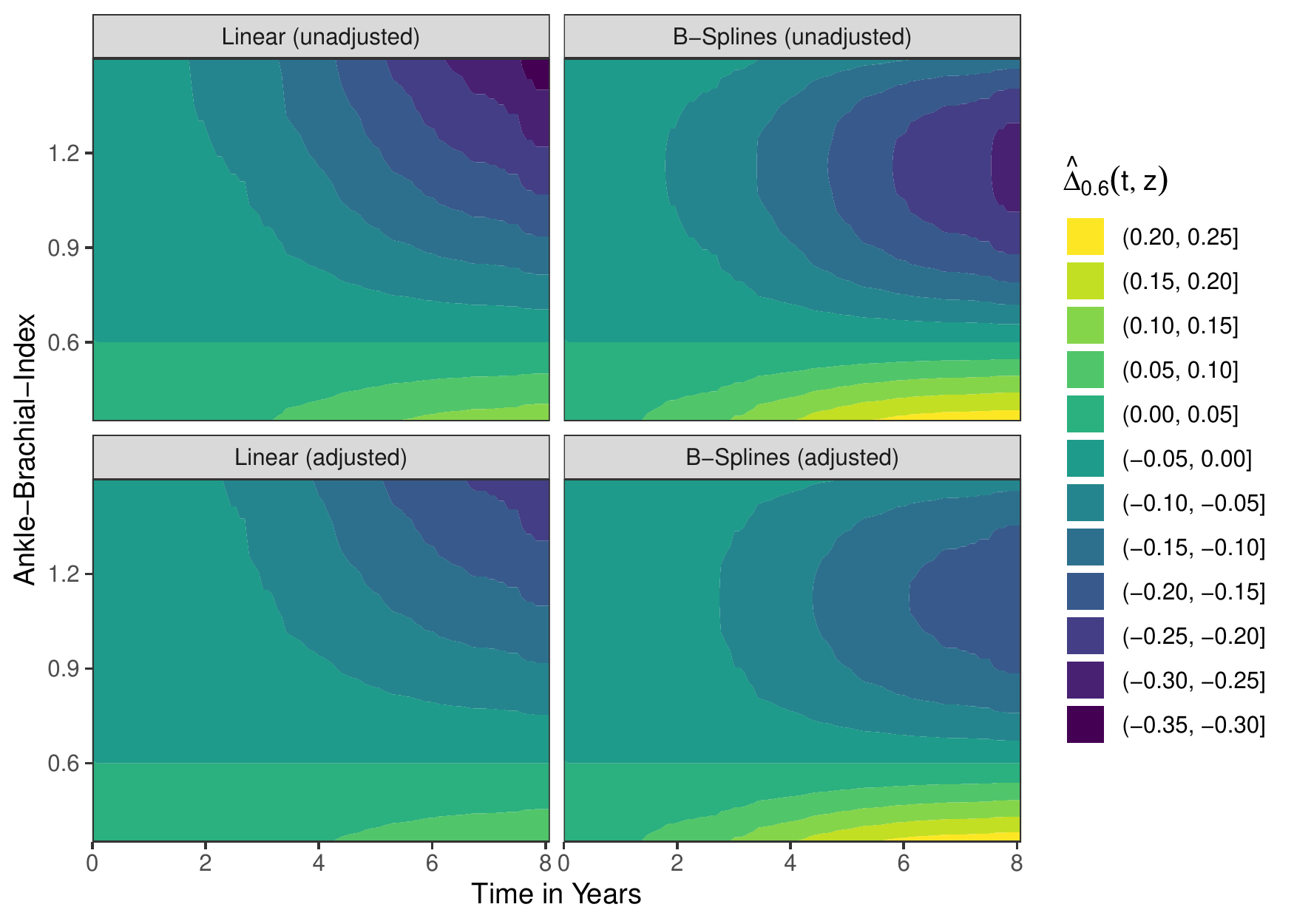}
	\caption{Survival contour plots displaying the causal effect of the ABI on survival using $\hat{\Delta}_{0.6}$, created using four different Cox proportional hazards regression models and g-computation.}
\end{figure}

\begin{figure}[!htb]
	\centering
	\includegraphics[width=0.84\linewidth]{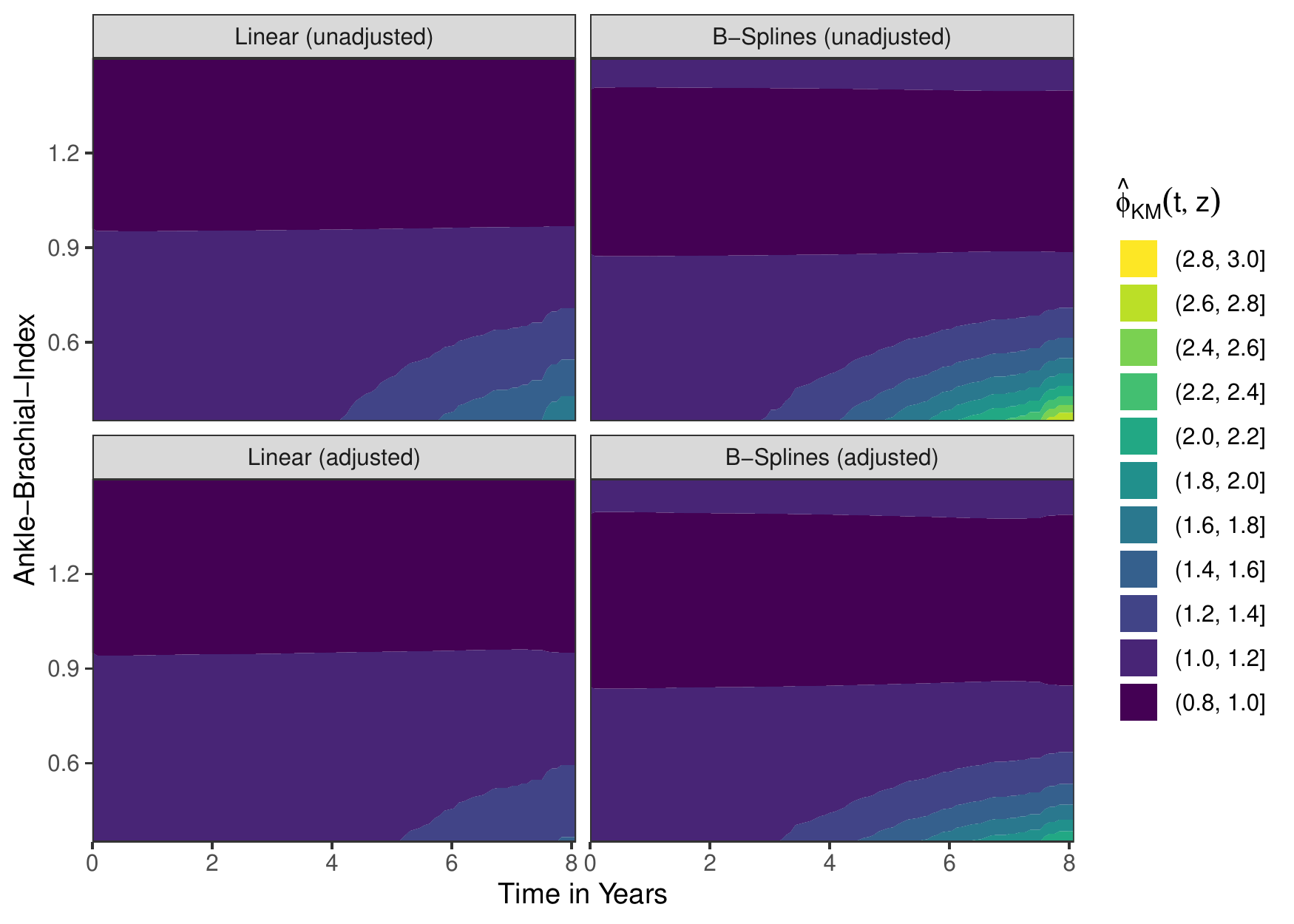}
	\caption{Survival contour plots displaying the causal effect of the ABI on survival using $\hat{\phi}_{KM}$, created using four different Cox proportional hazards regression models and g-computation.}
\end{figure}

\begin{figure}[!htb]
	\centering
	\includegraphics[width=0.84\linewidth]{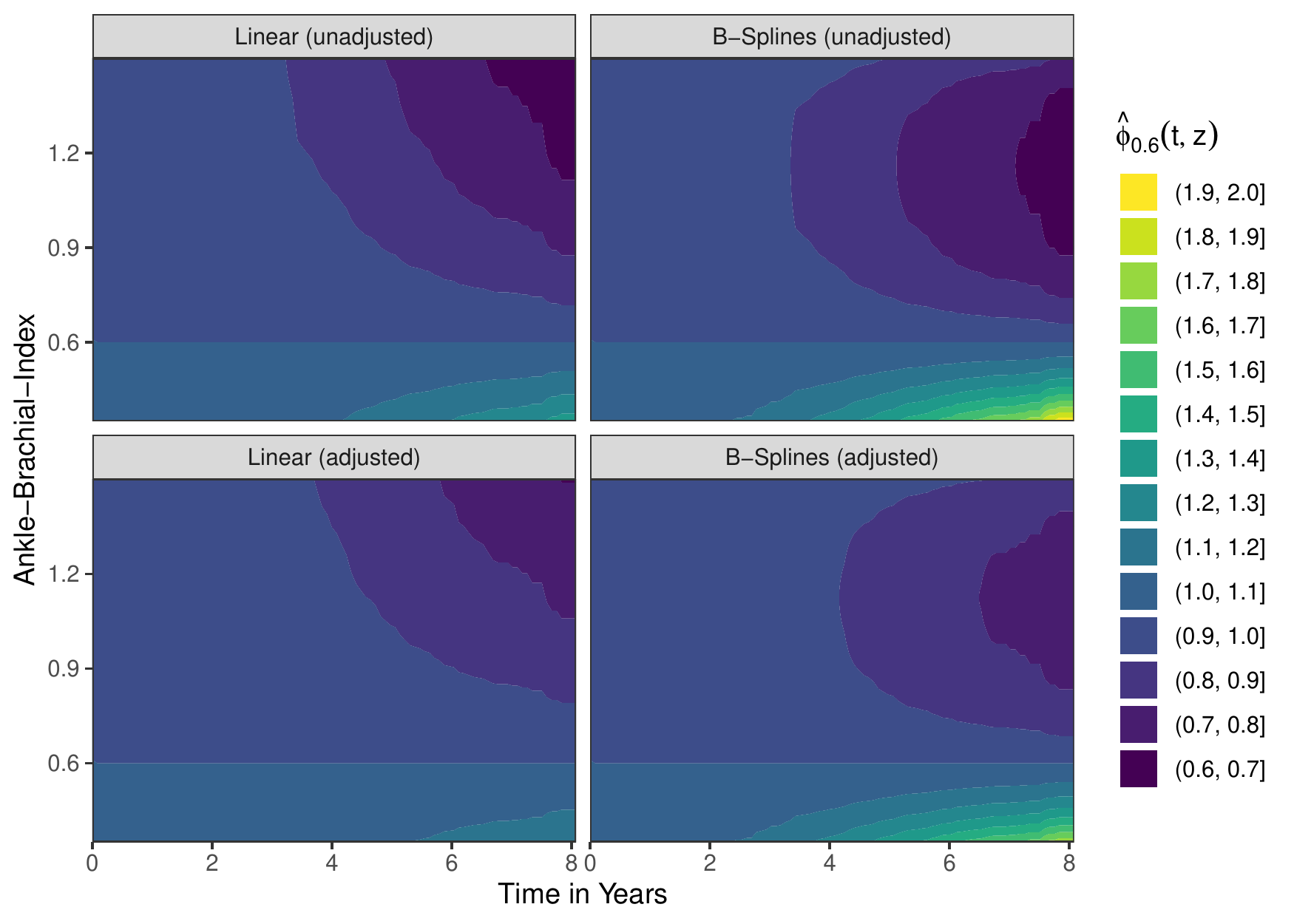}
	\caption{Survival contour plots displaying the causal effect of the ABI on survival using $\hat{\phi}_{0.6}$, created using four different Cox proportional hazards regression models and g-computation.}
\end{figure}

\begin{figure}[!htb]
	\centering
	\includegraphics[width=0.84\linewidth]{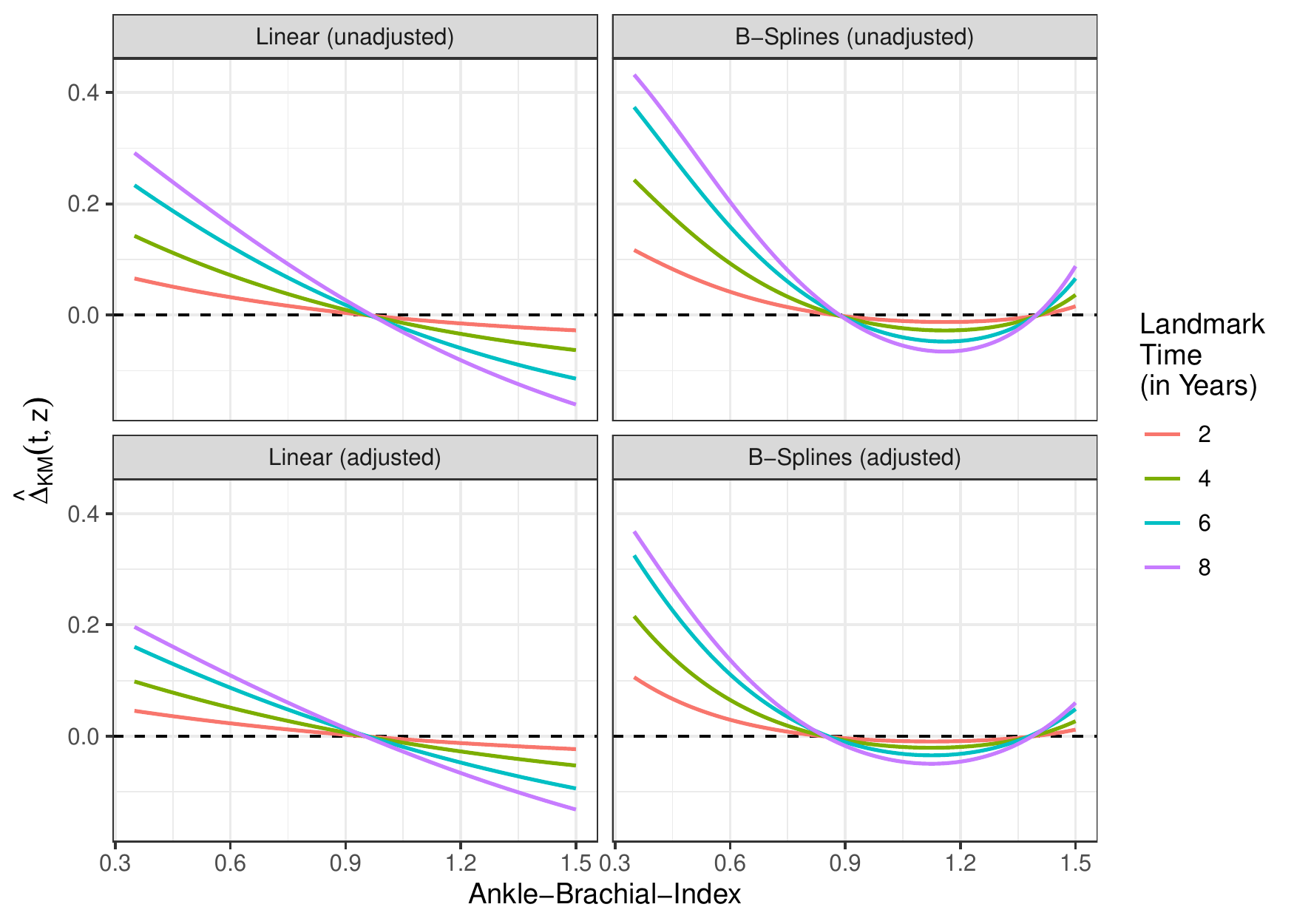}
	\caption{$\hat{\Delta}_{KM}$ at different points in time $t$ as a function of the ABI, created using four different Cox proportional hazards regression models and g-computation.}
\end{figure}

\begin{figure}[!htb]
	\centering
	\includegraphics[width=0.84\linewidth]{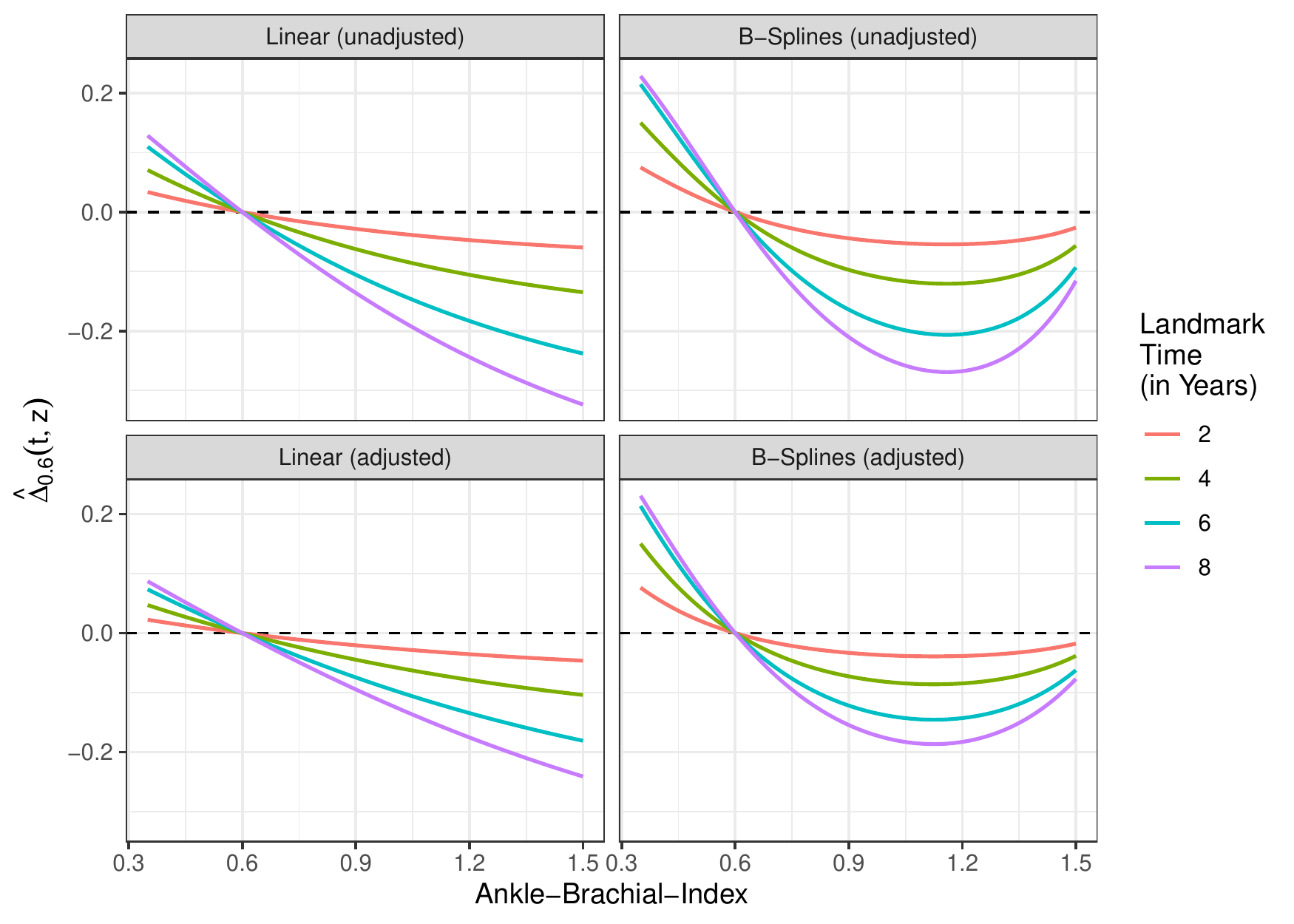}
	\caption{$\hat{\Delta}_{0.6}$ at different points in time $t$ as a function of the ABI, created using four different Cox proportional hazards regression models and g-computation.}
\end{figure}

\begin{figure}[!htb]
	\centering
	\includegraphics[width=0.84\linewidth]{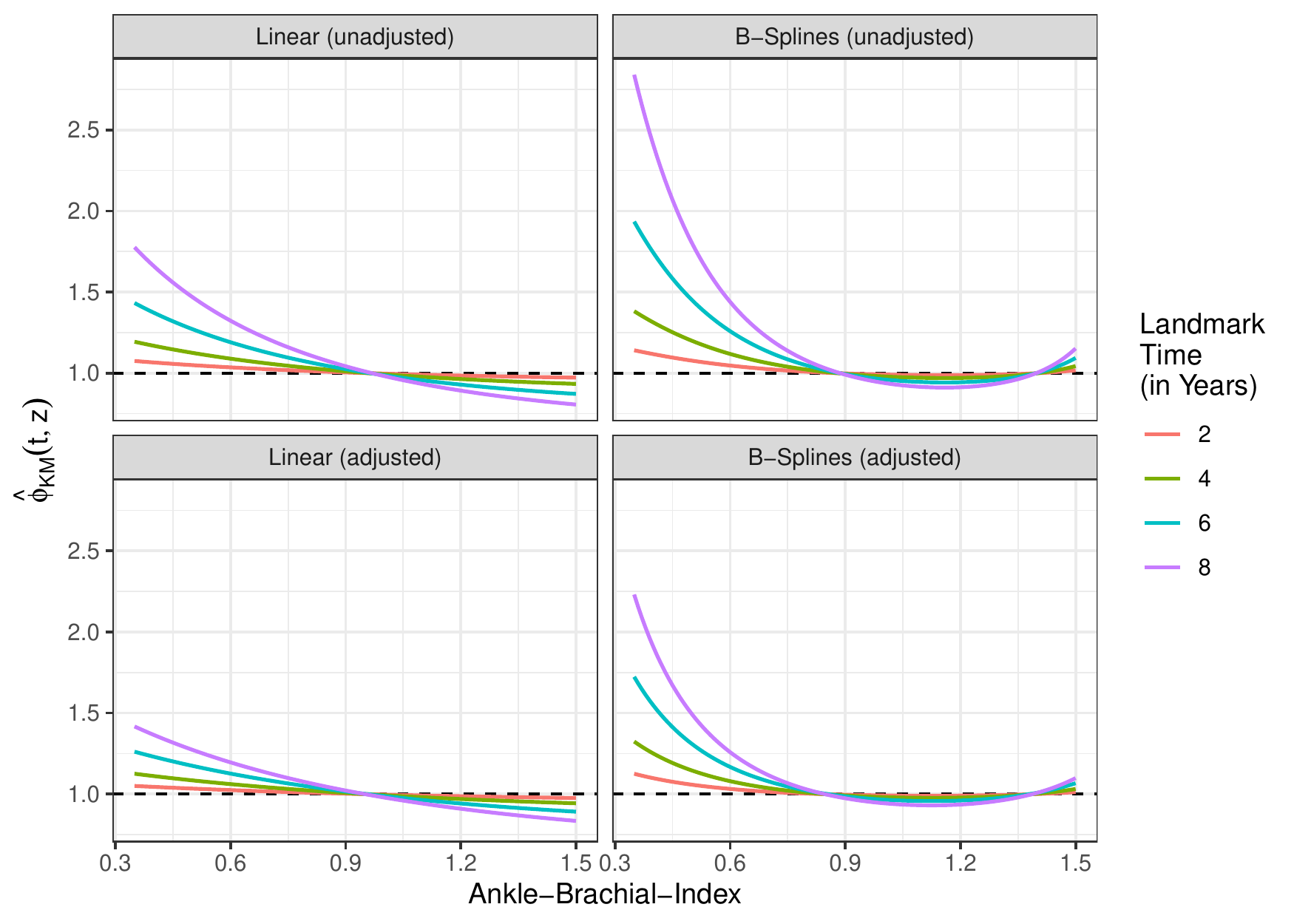}
	\caption{$\hat{\phi}_{KM}$ at different points in time $t$ as a function of the ABI, created using four different Cox proportional hazards regression models and g-computation.}
\end{figure}

\begin{figure}[!htb]
	\centering
	\includegraphics[width=0.84\linewidth]{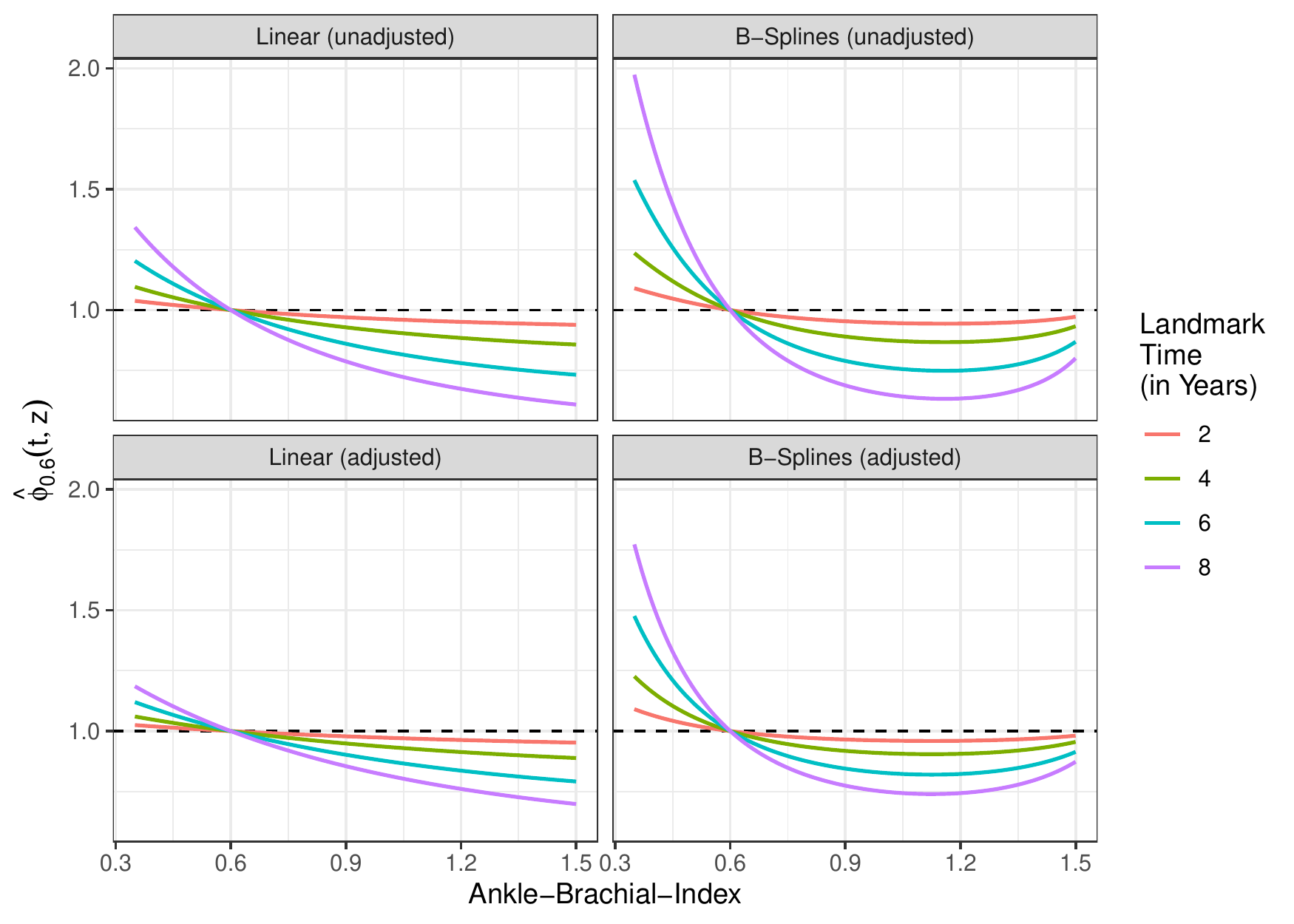}
	\caption{$\hat{\phi}_{0.6}$ at different points in time $t$ as a function of the ABI, created using four different Cox proportional hazards regression models and g-computation.}
\end{figure}

\FloatBarrier
\newpage

\section{Survival Area Plots with Non-Monotonic Effects}

As described in the main text, using a standard survival area plot it is impossible to depict non-monotonic effects of the continuous variable on the survival probability. Whenever there is a curved relationship between the continuous variable and the survival probability, some parts of the areas to be drawn will occupy the same space on the two-dimensional representation, making the plot unusable. In our illustrative example this is the case in model (4) and model (5), in which B-splines were used to model the effect of the ABI.
\par
\begin{figure}[!htb]
	\centering
	\includegraphics[width=0.84\linewidth]{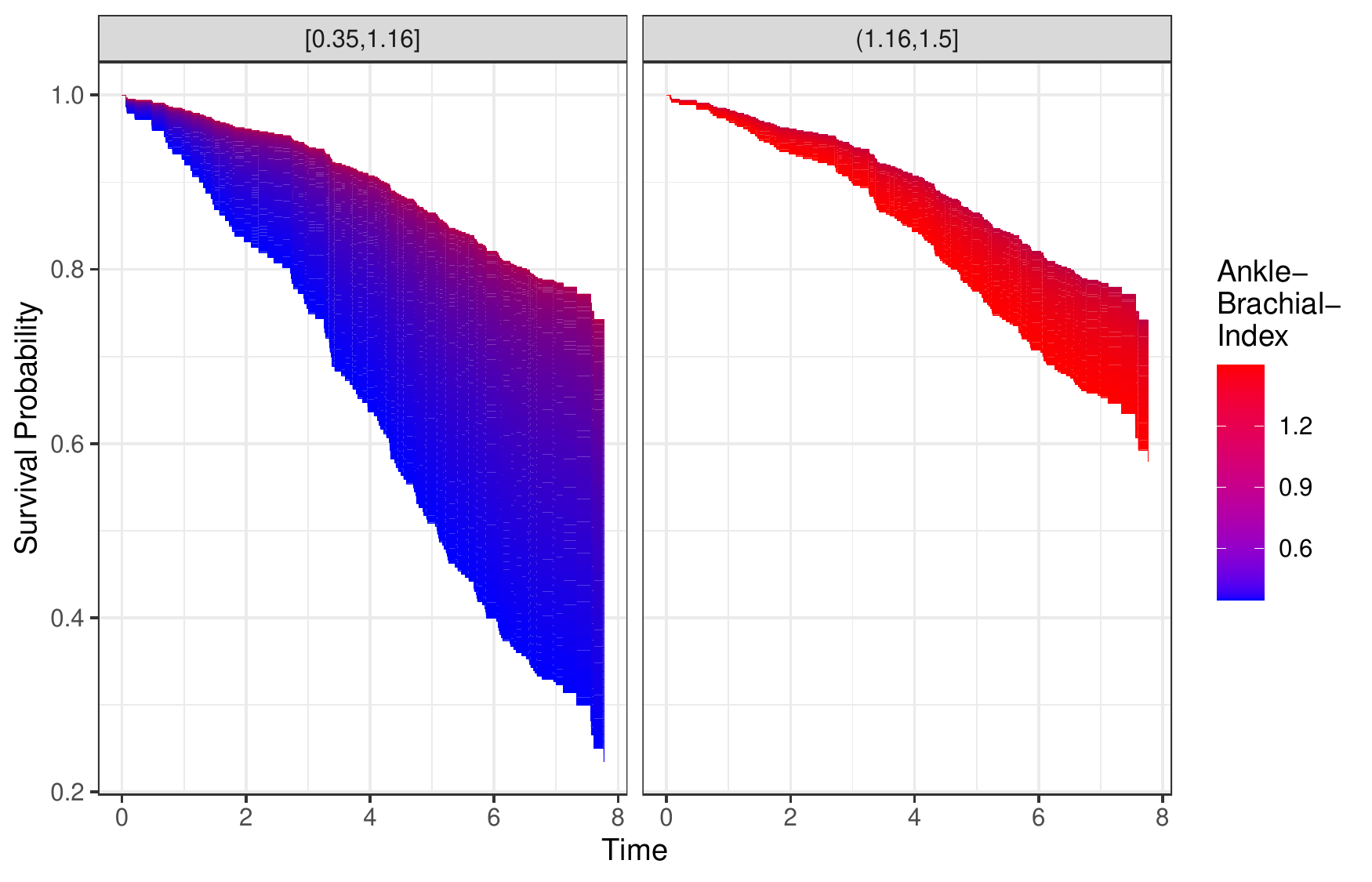}
	\caption{Faceted survival area plot displaying the causal effect of the ABI on survival as estimated using model (4).}
	\label{fig::surv_area_bs}
\end{figure}

\begin{figure}[!htb]
	\centering
	\includegraphics[width=0.84\linewidth]{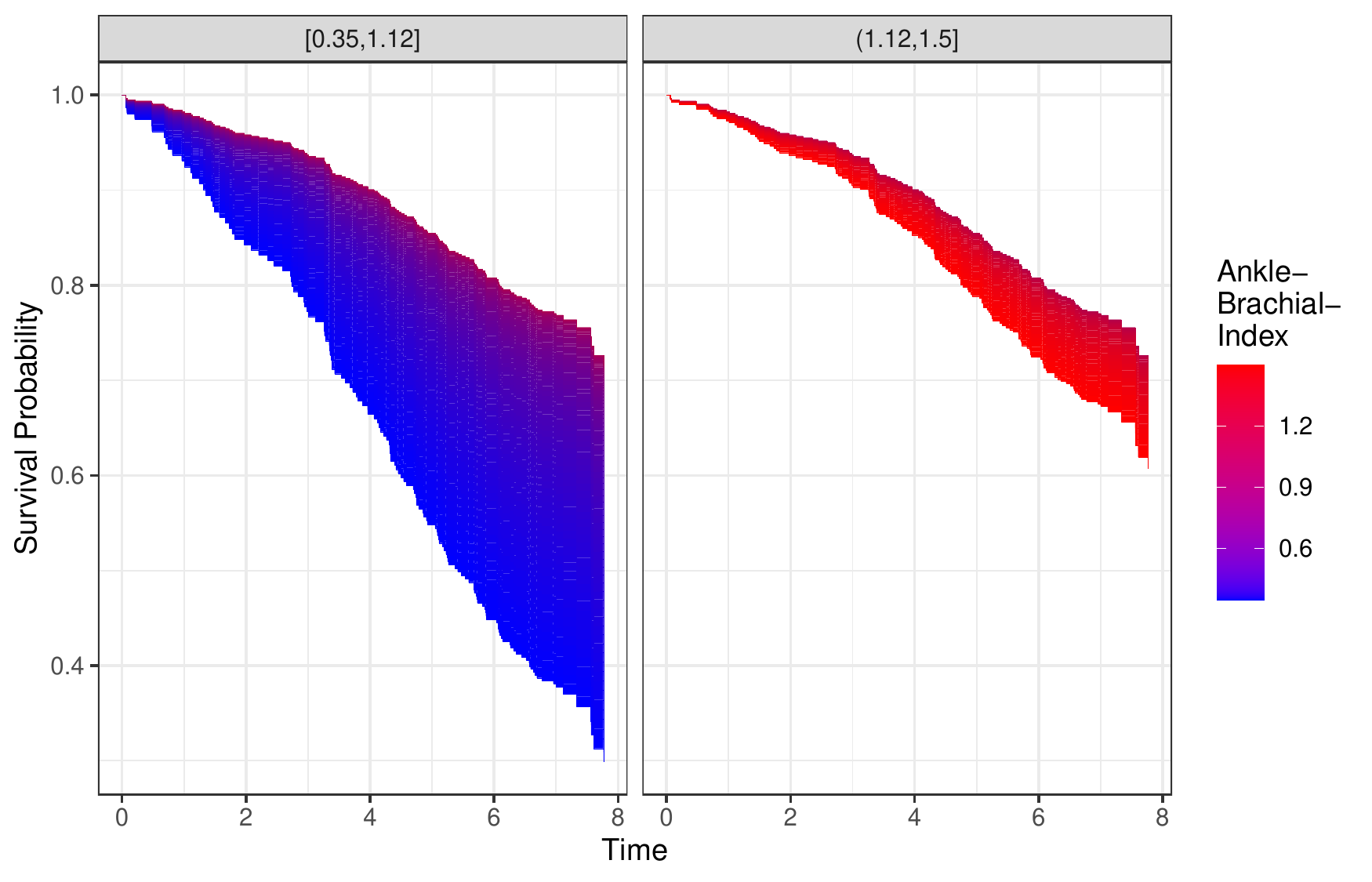}
	\caption{Faceted survival area plot displaying the causal effect of the ABI on survival as estimated using model (5).}
	\label{fig::surv_area_bs_adj}
\end{figure}

\medskip
The only possible way to still use survival area plots in this case is to divide the plot into multiple facets. Each facet only contains a survival area plot for different ranges of the covariate space in which its effect on the survival probability is either non-increasing or non-decreasing. Figure~\ref{fig::surv_area_bs} shows the resulting faceted plot when using model (4) and figure~\ref{fig::surv_area_bs_adj} shows the results when using model (5).
\par\medskip

The left facet of figure~\ref{fig::surv_area_bs} contains the survival area plot for values of the ABI between $0.35$ and $1.16$, while the right facet contains the survival area plot for the rest of the considered range of the ABI. This works because the survival probability strictly increases with higher ABI values in the left facet and it strictly decreases with higher ABI values in the right facet. If the relationship between the variable of interest and the survival probability is more complex, the plot would have to be divided into more facets. Plots with such relationships become difficult to understand very fast, which is why we recommend using different plotting strategies, such as survival contour plots, in those cases instead.
\par\medskip
It is important to note that this strategy of dividing the survival area plot into different segments only works if the shape of the relationship between the continuous variable and the survival probability is the same at each point in time. If there are interactions between time and the shape of the covariate, this strategy is not viable. Additionally, it requires the survival probability to be strictly non-increasing with increasing time over the whole range of the continuous variable.

\FloatBarrier
\newpage

\section{On the Assumption of Random Right-Censoring}

Throughout the article, we explicitly assumed that right-censoring is independent of any other variable. This assumption allowed us to ignore censoring in the definition of the target estimand and the description of the causal identifiability assumptions. In reality, however, multiple variables may have complex relationships with the occurrence of censoring. In this section, we give a short description about the implications of dependent right-censoring on the target estimand, required assumptions and estimation methodology, based on work by Hernán and Robins (2020).
\par\medskip
First, the target estimand needs to be slightly changed in this case. We are not just interested in the population-level counterfactual survival probability if the value of $Z$ was set to $z$ for every individual in the population. Instead we are interested in the population-level counterfactual survival probability if the value of $Z$ was set to $z$ for every individual in the population \emph{and} if they had remained uncensored. This is technically what we are interested in whether right-censoring is random or not, but if it is completely random it can be safely ignored. Let $C$ be a dichotomous indicator of right-censoring with $C = 0$ indicating the absence of censoring and $C = 1$ indicating that an observation is censored. Keeping the rest of the notation of section~2.1, the target estimand can be re-defined as:

\begin{equation}
	S_z(t) = E\left(I\left(T^{Z=z, C=0} > t\right)\right).
\end{equation}

Essentially, this means that we are no longer interested in the causal effect of a single treatment on $Z$, but instead that we are interested in the causal effect of the \emph{joint treatment} on $Z$ and $C$.
\par\medskip
This changes the assumptions required to make $S_z(t)$ identifiable. We still need to make the \emph{no interference}, the \emph{counterfactual consistency} assumption, the \emph{conditional exchangeability} assumption and the \emph{positivity} assumption, but those are no longer made on just $Z$, but instead have to be made for the joint treatment on $Z$ and $C$. This makes the estimation of $S_z(t)$ using g-computation more complicated, especially when the underlying causal structure is highly complex. Westreich et al. (2012) contains an applied example and more discussion about this topic.
\par\bigskip
\par\medskip
\textbf{Literature:} \\
Miguel A. Hernán and James M. Robins. \emph{Causal Inference: What If}. CRC Press, 2020. [section 8.4 and 17]
\\
Daniel Westreich, Stephen R. Cole, Jessica G. Young, Frank Palella, Phyllis C. Tien, Lawrence Kingsley, Stephen J Gange, Miguel A. Hernán. ``The parametric g-formula to estimate the effect of highly active antiretroviral therapy on incident AIDS or death''. In: \emph{Statistics in Medicine} 31.18 (2012).

\newpage

\section{Additional Results}

The following text includes a histogram of the ABI in the used getABI cohort, as well as the results of graphical tests of the proportional hazards assumption for all Cox models used and graphics used to identify the correct functional form of the ABI on survival.

\subsection{Distribution of the ABI}

\begin{figure}[!htb]
	\centering
	\includegraphics[width=0.84\linewidth]{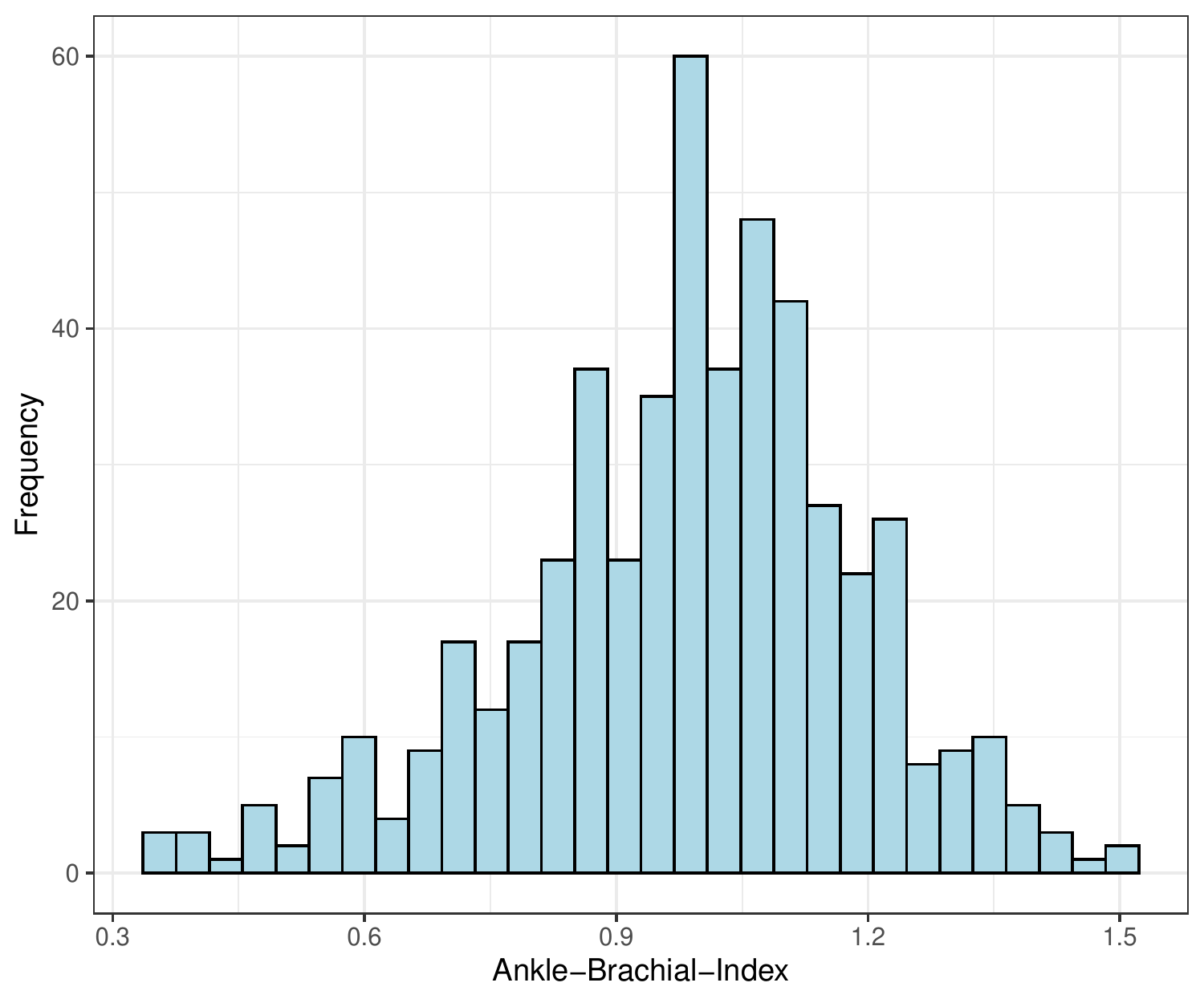}
	\caption{Histogram showing the distribution of the ABI in the used population.}
\end{figure}

\FloatBarrier

\subsection{Graphical Tests of the Proportional Hazards Assumption}

\begin{figure}[!htb]
	\centering
	\includegraphics[width=0.84\linewidth]{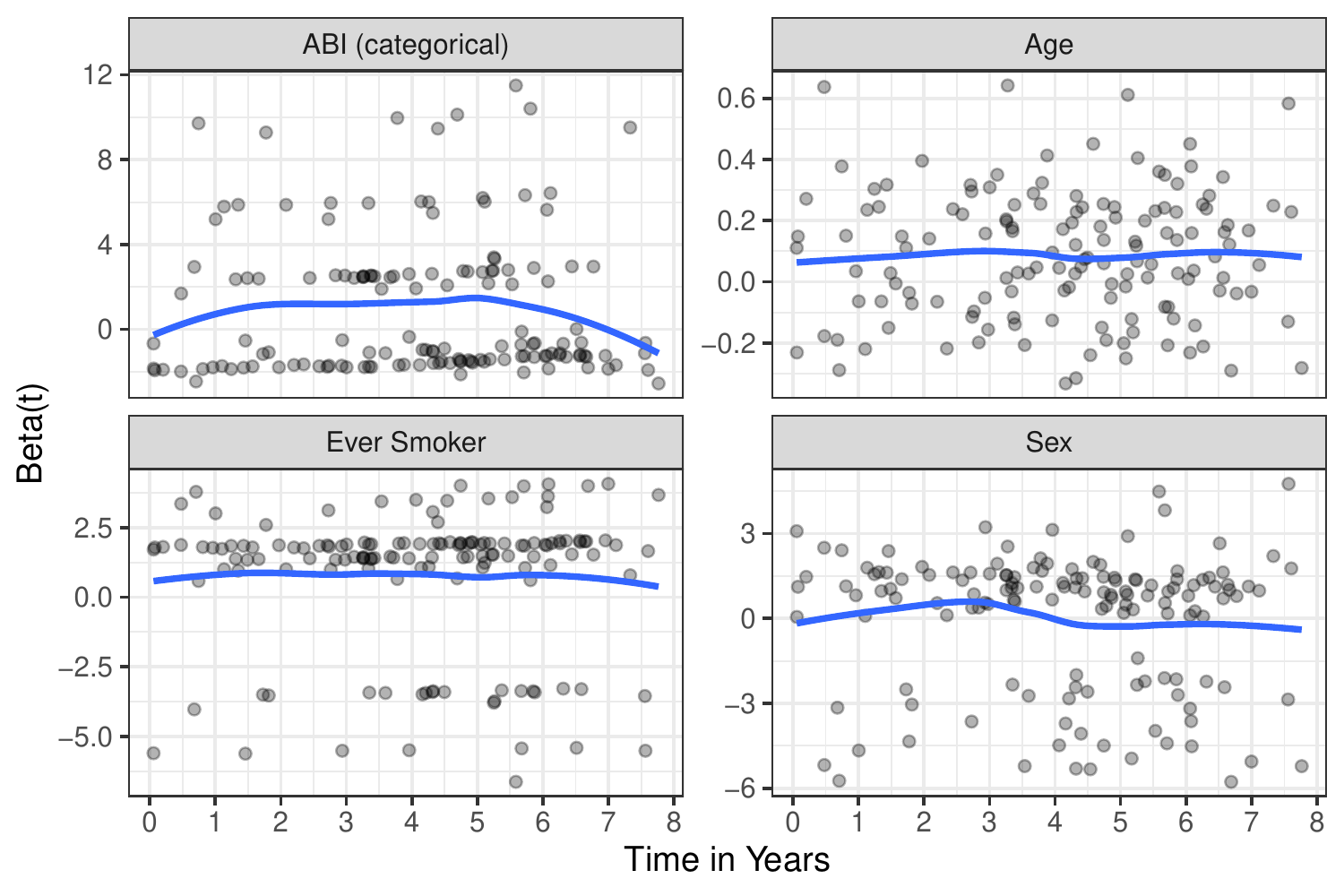}
	\caption{A graphical test for the proportional hazards assumption for each variable in Cox model (1). The blue lines are non-parametric locally weighted regressions.}
\end{figure}

\begin{figure}[!htb]
	\centering
	\includegraphics[width=0.84\linewidth]{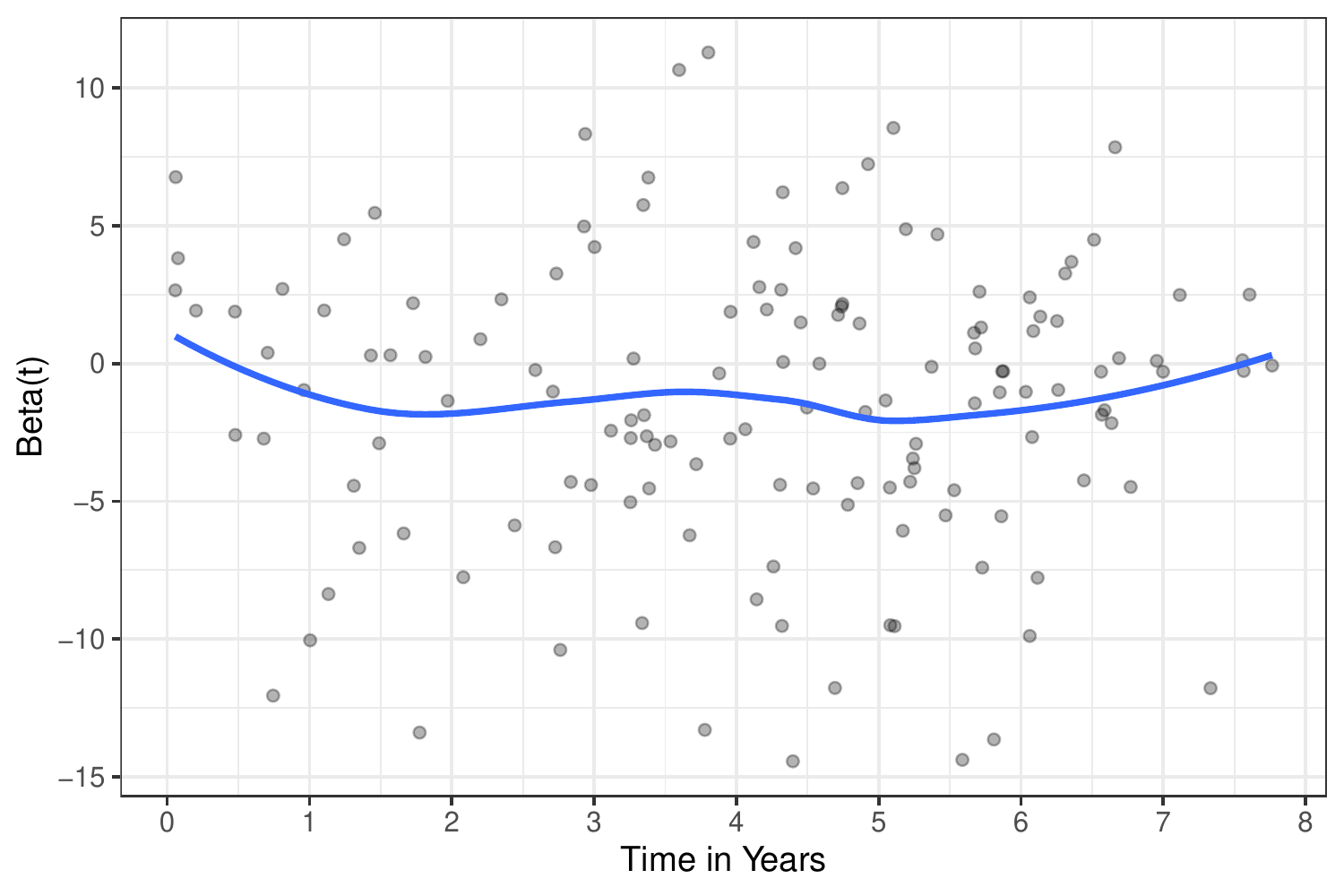}
	\caption{A graphical test for the proportional hazards assumption for the Ankle-Brachial-Index in Cox model (2). The blue line is a non-parametric locally weighted regression.}
\end{figure}

\begin{figure}[!htb]
	\centering
	\includegraphics[width=0.84\linewidth]{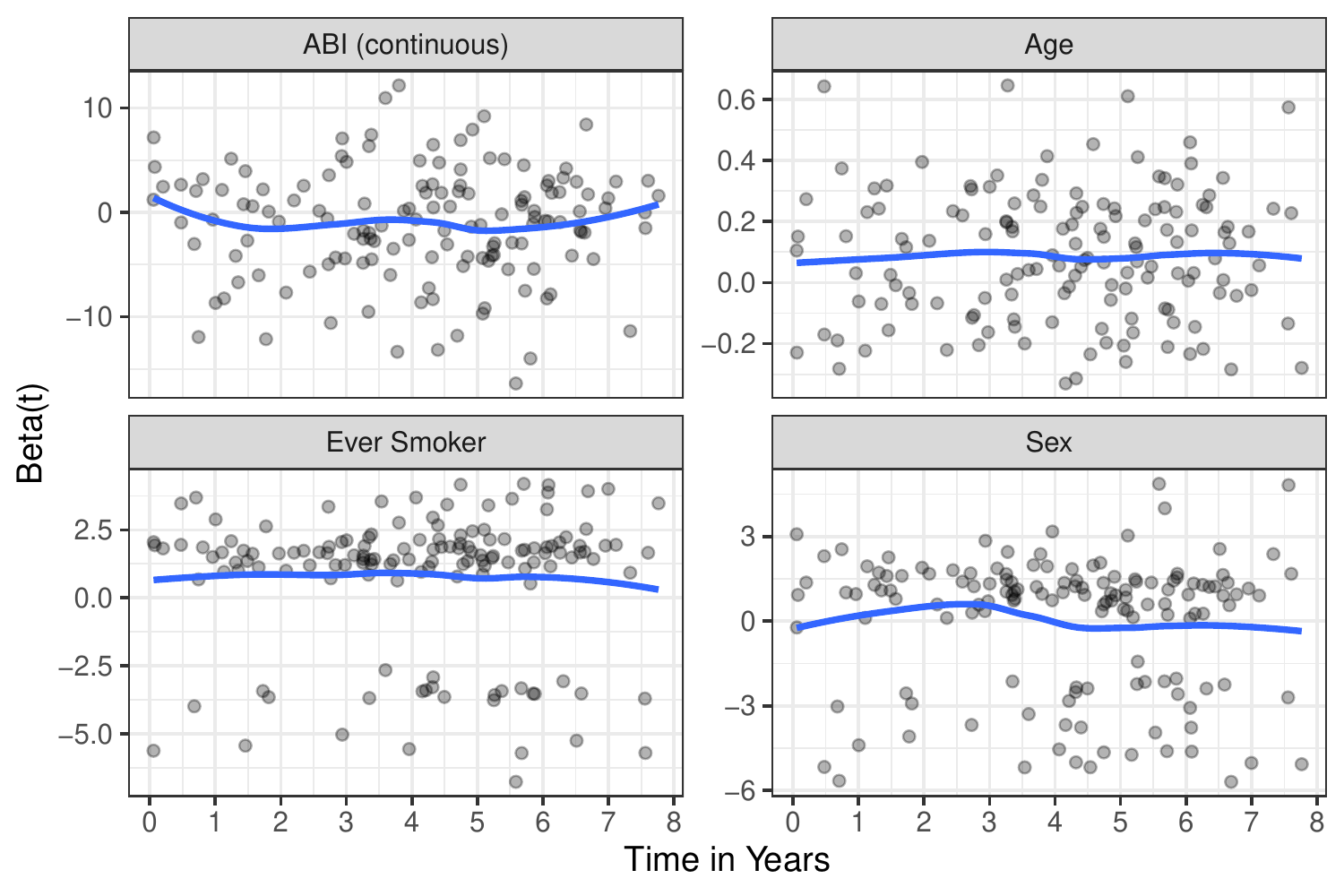}
	\caption{A graphical test for the proportional hazards assumption for each variable in Cox model (3). The blue lines are non-parametric locally weighted regressions.}
\end{figure}

\begin{figure}[!htb]
	\centering
	\includegraphics[width=0.84\linewidth]{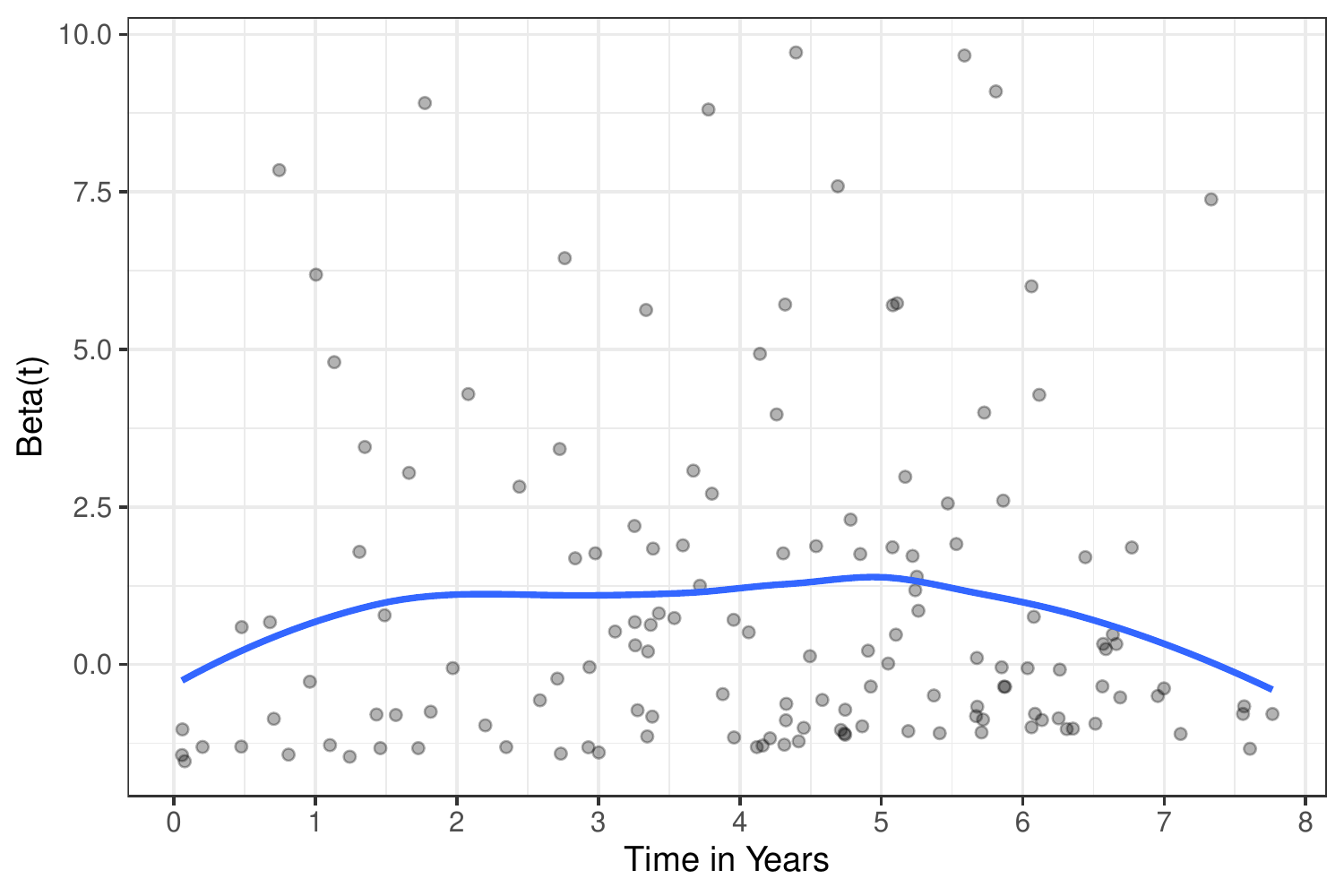}
	\caption{A graphical test for the proportional hazards assumption for the Ankle-Brachial-Index in Cox model (4). The blue line is a non-parametric locally weighted regressions.}
\end{figure}

\begin{figure}[!htb]
	\centering
	\includegraphics[width=0.84\linewidth]{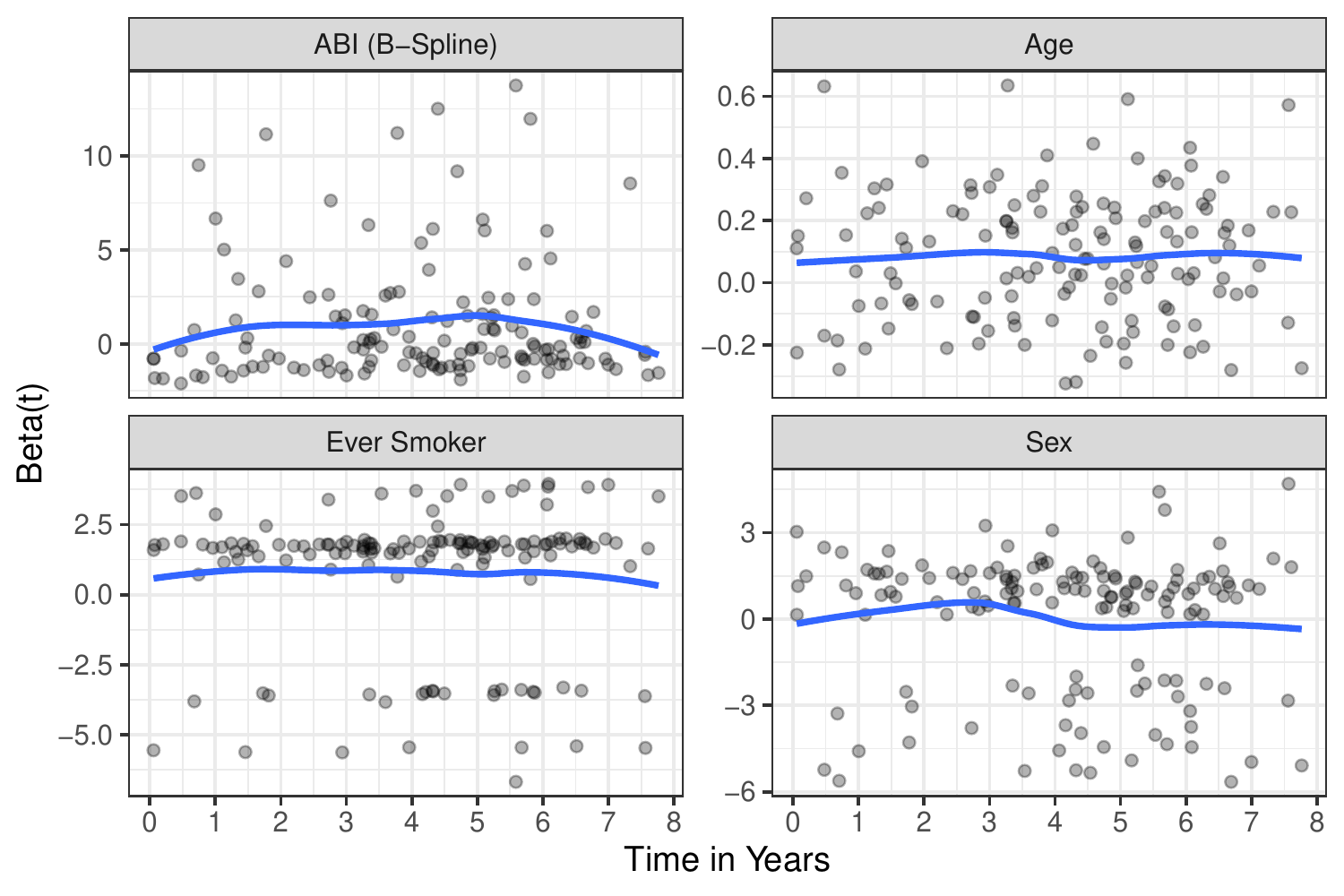}
	\caption{A graphical test for the proportional hazards assumption for each variable in Cox model (5). The blue lines are non-parametric locally weighted regressions.}
\end{figure}

\FloatBarrier

\subsection{On the Functional Form of the Relationship Between ABI and Survival}

\begin{figure}[!htb]
	\centering
	\includegraphics[width=0.84\linewidth]{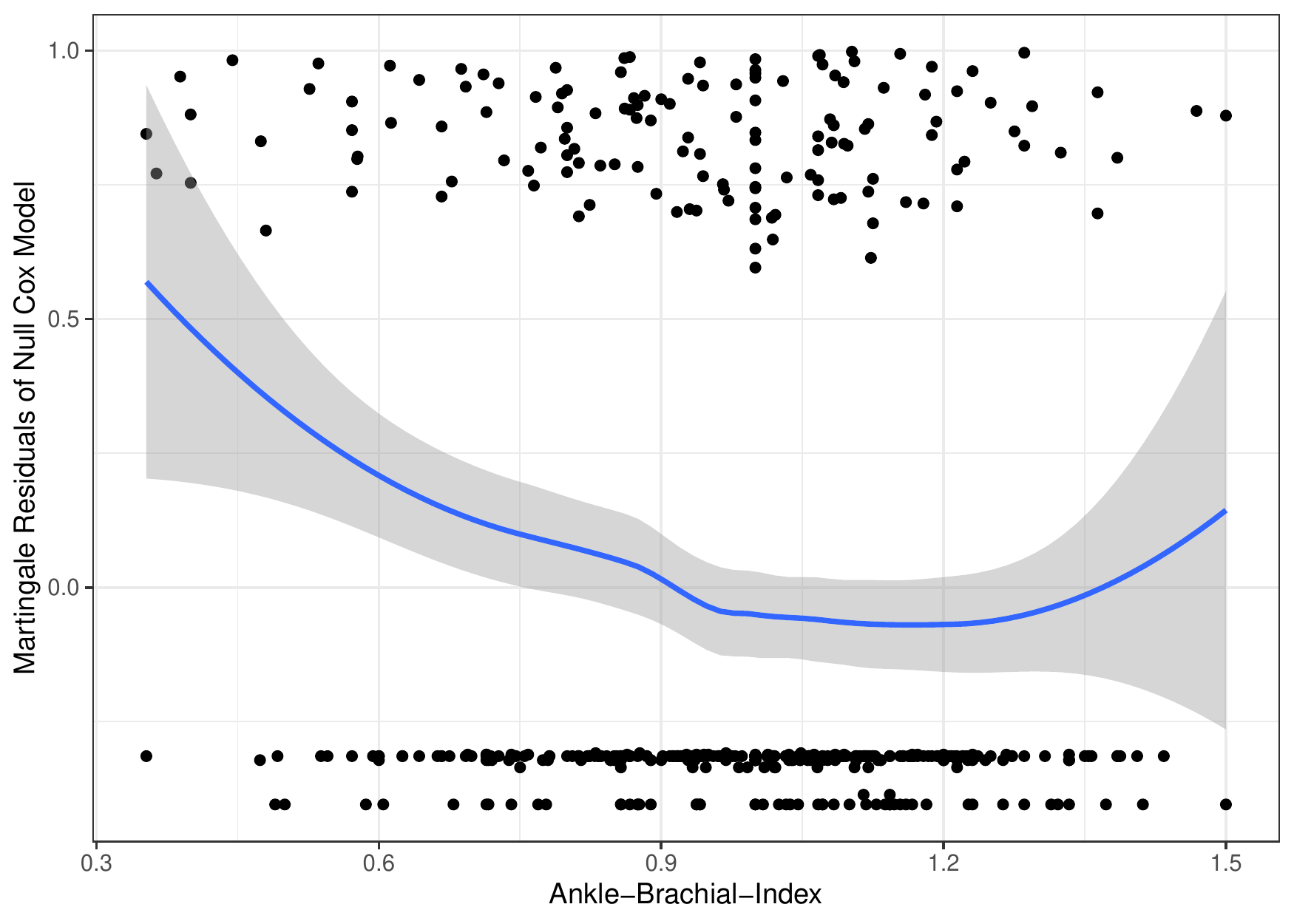}
	\caption{Martingale Residuals of an intercept-only Cox model versus the ABI. The blue line shown is a non-parametric locally weighted regression plus its 95\% confidence band.}
	\label{fig::ff_martingales}
\end{figure}

\begin{figure}[!htb]
	\centering
	\includegraphics[width=0.84\linewidth]{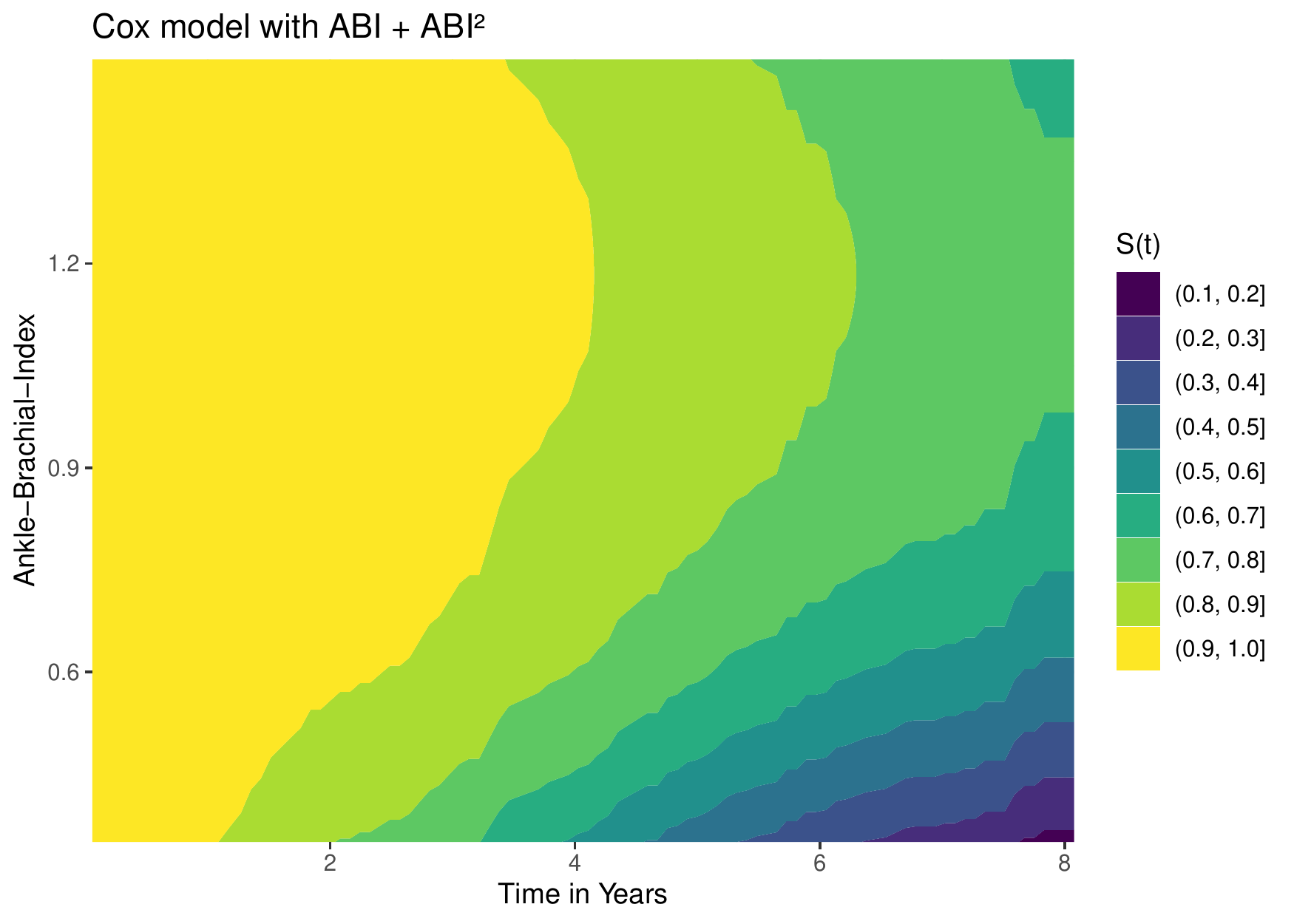}
	\caption{Survival contour plot created using a Cox model including the ABI and ABI$^2$.}
	\label{fig::ff_cube_1}
\end{figure}

\begin{figure}[!htb]
	\centering
	\includegraphics[width=0.84\linewidth]{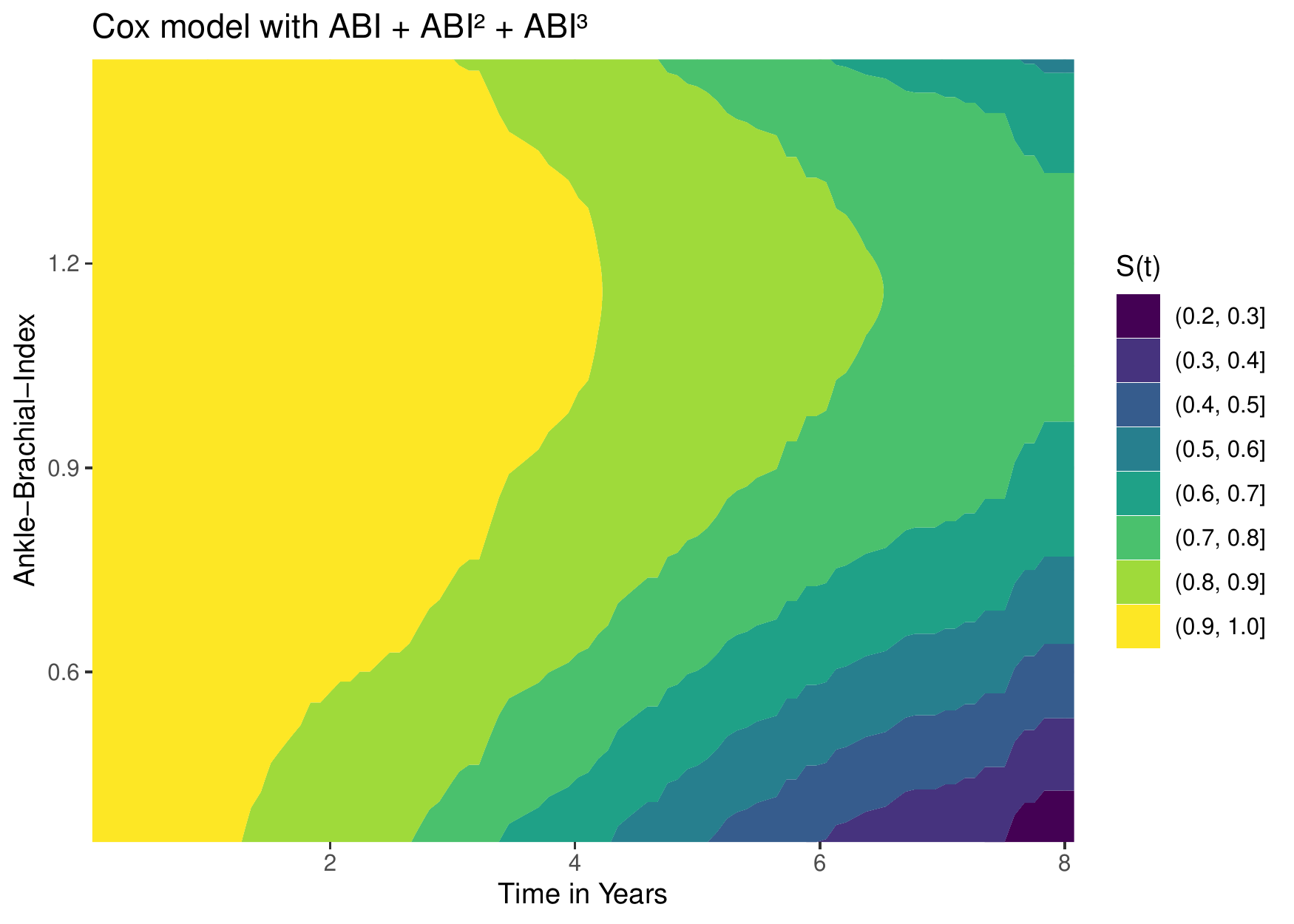}
	\caption{Survival contour plot created using a Cox model including the ABI, ABI$^2$ and ABI$^3$.}
	\label{fig::ff_cube_2}
\end{figure}

\begin{figure}[!htb]
	\centering
	\includegraphics[width=0.84\linewidth]{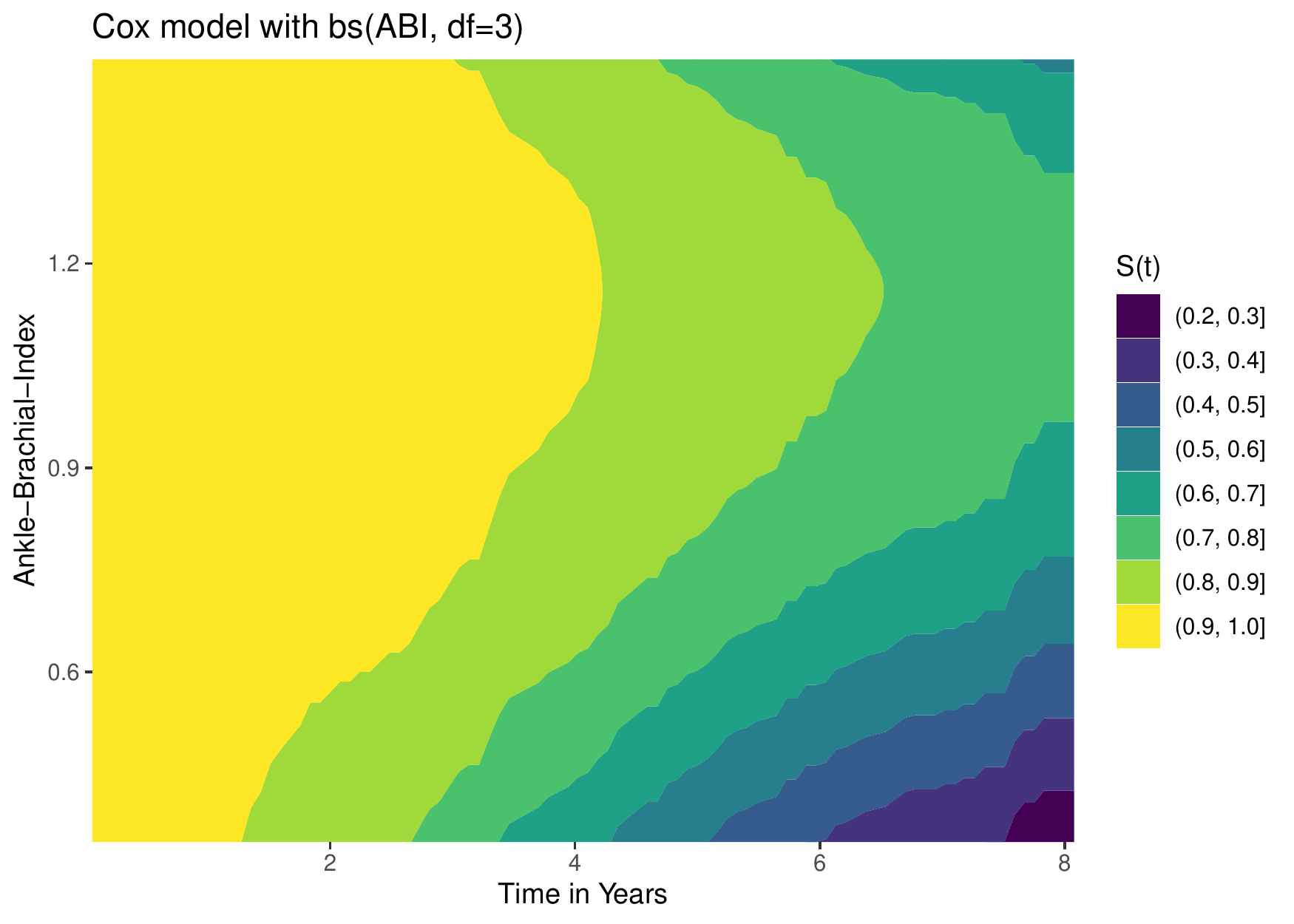}
	\caption{Survival contour plot created using a Cox model including the ABI modelled using B-splines with three degrees of freedom.}
	\label{fig::ff_bs_3}
\end{figure}

\begin{figure}[!htb]
	\centering
	\includegraphics[width=0.84\linewidth]{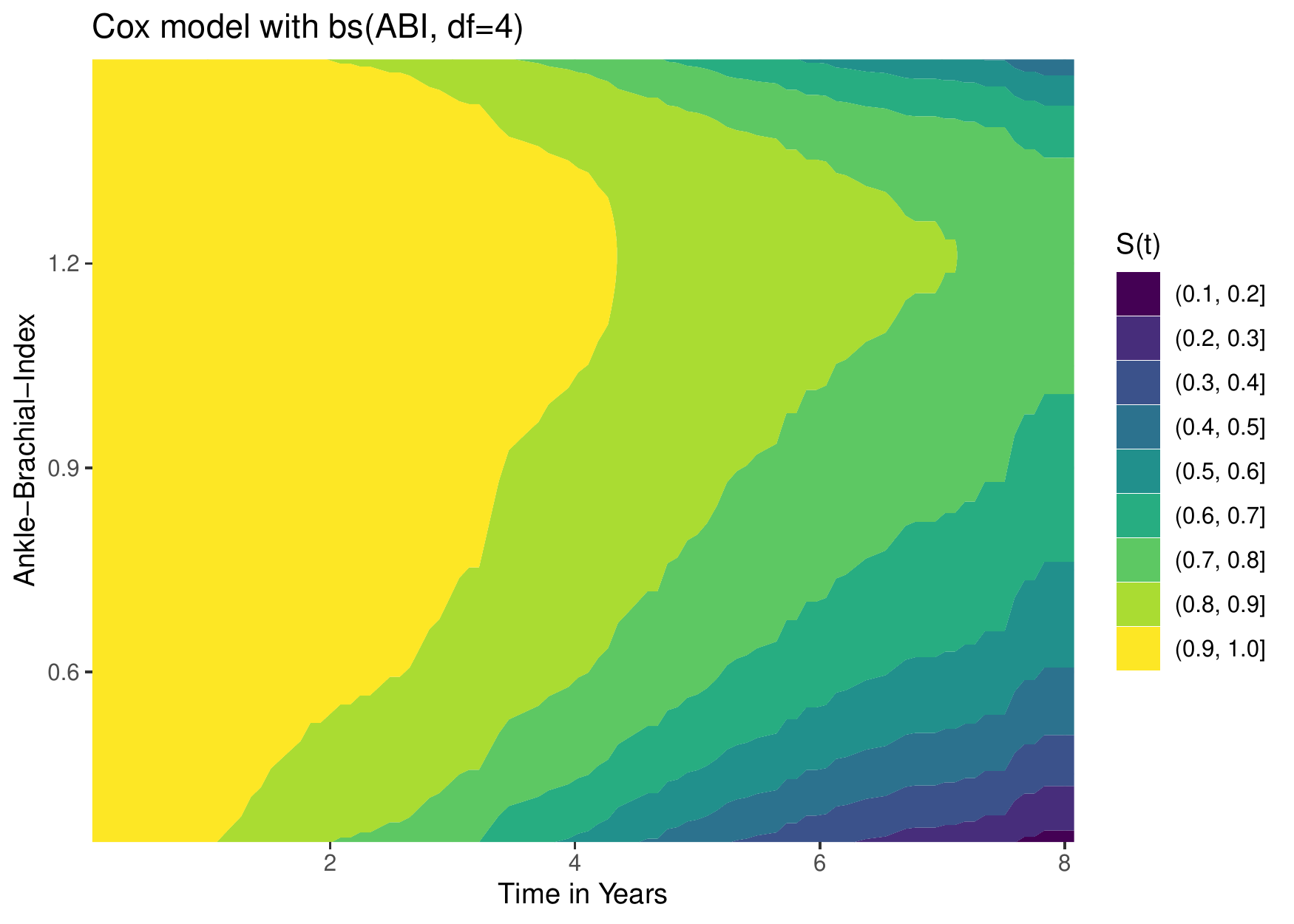}
	\caption{Survival contour plot created using a Cox model including the ABI modelled using B-splines with four degrees of freedom.}
	\label{fig::ff_bs_4}
\end{figure}

\begin{figure}[!htb]
	\centering
	\includegraphics[width=0.84\linewidth]{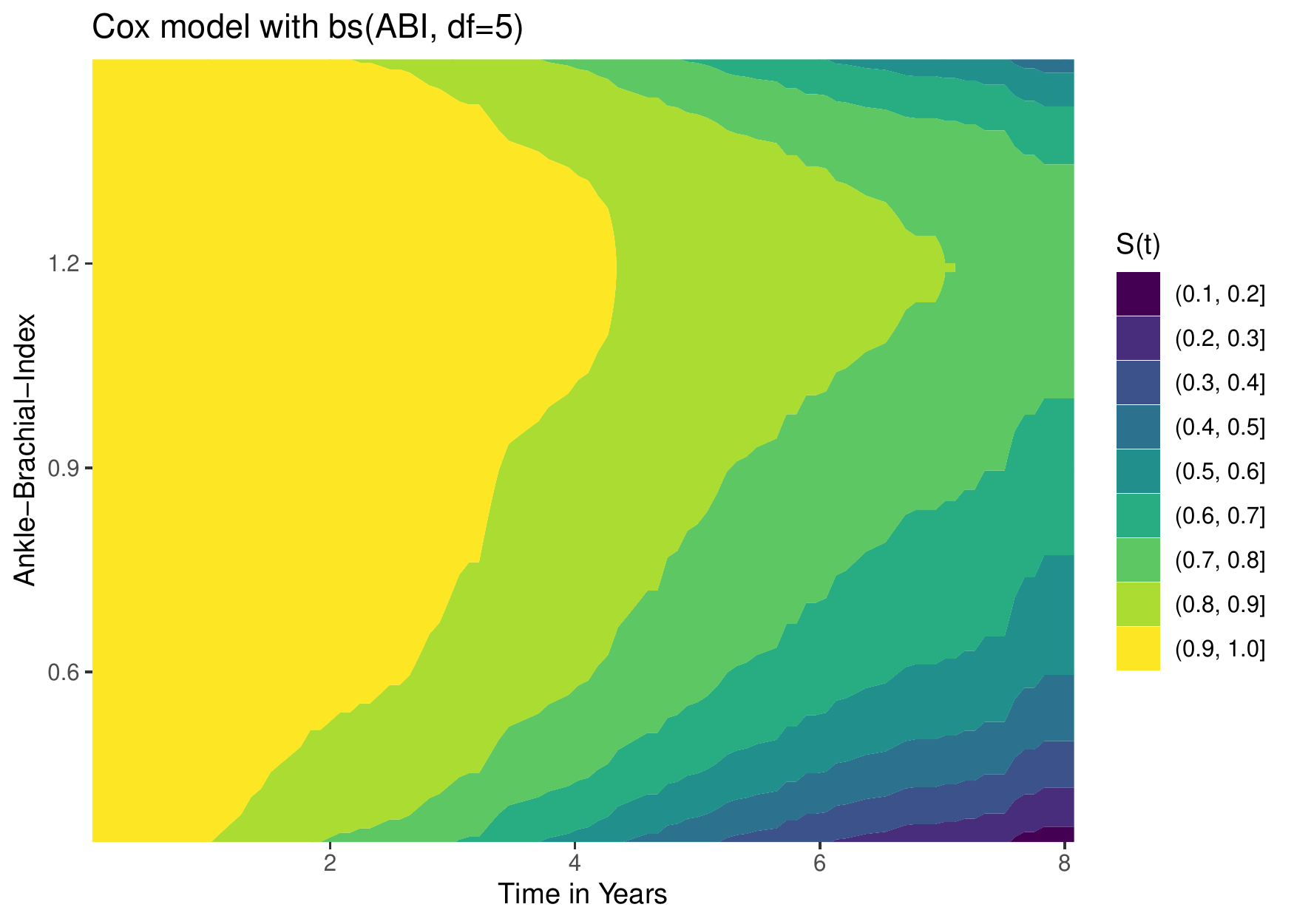}
	\caption{Survival contour plot created using a Cox model including the ABI modelled using B-splines with five degrees of freedom.}
	\label{fig::ff_bs_5}
\end{figure}

\begin{figure}[!htb]
	\centering
	\includegraphics[width=0.84\linewidth]{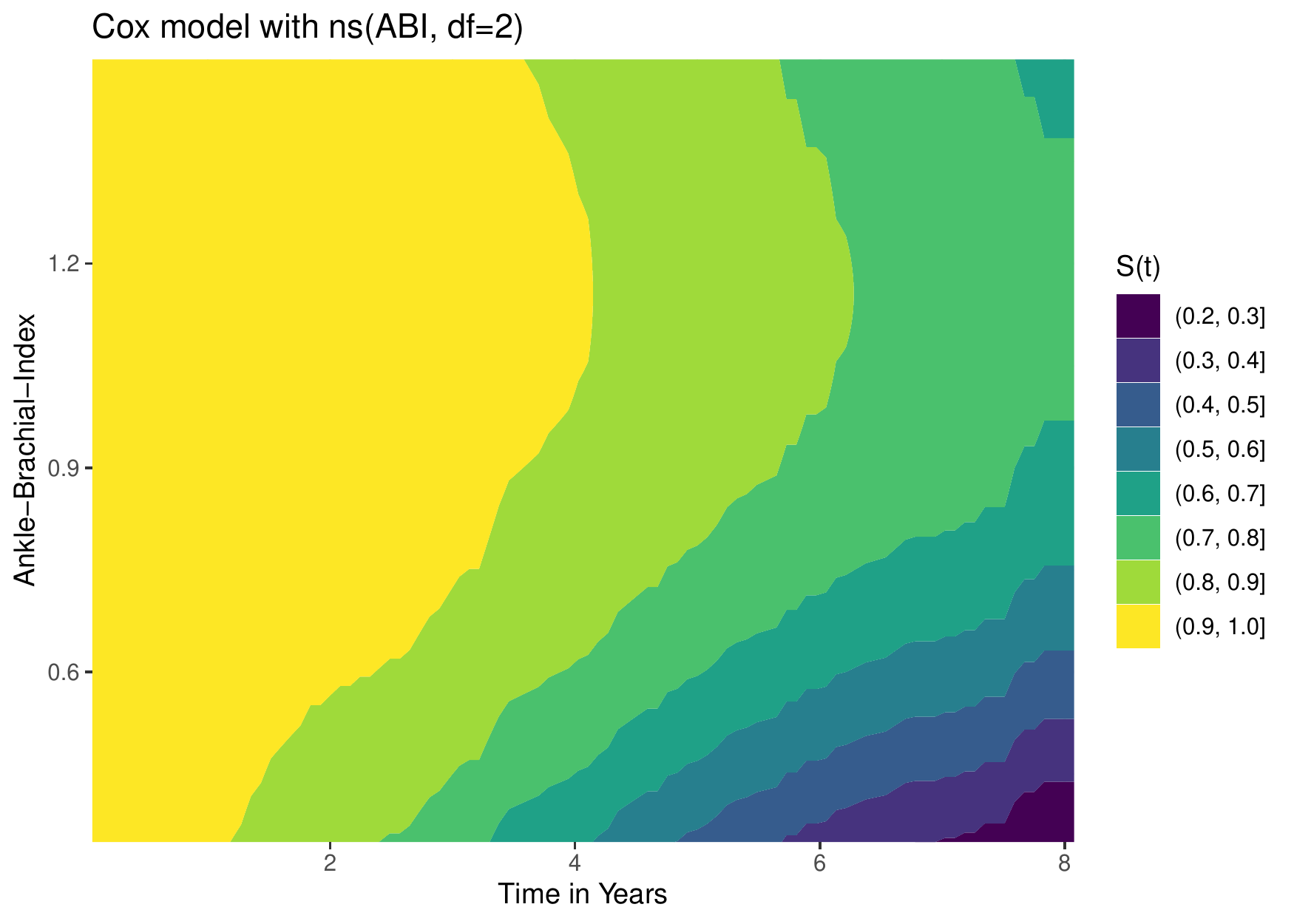}
	\caption{Survival contour plot created using a Cox model including the ABI modelled using natural splines with two degrees of freedom.}
	\label{fig::ff_ns_2}
\end{figure}

\begin{figure}[!htb]
	\centering
	\includegraphics[width=0.84\linewidth]{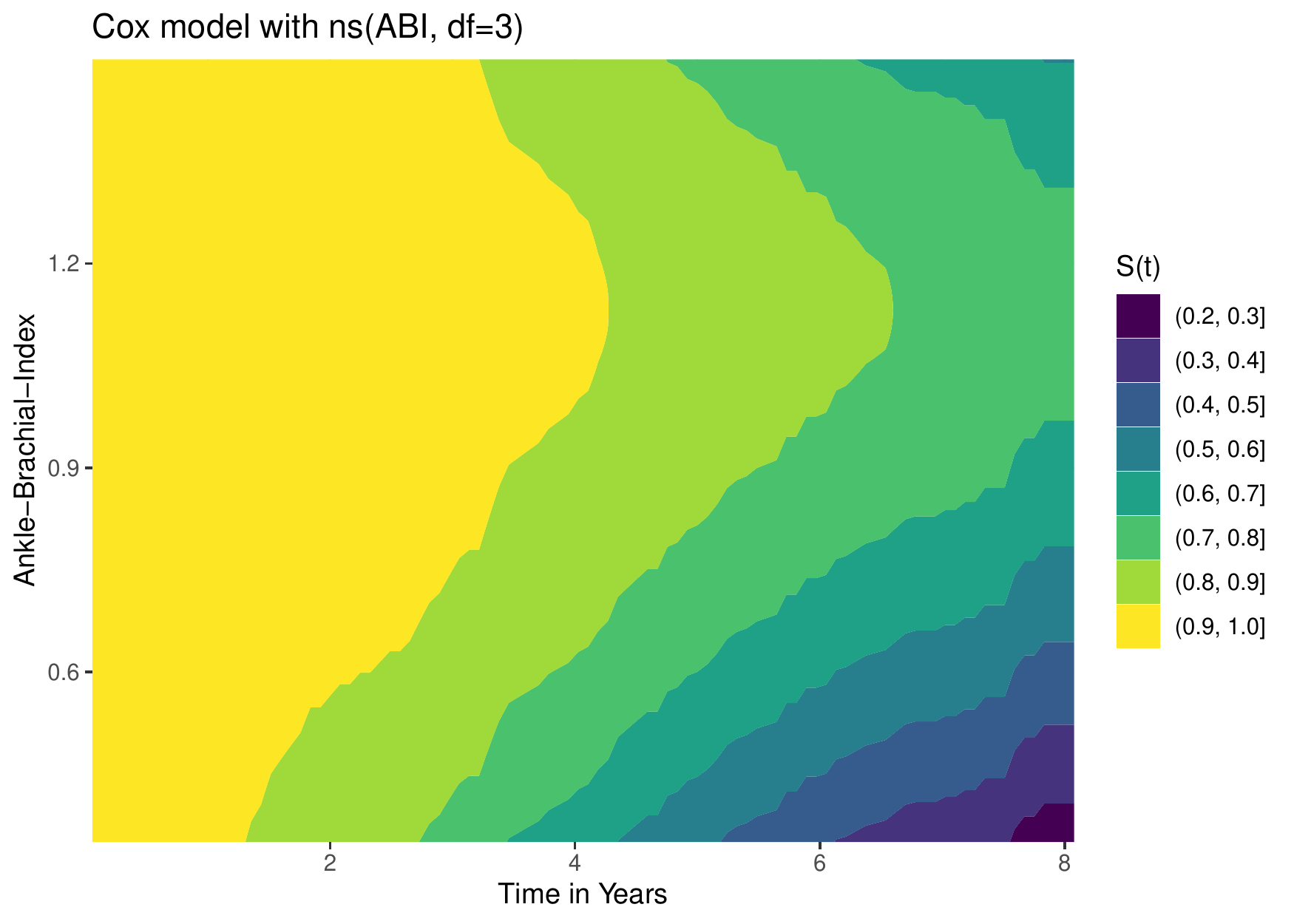}
	\caption{Survival contour plot created using a Cox model including the ABI modelled using natural splines with three degrees of freedom.}
	\label{fig::ff_ns_3}
\end{figure}

\begin{figure}[!htb]
	\centering
	\includegraphics[width=0.84\linewidth]{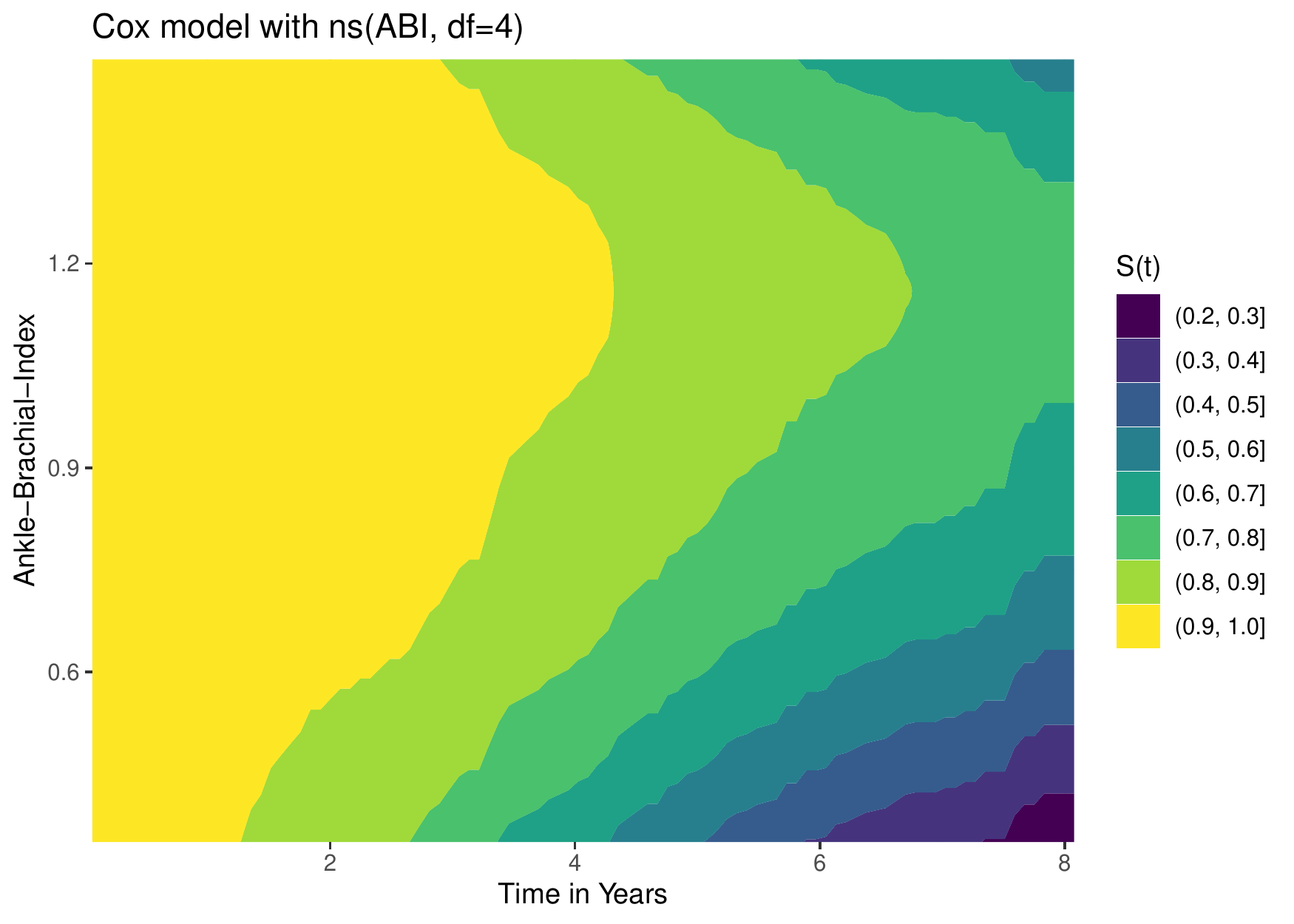}
	\caption{Survival contour plot created using a Cox model including the ABI modelled using natural splines with four degrees of freedom.}
	\label{fig::ff_ns_4}
\end{figure}

\begin{figure}[!htb]
	\centering
	\includegraphics[width=0.84\linewidth]{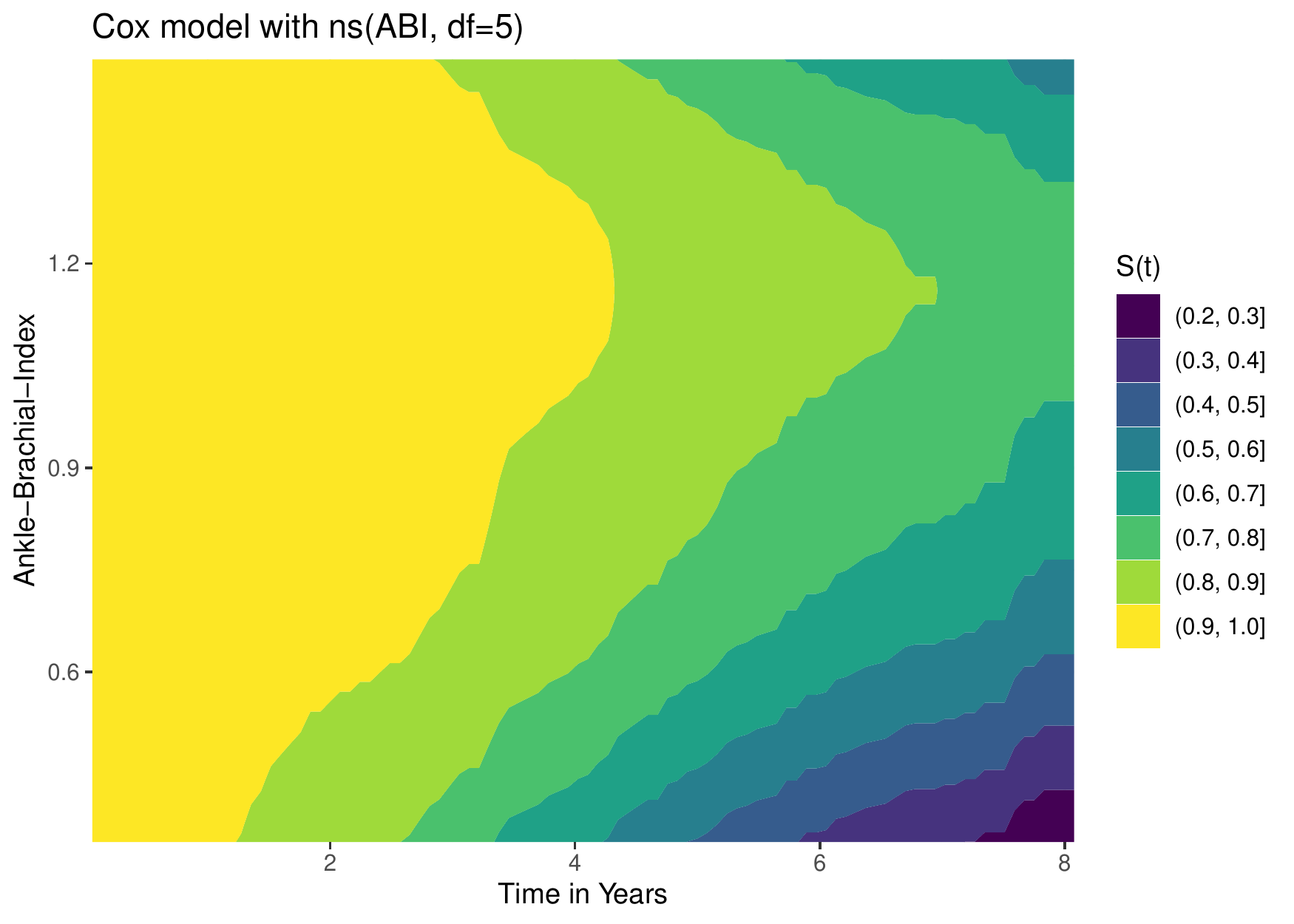}
	\caption{Survival contour plot created using a Cox model including the ABI modelled using natural splines with five degrees of freedom.}
	\label{fig::ff_ns_5}
\end{figure}

To choose the appropriate functional form of the relationship between the ABI and the survival time in our applied example, we relied on different graphical methods. First, we fitted an intercept-only Cox model and plotted the martingale residuals of this model against the observed values of the ABI using a LOESS smoother, as shown in figure~\ref{fig::ff_martingales}, which clearly indicated a non-linear relationship. Based on this graphical assessment we tried out different specifications, using simple polynomial terms (figures~\ref{fig::ff_cube_1} and \ref{fig::ff_cube_2}), using b-splines (figures~\ref{fig::ff_bs_3}--\ref{fig::ff_bs_5}) and using natural splines (figures~\ref{fig::ff_ns_2}--\ref{fig::ff_ns_5}) and compared the results graphically using survival contour plots. All of those methods resulted in clear non-linear relationships.

\end{document}